\documentstyle[11pt,epsfig]{article}
\newcommand{\E}{{\rm E}}

\newcommand \be  {\begin{equation}}
\newcommand \bea {\begin{eqnarray} \nonumber }
\newcommand \ee  {\end{equation}}
\newcommand \eea {\end{eqnarray}}
\begin{document}

\title{{\bf ``Nonlinear'' covariance matrix and portfolio theory
for non-Gaussian multivariate distributions}\thanks{We are grateful to
P.~Bak, P.M.~Bentler,
F.~Lacan, P.~Santa-Clara, B.~Urosevic and S.~Xu for helpful discussions.}}

\author{\bf D. Sornette\thanks{Institute of Geophysics
and Planetary Physics and Department of Earth and Space Science,
University of California, Los Angeles, California 90095 and
Laboratoire de Physique de la Mati\`ere Condens\'ee, CNRS UMR6622 and
Universit\'e des
Sciences, B.P. 70, Parc Valrose, 06108 Nice Cedex 2, France, E-mail:
sornette@cyclop.ess.ucla.edu}~,
P. Simonetti\footnote{now at
Servizio Risk Management, Banca Intesa, Via Boito 7, 20121 Milano,Italy}
\thanks{Department of Physics and Astronomy, University of Southern California,
Los Angeles, CA 90089-0484}~ and J.V. Andersen\thanks{Nordic Institute for
Theoretical Physics,
 Blegdamsvej 17, DK-2100 Copenhagen, Denmark}}

\date{University of California, Los Angeles \ \\ \ \\ {\normalsize First
Version: March 1998 \ \\ This Version: February 1999}}

\maketitle

\begin{abstract}
\noindent

This paper offers a precise analytical characterization of the distribution
of returns
for a portfolio constituted of assets whose returns are described by an
arbitrary joint multivariate distribution.
In this goal, we introduce a non-linear transformation that maps the
returns onto
gaussian variables whose covariance matrix provides
a new measure of dependence between the non-normal returns, generalizing
the covariance matrix into a non-linear fractional covariance matrix.
This nonlinear covariance matrix
is chiseled to the specific fat tail structure of the underlying marginal
distributions, thus
ensuring stability and good-conditionning. The portfolio distribution is
obtained
as the solution of a mapping to a so-called $\phi^q$ field theory in
particle physics, of which
we offer an extensive treatment using Feynman diagrammatic techniques and
large deviation
theory, that we illustrate in details for multivariate Weibull distributions.
The main result of our theory is that
minimizing the portfolio variance (i.e. the relatively ``small'' risks)
may often increase the large risks, as measured
by higher normalized cumulants. Extensive
empirical tests are presented on the foreign exchange market that validate
satisfactorily the theory. For ``fat tail'' distributions, we show that an
adequete prediction
of the risks of a portfolio relies much more on the correct description
of the tail structure rather than on their correlations.

\end{abstract}

\thispagestyle{empty}

\pagenumbering{arabic}

\newpage
\tableofcontents

\newpage

\section{Introduction}

Many problems in Finance, including risk management, optimal asset
allocation and
derivative pricing, require an understanding of the volatility and correlations
of assets returns. In practice, volatility and correlations are often
estimated from
historical data and the risk dimension is represented by the variance or
volatility
for a single asset and by the covariance matrix for a set of assets.
In portfolio optimization, the variance of the distribution of returns is
minimized
by inverting the covariance matrix of returns. However, the covariance
matrix is often
ill-conditionned and unstable [Litterman and Winkelmann, 1998].
There are several origins to this unstable behavior,
in particular the non-normality of the asset returns, i.e. large price
variations
are more probable than extrapolated from a normal estimation. For L\'evy or
power law
distributions with index less than two, the covariance matrix is not even
defined.

Option pricing and hedging relies on representing risk by a single scalar
measure, the volatility. In practice, the volatility has the complexity of
a time-dependent
fluctuating surface defined as a function of
the strike price and the time-to-maturity. Again, an important
origin of this complexity stem from the fat tail structure of the underlying
distributions as well as their non-stationarity.

The need for models that go beyond the Gaussian paradigm is vividly felt by
practionners,
regulatory agencies and is also advocated in the academic literature. Let
us mention
G\'eczy [1998] who proposes to use
non-normal multivariate distributions to
better assess factor-based asset pricing models\,:  indeed, some factor models
are strongly rejected when relying on the presumption that returns or model
residuals
are independent and identically distributed multivariate normal, while they
are no longer
rejected when fatter tail elliptic multivariate distributions are used.

The non-Gaussian nature of empirical return distributions
has been first addressed by generalizing the normal hypothesis to the
stable L\'evy law hypothesis [Mandelbrot, 1963,1997; Fama, 1965].
Gaussian and L\'evy laws are stable distributions under
convolution  and enjoy simple additivity properties
that allowed the generalization of Markovitz's porfolio theory in
a natural way [Samuelson, 1967; Arad, 1975; Bawa et al., 1979].
More recently, a further generalization
has been performed [Bouchaud et al., 1998]
to  situations where the marginal distributions of asset returns may have
different
power law behaviors for large positive and negative returns, with arbitrary
exponents
(possibly larger than $2$, i.e. not stable in the sense of L\'evy laws but
only in
a sense of large deviation theory).

This stable or quasi-stable property enjoyed by L\'evy and power laws, that
are instrumental in the generalization of Markovitz's theory,
presents however rather stringent restrictions. Indeed, these laws constitute
the only solutions obeying the consistencey property [Kano, 1994],
according to which
any marginal distribution of random vectors whose distribution belongs to a
specific
family also belongs to the same family. This consistency property is important
for the independence of the marginal variances on the order (number of assets
constituting a porfolio, for instance) of the multivariate
distributions (see the appendix A).
Thus, the generalization of portfolio theory from Gaussian to L\'evy
and then to power laws relies fundamentally on the consistency property
[Kano, 1994].

Elliptical distributions [Cambanis et al., 1981] provide a priori maybe the
simplest
and most natural hope for descrihing fat tails and for
generalizing portfolio theory to non-normal multivariate distributions,
solely on the
basis of the measure of a covariance matrix.
Elliptical distributions are defined as arbitrary density functions $F$ of the
quadratic form $(X-\langle X \rangle)' \Sigma^{-1} (X-\langle X \rangle)$,
where
$X$ is the unit column matrix of the $N$ asset returns and
$\Sigma$ is a $N \times N$ dependence matrix between the $N$ assets which
is proportional to
the covariance matrix, when it exists. It was argued that the results of the
CAPM extends to these elliptical distributions [Owen and Rabinovitch, 1983;
Ingersoll, 1987]
on the basis
that, since any linear transformation of an elliptical random vector is
also elliptical, the
risk measure $W' \Sigma W$ (where $W$ is the column vector of the asset
weights) is positively
proportional to the portfolio wealth variance. Thus the ranking of
portfolios by risk adverse
investors appears to retain their ordering.
However, Kano [1994] calls attention to the fact that, in
general, the density of a marginal distribution has not the same shape as
the function
$F$. This leads to an additional mispecification of the risk since it
should be
in principle captured both by the covariance matrix $\Sigma$ and the shape
and tail
structure of the distribution. It is thus not possible to quantify the risk
solely
by $W' \Sigma W$. In practice, the marginal variance of one of the
variable will depend on the number $N$ of assets used in the portfolio.
This is
clearly an unwanted property for the chosen multivariate distribution.
This inconsistency never arises for special cases such as the
multivariate t-distribution or for mixtures of normal distributions. In
appendix A,
we clarify the origin of the problem and show that, for arbitrary elliptic
distributions, the portfolio return distribution depends on the asset
weights $P$,
not solely via the quadratic form  $W' \Sigma W$\,: the weights $W$ also
control the shape as a whole of the portfolio return distribution
[Sornette, 1998].
Minimizing the variance $W' \Sigma W$ of the portfolio may actually
distord the portfolio return distribution such as to increase the
probability weight
far in the tail, leading to increased large risks. This result provides a
first cautionary note on any attempt to calibrate the portfolio risks by a
single
volatility measure.

This paper aims precisely at offering a novel approach to address
these questions. Our key
idea is to use a representation that strives to remain as parsimonious as
the Gaussian
framework while capturing the non-Gaussian ``fat tail'' nature of marginal
distributions and the
existence of nonlinear dependence. For this, a non-linear transformation
maps the returns onto
Gaussian variables whose covariance matrix provides
a new measure of dependence between the non-normal returns, generalizing
the covariance matrix into a non-linear fractional covariance matrix.
In our approach, the linear matrix calculation
of the standard portfolio theory is replaced by
Feynman's diagram calculations and large deviation theory
resulting from the fact that the portfolio wealth is
a nonlinear weighted sum of the Gaussian variables. Notwithstanding this
higher sophistication, all results can be derived analytically and thus can be
fully controlled. In the present
paper, we focus on the risk component of the problem, i.e. we assume symmetric
return distributions. In other words, the average returns are sufficiently
close to zero
that the fluctuations dominates. Even if not true in reality, this
approximation provides
a precise representation of real data at sufficiently small time scales,
corresponding
to Sharpe ratios less than one. This situation is reasonably accurate for
non-stock market
assets, which do not have a long-term trend. To test and validate our
theory, we
thus use empirical data from the Foreign exchange. We leave for a future
work the extension of
our theory to the cases where the average return cannot be neglected.

The paper is organized as follows. In section 2, we present our novel
method for
a full analytical characterization of the distribution of returns
for a portfolio constituted of assets with an arbitrary multivariate
distribution.
In this goal, we introduce a new measure of dependence between non-normal
variables, which
generalizes the covariance matrix. The proposed nonlinear covariance matrix
is chiseled to the specific fat tail structure of the underlying marginal
distributions, thus
ensuring stability and good-conditionning. In our formulation, the
multivariate
distribution is normal in terms of a set of reduced variables that are the
natural quantities to
work with. In section 3, we show how the calculation of
the distribution of returns for an arbitrary portfolio can be
formulated in terms of a functional integral approach. In the appendices,
we provide an extensive
synthesis of the tools that allow us to determine analytically the full
distribution of
returns and work out in details the calculations for multivariate Weibull
distributions.
In section 4, we use the information captured by the distribution of
returns of arbitrary portfolios to compare them with respect to their
moderate versus large risks. We develop especially our analysis for
uncorrelated assets with ``fat tails'' for
which we compare different portfolios such as the ones that minimize the
variance,
the excess kurtosis or higher normalized cumulants. We also compare these
portfolios with
those determined from the maximization of the expected Utility.  We show that
minimizing the portfolio variance (i.e. the relatively ``small'' risks)
may often increase the large risks, as measured by higher normalized
cumulants. In section 5, we offer
practical implementations and comparisons with data obtained from the Foreign
exchange market. We validate clearly the performance of the proposed nonlinear
representation. The comparisons between our theoretical calculations and
empirical estimations of the
cumulants of the portfolio return distributions are very satisfactory. We
find in
particular that, for a good representation of the risks of a portfolio,
tracking the correlations between assets is much less important
than precisely accounting for the ``fat tail'' structure of the distributions.
Section 6 concludes with future extensions.

\section{Non-linear mapping to normal distributions}

The first step of our approach is to perform a non-linear change of
variable, from the
return $\delta x$ of an asset over the unit time scale $\tau$ (taken as the
daily scale
for illustration below) onto the
variable $y(\delta x)$ such that the distribution of $y$ is normal. For
marginal
distributions, this is always possible. We first illustrate this change of
variables in simple cases and then proceed to the general multivariate
situation.

\subsection{The case of a single security}

\subsubsection{Non-centered case}

Let us consider the class of marginal distribution (probability density
function or pdf)
$p(\delta x)$ which can be parameterized as
\be
p(\delta x) d\delta x = C {f'(\delta x) \over \sqrt{|f(\delta x)|}} ~e^{-{1
\over 2}
f(\delta x)}~d\delta x~,
\label{ezrrsffgsghhs}
\ee
where $C$ is a normalizing constant and
$f'$ denotes the derivative of the function $f(\delta x)$. Note that $f$
must go to
$+\infty$ for $|\delta x| \to \infty$) at least logarithmically to ensure
normalization.
This distribution (\ref{ezrrsffgsghhs}) describes so-called Von Mises variables
[Embrechts et al., 1997]. This parameterization (\ref{ezrrsffgsghhs})
covers all cases where the pdf has a single maximum.
These are the relevant situations for essentially all securities. Note that
for $f(\delta x)/(\delta x)^2 \to 0$ for large $|\delta x|$, the pdf has
``fat tails'',
i.e. the pdf decays slower than a Gaussian for large $|\delta x|$.

Let us then define the variable $y$ by
\be
y = \rm{sign}(\delta x)~ \sqrt{|f(\delta x)|}~.
\label{rtgeejj}
\ee
The sign function is essential in order to keep possible correlations.
In the case of fat tails for which $\sqrt{|f(\delta x)|} << |\delta x|$ for
large
$|\delta x|$, the variable $y$ is the result of a ``contraction'' of
$\delta x$ (such that its pdf is normal).
Indeed, the pdf of $y$ is such that the conservation $p(y) dy = p(x) dx$ of
probability
holds and we obtain
\be
p(y) dy = {1 \over \sqrt{2 \pi}}~e^{-{y^2 \over 2}}~dy~.
\label{zaaqqqqw}
\ee
Obviously, the form (\ref{ezrrsffgsghhs}) has been chosen such that a change of
variable recovers the normal pdf through a simple transformation. We
stress that this parameterization
(\ref{ezrrsffgsghhs}) is fully general as any unimodal pdf can be put in
this form.

\subsubsection{Example\,: the modified Weibull distribution}

Consider the case of the modified Weibull
distribution where
\be
f(\delta x) = 2 \biggl({|\delta x| \over \chi}\biggl)^c~,
\label{qkjqklkqqq}
\ee
and thus
\be
p(\delta x) d\delta x = d\delta x~{1 \over 2\sqrt{\pi}}
{c \over \chi^{c \over 2}}~|\delta x|^{{c \over 2}-1} ~~
e^{-({|\delta x| \over \chi})^c}~~~~~{\rm for}~~
-\infty < \delta x < +\infty~.
\label{aeqgdfbbzre}
\ee
The case, where the exponent $c<1$, corresponds to a ``stretched''
exponential with a tail fatter
than an exponential and thus much fatter than a Gaussian, but still thinner
than a
power law. Stretched exponential pdf's have been found to provide a
parsimonious and
accurate fit to the full range of currency price variations at daily
intermediate time scales [Laherr\`ere and Sornette, 1998]. This
stretched exponential model is also validated theoretically by the recent
demonstration that
the tail of pdf's of products of a finite number of random variables
is generically a stretched exponential [Frisch and Sornette, 1997], in
which the exponent $c$ is
proportional to the inverse of the number of generations
(or products) in a multiplicative process.

In this case (\ref{qkjqklkqqq}), the change of variable (\ref{rtgeejj})
corresponds to
\be
y = \rm{sign}(\delta x)~|\delta x|^{c \over 2}~.
\label{rtgbn,jj}
\ee
For $c < 2$, $y(\delta x)$ has a negative (resp. positive) curvature for
$\delta x >0$
(resp. for $\delta x <0$), i.e. is
concave  (resp. convex). This convexity reflects the contracting
nature of the mapping that allows to obtain a normal distribution,
with a standard deviation $\xi = \chi^{c \over 2}$.
This change of variable (\ref{rtgeejj}) has already been briefly mentionned
in the literature.
In a footnote of [Jorgensen, 1987], we find mentionned that, for distributions
(\ref{aeqgdfbbzre}) with $c>2$, the change of variable (\ref{rtgbn,jj}) gives
a normal distribution. To our knowledge, the case $c<2$ has not been
investigated.

The pdf (\ref{aeqgdfbbzre}) is called ``modified'' Weibull distribution,
because it is
slightly different by the distinct power in the pre-exponential factor
from the standard Weibull pdf defined by
\be
p_W(\delta x) \sim  |\delta x|^{c-1} ~~
e^{-({|\delta x| \over \chi})^c}~,
\ee
such that its cumulative distribution is exactly $e^{-({|\delta x| \over
\chi})^c}$. This
difference accounts for the fact that the change of variable
(\ref{rtgbn,jj}) maps exactly the
modified Weibull pdf onto a Gaussian pdf. This property does not hold
exactly for the
standard Weibull distribution. Notice that the modified Weibull
distribution with $c=2$ is
nothing but the Gaussian distribution. This provides another incentive in
this definition
(\ref{aeqgdfbbzre}) as it retrieves exactly the Normal law as one of its
member.
From now on, we drop the term ``modified'' as we will only
study (\ref{aeqgdfbbzre}).

The Weibull distribution (\ref {aeqgdfbbzre})
is not stable under convolution, i.e. the distribution of weekly
returns is not exactly of the same form as the distribution of the daily
returns. However,
it presents an approximate stability in the sense that it is possible to define
an effective exponent $c_T$ and scale $\chi_T$ such that
the tail of the pdf of the distribution of return over $T$ days
is of the form (\ref {aeqgdfbbzre}) with $c$ replaced by $c_T$ and $\chi$
by $\chi_T$.
The analytical procecedure to derive this result and numerical tests are given
in the appendix B.

\subsubsection{Centered case}

We generalize (\ref{ezrrsffgsghhs}) further by the following
parameterization\,:
\be
p(\delta x) d\delta x = C {f'(\delta x) \over \sqrt{|f(\delta x)|}} ~e^{-{1
\over 2}
(\rm{sign}(\delta x)~\sqrt{|f(\delta x)|} - m)^2} ~d\delta x~,
\label{ezrrszgs}
\ee
where $m = \E[(\rm{sign}(\delta x_a)~\sqrt{|f(\delta x)|}] $ and $\E[x]$
is the expectation of $x$.
The parameterization
(\ref{ezrrszgs}) is such that, by the change of variable (\ref{rtgeejj}),
expression (\ref{ezrrszgs}) transforms into a standard centered Gaussian
distribution
\be
p(y) dy = { dy \over \sqrt{2\pi}}~~e^{-{(y - m)^2 \over 2}}~.
\ee

For the Weibull (stretched exponential) (\ref{aeqgdfbbzre}), this
corresponds to
\be
p(\delta x) d\delta x = d\delta x~{1 \over 2\sqrt{\pi}}
{c \over \chi^{c \over 2}}~|\delta x|^{{c \over 2}-1} ~~
e^{-{(\rm{sign}(\delta x)~|\delta x|^{c \over 2} - m)^2 \over
\chi^c}}~~~~~{\rm for}~~
-\infty < \delta x < +\infty~,
\label{arre}
\ee
where $m = \E[(\rm{sign}(\delta x)~|\delta x|^{c \over 2}]$.
The form (\ref{arre})
of the Weibull exponential corresponds to the following cumulative
distribution function
\footnote{The complementary error function is defined as
${\rm erfc}(u) \equiv {2\over \sqrt{\pi}} \int_u^\infty dv ~\exp(-{v^2})$,
and has the asymptotic behavior
${\rm erfc}(u) \to {1 \over \sqrt{\pi}} ~{1 \over u} ~e^{-u^2}$ for large
$u>0$.}
\be
P_<(\delta x) = {1 \over 2}~{\rm erfc}\biggl({|\delta x|^{c \over 2} + m
\over \chi^{c \over 2}}
\biggl)~.
\label{redddd}
\ee

\subsection{General case of several assets}

Let us assume that the returns of $N$ assets over a period of
$T$ time intervals are measured\,: $\delta x_1^{(1)}, \delta
x_2^{(1)},...,\delta x_T^{(1)}$
for the first asset, $\delta x_1^{(2)}, \delta x_2^{(2)},...,\delta x_T^{(2)}$
for the second asset, up to $\delta x_1^{(N)}, \delta x_2^{(N)},...,\delta
x_T^{(N)}$
for the last asset. From these, the $N$ marginal distributions
of the returns for each asset are determined empirically either using a
parametric
or non-parametric representation. Call $P_j(\delta x)$ the marginal
distribution
of the $j$th asset. In general, $P_j(\delta x)$ is non-gaussian and
exhibits fat tails,
for instance described by a Weibull distribution. But we do not restrict
ourselves to this special
case and consider a general form for the $P_j(\delta x)$'s.

\subsubsection{Non-linear mapping}

As for the one security case discussed previously using (\ref{rtgeejj}),
each marginal distribution $P_j(\delta x)$ is mapped onto a gaussian
distribution by
performing a change of variable that is specific to each marginal
distribution.
We discuss here the
symmetric case.
For a given asset $j$, the new variable is such that
\be
P_j(\delta x) d \delta x = {1 \over \sqrt{2 \pi}} ~e^{-{y^2 \over 2}}~dy~.
\label{thjfklkl}
\ee
By construction, the variable $y$ follows
a Gaussian distribution of mean~$0$ and variance~$1$. Introducing the
cumulative distribution
$F_j(x)$
\begin{equation}
F_j(\delta x) = \int_{-\infty}^{\delta x} P_j(x')\,dx' ~,
\label{EQ:cumul}
\end{equation}
the differential equation (\ref{thjfklkl}) for $y(\delta x)$
is integrated into a general implicit equation
giving $y$ as a function of $\delta x$\,:
\be
F_j(\delta x) = {1 \over 2} \biggl[1 + {\rm erf}\left({y \over \sqrt{2}}
\right)\biggl]~.
\label{ghfkqwl}
\ee
For a Weibull distribution (\ref{aeqgdfbbzre}), expression (\ref{ghfkqwl})
reads
\be
e^{-({|\delta x| \over \chi})^c} = {\rm erfc}({y \over \sqrt{2}})~.
\ee

We can rewrite (\ref{ghfkqwl}) as
\begin{equation}
y(\delta x)=\sqrt{2}~~{\rm erf}^{-1}\left(2F_j(\delta x)-1 \right)
\label{EQ:transform}
\end{equation}
where erf$^{-1}$ is the inverse of the error function. If $P_j(\delta x)$
is of the form
(\ref{ezrrsffgsghhs}), the solution of (\ref{EQ:transform}) retrieves the
expression
(\ref{rtgeejj}).

We perform this change of variable
(\ref{thjfklkl},\ref{ghfkqwl},\ref{EQ:transform})
for each of the $N$ distributions
$P_j(\delta x)$. This leads to the transformed measurements $y_1^{(1)},
y_2^{(1)},...,y_T^{(1)}$
for the first asset, $y_1^{(2)}, y_2^{(2)},...,y_T^{(2)}$
for the second asset, up to $y_1^{(N)}, y_2^{(N)},...,y_T^{(N)}$
for the last asset.

\subsubsection{Nonlinear covariance matrix}

Having thus defined the variables $y^{(j)}$ ($j=1...N$) from the $N$ assets,
we construct the covariance matrix
\be
V = \E[{\bf y} ~{\bf y}'] ~,
\label{klgoioc}
\ee
with element $V_{ab}$ defined by
\be
V_{ab} = {1 \over T} \sum_{i=1}^T  y_i^{(a)}~y_i^{(b)}~.
\label{iklksfksl}
\ee
In (\ref{klgoioc}), we note for short
${\bf y}$ as the unicolumn matrix of size $N$ with elements $y^{(j)}$ equal
to some
realization.

For distributions that can be parameterized as (\ref{ezrrszgs}),
we have seen that the change of
variable (\ref{EQ:transform}) becomes expression (\ref{rtgeejj}).
The covariance matrix (\ref{klgoioc}) of the set of $y_a$'s then
corresponds to the following dependence
matrix for the asset returns $\delta x_a$\,:
\be
V_{ab} \equiv \E\biggl[ \biggl(\rm{sign}(\delta x_a) \sqrt{|f(\delta x_a)|}
- m_a\biggl)~\biggl(
\rm{sign}(\delta x_b)~\sqrt{|f(\delta x_b)|}  - m_b\biggl) \biggl]~,
\label{defeedqagr}
\ee
where
$m_a = \E[\rm{sign}(\delta x_a) ~\sqrt{|f(\delta x_a)|}]$. The
distribution of the $y_a$ is multivariate Gaussian with covariance matrix
$V_{ab}$.

The covariance matrix (\ref{klgoioc},\ref{defeedqagr}) measures the
covariance between
the assets described by the ``effective'' returns $y$. This definition
(\ref{klgoioc},\ref{defeedqagr}) amounts to a non-standard measure of the
covariance
in terms of the variables $\delta x$'s. We use
the name ``nonlinear covariance'' to recall that the
change of variable $\delta x \to y$ for fat tail pdf's $P_j(\delta x)$ lead
to a contraction,
for instance a concave power law (\ref{rtgbn,jj}) for $c<2$. In this case,
the covariance
of $y$ amounts to estimate the covariance of fractional powers of the returns.

\subsubsection{Multivariate representation of the joint distribution of
asset returns}

No approximation has been made until now. We have characterized the
marginal distributions of the assets, defined new variables in terms of
which the marginal
distributions are Gaussian and we have calculated the covariance matrix of
these
new variables as the new measure of the dependence between the asset returns.

Here comes the simplifying approximation. If the marginal distributions in
terms of the
new variables $y^{(j)}$ are gaussian, nothing imposes a priori that the
multivariate
distribution of these variables is also a multivariate Gaussian distribution.
From our construction, it is only guaranted that the projections of the
multivariate
distribution onto each $y^{(j)}$ variable are Gaussian.

However, a standard theorem from Information Theory [Rao, 1973] tells us that,
conditioned on the sole knowledge of the covariance matrix,
the best representation of a multivariate distribution is the Gaussian. In
other words, the multivariate normal distribution contains the least possible
a priori assumptions in addition to the covariance matrix (and the vector
of the
means when they are non-zero), i.e. is the most likely representation of
the data.

This implies that, conditionned on the sole knowlege of the covariance
matrix (\ref{klgoioc},\ref{iklksfksl}), the best parametric, although not
exact,
representation of
the multivariate distribution ${\hat P}({\bf y})$ is
\begin{equation}
  {\hat P}({\bf y}) = (2\pi)^{-N/2}\,\vert V\vert^{-1/2}
      \exp\left({-{{\textstyle{1\over2}}}\, {\bf y}'\, V^{-1}\, {\bf y}}
\right)~,
  \label{EQ:ngaus}
\end{equation}
where $\vert V \vert$ is the determinant of $V$.

Note that our ``nonlinear covariance'' approach
is very different from the usual approximation which assumes that the asset
returns
$\delta x_i^{(j)}$ themselves follow a multivariable Gaussian distribution.
In particular, the non-linear change of variable ensures that the fat tail
structure
of the marginal distribution is fully described and, in addition, leads to
a novel more stable measure of the covariances. The normal structure of the
multivariate distribution in terms of the $y$ variables does {\it not} lead
to a Gaussian multivariate distribution for the returns.

From the expression (\ref{EQ:ngaus}), we obtain
the explicit expression of the multivariate distribution of the asset
returns by
replacing the $y_i^{(j)}$ as a function of
$\delta x_i^{(j)}$ as given by (\ref{EQ:transform}) for each asset $j$ and use
the identity
\be
P({\bf x}) = {\hat P}({\bf y}) ~{d{\bf y} \over d{\bf x}}~,
\ee
where ${d{\bf y} \over d{\bf x}}$ is the jacobian of the transformation from
${\bf x} \to {\bf y}$. The Jacobian is the determinant of a diagonal matrix
whose
$j$th element is simply
\be
{dy^{(j)} \over d \delta x^{(j)}} = \sqrt{\pi}~P_j(\delta x^{(j)})
e^{(y^{(j)})^2/2}
\ee
as seen from (\ref{EQ:transform}). This finally yields the
following approximation for the multivariate distribution $P({\bf x})$
of the column vector ${\bf x}$ of $N$ returns of a given realization of the
$N$ asset returns\,:
\be
P({\bf x})= \vert V \vert^{-1/2}
          \exp\left(-{{\textstyle{1\over2}}}\,{\bf y}'\,(V^{-1}-I)\,{\bf y}
              \right)
              \prod_{j=1}^N P_j(x^{(j)})~,
\label{EQ:pca}
\ee
where $V$ is again the covariance matrix for ${\bf y}$ and $I$ is the
identity matrix.

This ``nonlinear covariance'' approximation is exact for distributions
with uncorrelated variables, in which case $V=I$.
It is also exact for a Gaussian distribution modified by
monotonic one-dimensional variable transformations for any
number of variables; or equivalently, multiplication by a non-negative
separable function.
Note that the multivariate distribution (\ref{EQ:pca}) obeys automatically
the condition
that the corresponding marginal distributions \footnote{i.e. the
monovariate unconditional
distributions derived from the multivariate distribution by unconditionally
integrating over
all variables except one.} are of the same analytic form. This
corresponds to the consistency condition [Kano, 1994] discussed in the
introduction.
This approach leading to (\ref{EQ:pca}) has also been introduced
independently for the
analysis of particle physics experiments by Karlen [1998].

\subsubsection{Multivariate Weibull distributions}

Let us now apply this approach to Weibull distributions.
The $N$ assets are assumed to have their returns $\delta x_a$ distributed
according to the {\it unconditional} marginal distributions given by
expression (\ref{ezrrszgs}).
With the change of variable (\ref{rtgeejj}), we construct
the covariance matrix $V$ defined by (\ref{defeedqagr}).
This defines a signed and fractional dependence matrix for the asset returns.
Since $V_{aa} = \E[ [(\delta x)^2]^{c \over 2}]$, we see that $[V_{aa}]^{2
\over c}$ has the
same ``dimension'' as the usual variance but is fundamentally different in that
\be
\E\biggl[(\delta x)^2\biggl]^{c \over 2} \neq
 \E\biggl[ [(\delta x)^2]^{c \over 2}\biggl]~,~~~~~~{\rm in~general}~.
 \ee
The expectation of a power is in general different from the power of the
expectation.

The multivariate Weibull exponential distribution of the $N$ asset returns
is then given by
expression (\ref{EQ:pca}) that we write explicitely
$$
P(\delta x_1, \delta x_2, ..., \delta x_N) \prod_{i=1}^N d\delta x_i
\propto \prod_{i=1}^N \biggl({c \over 2} |\delta x_i|^{{c \over 2}-1}~
d\delta x_i \biggl)
$$
\be
\exp \biggl(-{1 \over 2} ~
\sum_{i=1}^N \sum_{j=1}^N  ~V^{-1}_{ij} [\rm{sign}(\delta x_i)~|\delta
x_i|^{c \over 2}-m_i)~
(\rm{sign}(\delta x_j)~|\delta x_j|^{c \over 2} -m_j)\biggl) ~.
\label{dispqgqdgaqooob}
\ee
This expression obeys the requirement that
the marginal (monovariate) distributions are Weibull exponentials.

\section{The distribution $P_S(\delta S)$ of portfolio wealth variations}

\subsection{Theoretical formulation}

We now restrict our study to
symmetric multivariate distributions. As a consequence, all odd
moments (in particular the first moment which is the expected return)
and all odd cumulants are vanishing. This means that we are focusing on the
aspect of risk
embedded in the variance and in all higher even order cumulants. There is
no trade-off between
risk and return since the expected return is zero. We thus focus
exclusively on the risk dimension
of the portfolio. In another paper, we will
present our results obtained for non-symmetric distributions. The results
presented
below are already sufficiently rich that we feel compelled to separate the
discussion of
symmetric and non-symmetric distributions so as not to make the presentation
unecessarily cumbersome.

To a very good approximation, it is harmless and much simpler to replace
the returns $\delta x_i(t)$ by $(p_{i}(t+1) - p_i(t))/p_i(t)$ where
$p_i(t)$ is the price of asset $i$ at time $t$. Over reasonable large time
intervals
(e.g a year), one can neglect the variation of the denominator in comparison
to the variation of the numerator $\delta p_i(t) \equiv p_{i}(t+1) - p_i(t)$.

The total variation of the value of the portfolio made of $N$ assets between
time $t-1$ and $t$ reads
\be
\delta S(t) = \sum_{i=1}^N w_i \delta p_i(t) ~,
\label{sumport}
\ee
where $w_i$ is the weight of the $i$th asset
in the portfolio. By normalization, we have $\sum_{i=1}^N w_i = 1$.

In terms of the variables $y_i$'s defined by (\ref{rtgeejj}) (resp.
(\ref{rtgbn,jj}) for the Weibull exponential case), the
expression (\ref{sumport}) reads
\be
\delta S(t) = \sum_{i=1}^N w_i ~\rm{sign}(y_i)~f^{-1}(y_i^2)~,
\label{sumpqgqqgort}
\ee
where f$^{-1}$ is the inverse of $f$ (suitably defined to avoid double
valuedness) and
respectively
\be
\delta S(t) = \sum_{i=1}^N w_i ~\rm{sign}(y_i)~\chi_i~ |y_i|^{2 \over c}~,
\label{sumpqgort}
\ee
for the Weibull distribution for which $f_i(\delta p_i) = |{\delta p_i
\over \chi}|^c$.

All the properties of the portfolio are contained in
the probability distribution $P_S(\delta S(t))$ of $\delta S(t)$. We would
thus like
to characterize it,
knowing the multivariate distribution of the $\delta p_i$'s (or
equivalently the multivariate
Gaussian distribution of the $y_i$'s) for the different assets. The
general formal solution reads
\be
P_S(\delta S) = C \prod_{i=1}^N \biggl( \int dy_i \biggl) ~ e^{-{1 \over 2}
~{\bf y}' V^{-1} {\bf y}}
~\delta\biggl(\delta S(t) - \sum_{i=1}^N w_i
\rm{sign}(y_i)~f^{-1}(y_i^2)\biggl) ~ .
\label{soqglut}
\ee
Taking the Fourier transform
$\hat P_S(k) \equiv \int_{-\infty}^{+\infty} d\delta S ~P_S(\delta S)~
e^{-ik \delta S}$ of
(\ref{soqglut}) gives
\be
{\hat P}_S(k) = \prod_{i=1}^N \biggl( \int dy_i \biggl) ~ e^{-{1 \over 2} ~
{\bf y}' V^{-1} {\bf y} +
ik~\sum_{i=1}^N ~w_i ~\rm{sign}(y_i)~f^{-1}(y_i^2)} ~ .
\label{soqqgglut}
\ee

In the sequel, we
present a systematic calculation method of (\ref{soqqgglut}), that can be
applied in principle to a large variety of functional forms for $f$.
In this paper, we restrict our analysis to the Weibull case, because it is
probably one
of the simplest non-trivial situation that allows us
to make apparent the power of this approach. In addition, as will be shown by
extended comparisons with empirical data, the Weibull representation provides
a very reasonable description of the ``fat tail'' structures of empirical
distributions.
For the Weibull case, the term $f^{-1}(y_i^2)$ is replaced by $\chi
|y_i|^{2 \over c}$.

A further simplification of notation occurs when the exponent $c$ is given by
\be
c=2/q ~, ~~~~~~{\rm with}~q~~{\rm integer~ and~ odd}~.
\label{jqkqllklw}
\ee
Then,
\be
{sign}(u_i)~|u_i|^{2 \over c} = {sign}(u_i)~|u_i|^{q} = u_i^{q}~,
\ee
i.e. the sign function disappears. Expression
(\ref{soqqgglut}) then becomes
\be
{\hat P}_S(k) = {1 \over (2\pi)^{N/2} \det{V}^{1/2} }
 \prod_{i=1}^N \biggl( \int dy_i \biggl) ~ e^{-{1 \over 2} ~{\bf y}'V^{-1}
{\bf y} +
ik~\sum_{i=1}^N ~w_i ~\chi_i y_i^q} ~ .
\label{soqqgglurft}
\ee
We can absorb the terms $\chi_i$ by introducing the new variable $u_i^q =
\chi y_i^q$,
which shows that the covariance elements (in the formulation where
the $\chi_i$ do not appear) are proportional to $\chi^{2/q}$.

Expressions (\ref{sumpqgqqgort},\ref{sumpqgort}) together with
(\ref{soqqgglurft}) exemplify the origin of the complexity introduced by the
non-Gaussian nature of the distributions, namely the portfolio wealth is a
nonlinear
function of the variables $y$ transformed from
the asset returns and its distribution is expressed as a non-Gaussian
multivariate
integral, which is a priori not obvious to estimate.
We notice however that (\ref{soqqgglurft}) is similar to mathematical objects
studied in another context: it is the same as
 the partition function of a $\phi^q$ field theory studied in particle physics,
with $N$ components and imaginary coupling coefficients $ikw_i$. {\bf y} is
like a Gaussian
field with interactions described by the second nonlinear term in the
exponential. Notice that
the case $q=3$
corresponding to an exponent $c=2/3$ is particularly interesting, because
this value for
$c$ seems realistic empirically (see below the empirical section)
and because the $\phi^3$ theory is the simplest non-trivial case leading to
fat tails.

We shall characterize the portfolio distribution $P(\delta S(t))$ by its
cumulants $c_n$
defined by
\be
\hat{P}_S (k) =  \exp\biggl(\sum_{m=1}^{+\infty} \frac{c_m}{m!} (i k)^m
\biggl)~.
\label{fgjqkkkw}
\ee
The first cumulant $c_1$ is the mean, the second cumulant $c_2 \equiv
E[(x-\E[x])^2]$
is the variance,
the third cumulant $c_3$ defines the skewness $c_3/c_2^{3/2}$, the fourth
cumulant $c_4$
defines the excess kurtosis $\kappa \equiv c_4/c_2^2$. \footnote{The
kurtosis is defined as
the ratio of the fourth centered moment over the square of the variance
$\E[(x-\E[x])^4]/\E[(x-\E[x])^2]^2 = 3 + c_4/c_2^2$. In this definition,
Gaussian distributions
have a kurtosis equal to $3$ while the excess kurtosis is zero.}
For Gaussian distributions, all cumulants of order larger than two are
zero. The cumulants
are thus convenient ways to quantify the departure from normality.
Furthermore,
provided mild regularity conditions hold, they are equivalent to the
knowledge of the full
distribution. Since we deal here exclusively with symmetric distributions,
all odd-order cumulants
are vanishing and it is sufficient to calculate the even-order cumulants.

\subsection{The Gaussian case $c=2$}

When $c=2$, the multivariate asset return distribution is Gaussian. The term
$\rm{sign}(y_i)~|y_i|^{2 \over c}$ becomes simply $y_i$
 and the integrals in
(\ref{soqqgglut}) are standard Gaussian integrals that can be evaluated
using the identity
\be
\prod_{i=1}^N (\int dx_i) \exp \biggl(-{1 \over 4} x_i A_{ij}^{-1} x_j +
y_i x_i \biggl) =
\sqrt{\pi^N \over det(A_{ij})} \exp (y_i A_{ij} y_j)~.
\label{identt}
\ee
Applied to (\ref{soqqgglut}) and after taking the inverse Fourier
transform, we get
\be
P(\delta S(t)) \propto \exp \biggl(-{[\delta S(t)]^2 \over 2 W' V W}\biggl)  ~.
\label{distriss}
\ee
which recovers the classical result that the distribution of portfolio
wealth variations
$\delta S$ is uniquely and fully characterized by the sole value of its
variance $c_2$,
expressed in terms of the covariance
matrix of the asset price returns and their weight $W$ in the portfolio
[Markovitz, 1959, Merton, 1990]\,:
\be
c_2 = W' V W~.
\ee

We now turn to the more general case
$c \neq 2$ for which the evaluation of the integral (\ref{soqqgglut}) is
much more involved.

\subsection{The case of uncorrelated assets}

We first consider the case where the covariance matrix is diagonal, which
corresponds to
an absence of correlations between assets. Then, the portfolio distribution is
only sensitive to the intrinsic risks presented by each asset.
We will show in the next subsection
how to implement a perturbative diagrammatic expansion in the general
correlated case.

For notational convenience, we adopt a change of variable which absorbs the
$\chi$'s in the exponential term of (\ref{soqqgglurft})\,: the term
$\sum_{i=1}^N ~w_i ~\chi_i y_i^q$ becomes $\sum_{i=1}^N ~w_i ~ y_i^q$. As a
consequence, the new $y_i$ variables acquire a non-unit variance equal to
$\chi_i^c$,
with $c=2/q$.

In the diagonal case, the covariance matrix is given by
\be
V={\rm diag}\{d_i\}~, ~~~~~~~{\rm where}~d_i=\chi_i^{2/q}~.
\label{fjqkmqmkqmnqm}
\ee
The relationship between the variances of these $y$ variables and the variances
of the asset returns is obtained by noting that $d_i^q = \chi_i^2$ which is
proportional to
the asset return variance. This illustrates that $d_i$ plays the role of a
``nonlinear variance''.

\subsubsection{Cumulants}

In the diagonal case,
the multiple integral over the multiple asset contributions in
(\ref{soqqgglurft})
becomes the product of single integrals of the kind
\be
I_i = \int_{-\infty}^{+\infty} du ~ e^{-\frac{u^2}{2 d_i}+ i k w_i u^{q_i}}~.
\ee
We consider the general case where we allow the exponent $c_i=2/q_i$ to
vary from asset
to asset.
From this integral, we derive in appendix C the exact and explicit
expression of the
general $2r$-th cumulant of $P(\delta S)$\,:
\be
c_{2 r} =  \sum_{i=1}^N C(r,q_i)(w_i^2 d_i^{q_i})^{r}~,
\label{gghJD/D}
\ee
where
\be
C(r,q) = (2 r)!~2^{q r} \left\{\sum_{n=0}^{r-2} (-1)^n
\frac{\Gamma\left((r-n)q+\frac{1}{2}\right)}{(2r-2n)! \pi^{1/2}}
\left[\frac{\Gamma\left(q+\frac{1}{2}\right)}{2!\pi^{1/2}}\right]^n\,\,\,
- \frac{(-1)^r}{r}
\left[\frac{\Gamma\left(q+\frac{1}{2}\right)}{2!\pi^{1/2}}\right]^r\right\}~,
\l
abel{jfzjmg}
\ee
and $\Gamma$ denotes the Gamma function.

The expression (\ref{gghJD/D}) is
valid even when the $q_i$'s are real and the interaction term in
(\ref{soqqgglurft}) is proportional to ${\rm sign}(u_i) |u_i|^{q_i}$ and thus
applies to arbitrary Weibull distributions.

For $r=1$ (variance) and $r=2$ (fourth order cumulant), the expression
(\ref{jfzjmg}) gives
\be
C(1,q) = (2^q/\sqrt{\pi}) \Gamma(q+1/2)
\ee
\be
C(2,q) = (2^{2q}/\sqrt{\pi})~\Gamma(2q+1/2) - (3~2^{2q}/\pi)
[\Gamma(q+1/2)]^2~.
\ee

Note that the dependence of the cumulants in the $d_i$'s as given in
(\ref{gghJD/D}) enters through their
$q$th power $d_i^q$ since, as we noticed
before, $V_{ab}$ is proportional to the moment
$\langle [(\delta x)^2]^{q} \rangle$. As already mentionned,
 $[V_{ab}]^{q}$ (with $q={2 \over c}$) has the
same ``dimension'' as the usual variance.
The dependence of the cumulants $c_{2 r}(q)$ on the weights $w_i$'s
enters only through the terms $w_i^2 d_i^q$ which are the same for all
cumulant orders.

\subsubsection{Tails of the portfolio distribution for $c<1$ \label{sectijj}}

We now characterized the extreme tails of the distribution of wealth
variations of the total
portfolio constituted of $N$ assets.  We present results for the case where
all assets have the same exponent.
It is easy to reintroduce the dependence of $c_i=2/q_i$ in the formulas, as
it will
become necessary when comparing to the empirical data in section 5.

For this, we use (\ref{gghJD/D}) in (\ref{fgjqkkkw})
to get
\be
\hat{P}_S (k) =  \exp\biggl(
\sum_{m=1}^{+\infty}  {(-k^2)^m \over (2m)!} C(m,q) \sum_{i=1}^N  (w_i^2
d_i^q)^{m}\biggl) ~.
\ee
From the expression (\ref{jfzjmg}) of $C(m,q)$ (see also Appendix C), we
see that
\be
\ln C(m,q) \to_{m \to + \infty}~~~~ ({q \over 2} - 1) ~2m \ln 2m + {\cal
O}(2m)~.
\ee
Two cases must be distinguished.
\begin{enumerate}
\item $q>2$\,: $C(m,q)$ increases faster than exponentially for large
cumulant orders $2m$.
As a consequence, the tail of the distribution $P_S(\delta S)$ is
controlled by the
large cumulants.
\item $q \leq 2$\,: $C(m,q)$ decreases with $m$ and we cannot derive the
structure of the
tail from the large cumulants.
\end{enumerate}

For $q>2$, i.e. $c<1$ (stretched Weibull distributions),
the behavior of large order cumulants embodies the structure of the tail of
$P_S(\delta S)$.
From (\ref{gghJD/D}), we see that
\be
c_{2 m}(q) \to_{m \to + \infty}~~~  C(m,q) \biggl[{\rm Max}_i ~~w_i^2 d_i^q
\biggl]^{m}~,
\label{gghqqJDqq/D}
\ee
i.e. the high-order cumulants of the portfolio distribution are controlled
by the
single asset that maximizes the product $w_i^2 d_i^q$ of the square of the
weight with
the variance of the return. We call this maximum $w_{\rm max}^2 d_{\rm max}^q$.
Keeping only this term (\ref{gghqqJDqq/D}) for the
expression of the cumulants $c_{2 m}(q)$ of large order, we see that
$\hat{P}_S (k)$ takes the form of the cumulant expansion for a single
Weibull distribution
with $d_i^{q}$ replaced by $w_{\rm max}^2 d_{\rm max}^q$.
We then deduce the expression for the extreme tail
\be
P_S(\delta S) \to_{|\delta S| \to \infty}  \exp \biggl[-\biggl( {|\delta S|
\over w_{\rm max} d_{\rm max}^{q/2} } \biggl)^c \biggl] =
\exp \biggl[ -\biggl( {|\delta S|
\over w_{\rm max} \chi_{\rm max} }\biggl)^c\biggl]~~~~~{\rm
where}~\chi_{\rm max} \equiv
d_{\rm max}^{q/2} ~.
\label{jkqklql}
\ee
For $c<1$, we have thus shown that the extreme tail of the portfolio
distribution is
of the same functional form as the distribution of the individual assets with
a characteristic decay rate $w_{\rm max} \chi_{\rm max}$ controlled by
a single asset. The selection of this asset is a function of
the parameters of the asset distributions and of the portfolio weights.
Note that, for $w_{max}$ of order $1/N$,
\be
P_S(\delta S) \sim  \exp \biggl[- N^c \biggl( {|\delta S| \over \chi}
\biggl)^c\biggl]~,
\label{jqkqk}
\ee
where $\chi$ is a coefficient independent of $N$. The extreme tail of
$P_S(\delta S)$ thus decays
as the exponential of a power of $N$ smaller than one.

This formulation of the tail of the distribution $P_S(\delta S)$
of the portfolio wealth $\delta S$ allows one to characterize the risk by
the single
scale parameter $\chi$. This is in the same spirit as the
portfolio theory for power law distributions [Bouchaud et al., 1998]
in which all the different risk dimensions are encapsulated by the scale
parameter of the power law distribution of the portfolio wealth variations.

When the exponents $c_i = 2/q_i$ are distinct, the tail of the portfolio
distribution is
controlled by the asset with the smallest $c$, i.e. largest $q$.

\subsubsection{Tails of the portfolio distribution for $c>1$}

For $q<2$, i.e. $c>1$,
the above derivation does not hold. However, we can use
the extreme deviation theorem [Frisch and Sornette, 1997] to determine
the shape of the extreme tail of the portfolio wealth distribution.
This theorem applies only for the exponent $c > 1$ and the stretched
Weibull case
previously analyzed is excluded from this analysis.
This is due to the existence of a log-convexity condition
$f_i''(\delta x_i)>0$, where $f''$ is the second derivative of $f$ defined
in (\ref{ezrrsffgsghhs}), as is shown in the Appendix D and in [Frisch and
Sornette,1997].

The Appendix D generalizes the extreme deviation theorem [Frisch and
Sornette, 1997]
to the case of non-identically distributed random variables. The result is
\be
P_N(\delta S) \to_{{\rm large}~\delta S} ~~~~{\pi^{N-1 \over 2} \over X
~\prod_{j=1}^N w_j \chi_j}~
\biggl[{2 \over c(c-1)} ({\delta S \over \chi})^{2-c} \biggl]^{N-1 \over 2}~
\exp \biggl( -{N \over 2(c-1)} ({\delta S \over {\hat \chi}})^c \biggl)~,
\label{eezesqqdfwx}
\ee
where
\be
{\hat \chi}^c
= {(\sum_{j=1}^N w_j \chi_j )^c \over c {(\sum_{j=1}^N w_j \chi_j )^2
\over N \sum_{j=1}^N w_j^2 \chi_j^2} + c - 2}~.
\ee
This result is valid for $c>1$. Note that the extreme tail of $P_N(\delta S)$
decays as the exponential of the number $N$ of assets. This result gives the
correct cross-over to the result (\ref{jqkqk}) obtained for $c<1$.

This formulation of the tail of the distribution $P_S(\delta S)$
of the portfolio wealth $\delta S$ again allows one to characterize the
risk by the single
scale parameter $\chi$, in a similar spirit as for power laws [Bouchaud et
al., 1998], as
already mentioned.
The minimization of the risk in the tail is thus completely treated by
finding the weights $w_i$
that  minimize $\chi$. Our theory below is more complete however, since the
determination
of all cumulants  give us a characterization of the complete distribution
and not only
of its tail.

\subsection{The general case of correlated assets}

\subsubsection{Cumulants of the portfolio distribution with $N$ assets}

We assume that all assets are characterized by Weibull distribution with
the same
exponent $c=2/q$. Our goal is to compute the characteristic function
\be
\hat{P}^q_S(k) = {1 \over (2\pi)^{N/2} \det{V}^{1/2} } \int\left(\prod_i^N
du_i\right)
e^{-\frac{1}{2} U V^{-1} U  + i k \sum_i w_i u_i^q} ~.
\label{hgqjkjqld}
\ee
For this, we introduce a perturbation analysis and
the functional generator [Brezin et al., 1976; Sornette, 1998] defined by
\be
\hat{P}^q_S(k,J_i) = {1 \over (2\pi)^{N/2} \det{V}^{1/2} }
\int\left(\prod_i^N du_i\right)
e^{-\frac{1}{2} u V^{-1} u  + i k \sum_i w_i u_i^q + \sum_i J_i u_i}~.
\ee
When the integral is a Gaussian ($k=0$), we get
\bea
\hat{P}^q_S(0,0) &=& 1\\
\hat{P}^q_S(0,J_i) &=&  e^{\frac{1}{2} J V J}~.
\eea
With the relationship
\be
f\left(\frac{\delta}{\delta J_i}\right) \int\left(\prod_i^N du_i\right)
e^{-\frac{1}{2} U V^{-1} U + \sum_i J_i u_i} =
\int\left(\prod_i^N du_i\right)
e^{-\frac{1}{2} U V^{-1} U  + \sum_i J_i u_i} f(u_i)~,
\ee
we can formally  express the characteristic function as
\be
\hat{P}^q_S(k) =
\left. e^{i k \sum_i w_i \frac{\delta^q}{\delta J_i^q}} ~~e^{\frac{1}{2} J V J}
\right|_{J_i=0}~.
\label{tujgkj}
\ee
This formulation is the most natural for a perturbative analysis of
(\ref{hgqjkjqld}).

In appendix E, we use the Feynman diagram method to calculate
\be
\hat{P}^q_S(k) =
\left. \left[1+ i \frac{g_q}{q!}\sum_i w_i \frac{\delta^q}{\delta J_i^q} +
\frac{1}{2} \left(\frac{i g_q}{q!}\right)^2
\sum_{i,j} w_i w_j \frac{\delta^q}{\delta J_j^q} \frac{\delta^q}{\delta
J_i^q + ...}
\right] e^{\frac{1}{2} J V J} \right|_{J=0}~,
\ee
where we have defined the auxiliary coupling constant
\be
g_q \equiv q! k~.
\label{couplingconstant}
\ee

Up to second order in $k$ (which will thus provide the variance of
$P_S(\delta S)$), we
get
\be
\hat{P}^q_S(k) =
\left[1-g_q^2 \sum_{l=0}^{(q-1)/2} \frac{1}{(q-2l)!}
\frac{1}{(2!)^{2l +1}} \frac{1}{(l!)^2}
\sum_{i,j} w_i \left(V_{ii}\right)^l
\left(V_{ij}\right)^{q-2l} \left(V_{jj}\right)^l w_j \right]~.
\ee

The diagrammatic expansion used in appendix E becomes very useful for
higher order
terms in $g_q$ to obtain a systematic classification of their proliferation.
A well-known result of diagrammatic perturbation theory [Veltman, 1995]
tells us that neglecting
the disconnected diagrams, which appear beyond second order, corresponds to
compute
the logarithm of the characteristic function. Therefore, the set of
connected diagrams
at $m$-th order give us directly the $m$-th cumulant coefficient $c_m$
defined by
expression (\ref{fgjqkkkw}).

We give now the expressions of the second $c_2(3)$ and
fourth  cumulant $c_4(3)$ corresponding to $q=3$, i.e. $c=2/3$.
The second cumulant is read from expression (\ref{qml,klSD})
in the appendix
\be
c_2(3) = \sum_{ij} \biggl[ 6 w_i \left(V_{i j}\right)^3 w_j +
9 w_i V_{ii} V_{ij} V_{jj} w_j  \biggl]~,
\label{qml,kqfgflSD}
\ee
where  $V_{ij}$ is the $i,j$ element of the covariance matrix $V$
between the variable $y_i$ of asset $i$ and the variable $y_j$ of asset
$j$. The sum is
over all pairs of assets.
The fourth order cumulant is
\begin{eqnarray}
c_4(3) &=& 4! (3!)^4 \sum_{i_1,i_2,i_3,i_4} w_{i_1} w_{i_2} w_{i_3} w_{i_4}
\left\{ \frac{1}{2^4}
V_{i_1 i_2}^2 V_{i_1 i_3} V_{i_2 i_4} V_{i_3 i_3} V_{i_4 i_4} +
\right. \nonumber\\
& & \,\,\,\, \frac{1}{2^3}
V_{i_1 i_2}^2 V_{i_1 i_3} V_{i_2 i_3} V_{i_3 i_4} V_{i_4 i_4} +
\frac{1}{2^4}
V_{i_1 i_2}^2 V_{i_1 i_3} V_{i_2 i_4} V_{i_3 i_4}^2 + \nonumber\\
& &\,\,\,\,\left.
\frac{1}{3! 2^3}
V_{i_1 i_2} V_{i_1 i_3} V_{i_1 i_4} V_{i_2 i_2} V_{i_3 i_3} V_{i_4 i_4} +
\frac{1}{4!}
V_{i_1 i_2} V_{i_1 i_3} V_{i_1 i_4} V_{i_2 i_3} V_{i_2 i_4} V_{i_3 i_4}
\right\}\,\,.
\label{jjjdjhshhs}
\eea
The sum is carried over all possible quadruplets of assets.

This result generalizes for higher order $m$-th cumulants and for arbitrary $q$
as
\be
C_m(q)= m! (q!)^m \sum_{i_1,\ldots,i_m} w_{i_1}\cdots w_{i_m}
\sum_{{\cal G}_m (q)}
\frac{1}{S(\{l_r\},\{n_{rs}\})} \prod_{r=1}^m \left(V_{i_r i_r}\right)^{l_r}
\prod_{r<s=1}^m \left(V_{i_r i_s}\right)^{n_{rs}}
\ee
where the sum is carried over the set ${\cal G}_m(q)$
of  all the topologically inequivalent connected
diagrams with $m$ vertices of $q$ legs as defined in the Appendix E and
the symmetry factor $S$ is of the form (see Appendix E)
\be
S(\{l_r\},\{n_{rs}\})= (2!)^{\sum_{r=1}^m l_r}
\prod_{r=1}^m l_r! \prod_{r<s=1}^m n_{rs}! S_v(\{l_r\},\{n_{rs}\})\,\,.
\ee

\subsubsection{Portfolio with two assets}

As an illustration, for the case of two assets ($N=2$),
Eqs.(\ref{qml,kqfgflSD},\ref{jjjdjhshhs}) reduce to
\be
c_2(3) = 15 V_{11}^3 w_{1}^2 + 15 V_{22}^3 w_{2}^2
+ (V_{11} V_{22})^{3 \over 2} ~[6 \rho_{12}^3 + 9 \rho_{12}] ~w_1 w_2 ~,
\label{hqkqk}
\ee
where the correlation coefficient is defined by
\be
\rho_{12} \equiv {V_{12} \over \sqrt{V_{11} V_{22}}}~.
\label{jqjfnmqnqmq}
\ee

The fourth cumulant is
$$
{c_4(3) \over 4! (3!)^4} =  a V_{11}^6 w_{1}^4 + a V_{22}^6 w_{2}^4
+ \biggl[(a- {1 \over 2^4} - {1 \over 4!}) V_{11} V_{12} V_{22}
+ ({1 \over 2^4} + {1 \over 4!}) V_{12}^3\biggl]
(V_{11}^3 w_{1}^3 w_{2} + V_{22}^3 w_{1} w_{2}^3)
$$
\be
+2 \biggl( (a - {1 \over 4!}) V_{11}^2 V_{12}^2 V_{22}^2 + {1 \over 4!} V_{11}
V_{12}^4 V_{22}\biggl) w_1^2 w_2^2~,
\label{kjlskskl}
\ee
where
\be
a \equiv {1 \over 2^4} + {1 \over 2^3} + {1 \over 2^4} + {1 \over 3! 2^3}
+ {1 \over 4!}~.
\ee

If the two assets are not correlated, $V_{12} = \rho_{12} = 0$, expressions
(\ref{hqkqk}) and (\ref{kjlskskl}) recover the non-correlated expression
(\ref{gghJD/D}) as they should, since $4! (3!)^4~a = C(2,3) = 9720$.

For $V_{11} = V_{22} = V$, the expression (\ref{kjlskskl}) of the fourth
cumulant simplifies
into
\be
{c_4(3) \over 4! (3!)^4 V^6} =  a  [w_{1}^4 + w_{2}^4]
+ \rho_{12} [a- ({1 \over 2^4} + {1 \over 4!}) (1 - \rho_{12}^2) ] (w_{1}^3
w_{2} + w_{1} w_{2}^3)
+ 2 \rho_{12}^2 [a - {1 \over 4!}(1 - \rho_{12}^2) ] w_1^2 w_2^2~.
\ee


\section{Portfolio optimization for uncorrelated assets}

For non-Gaussian distributions, all the cumulants and not only the variance
of the distribution of the portfolio wealth variations must
be considered as relevant measures of risk.
The variance as well as all higher order cumulants depend on the weights
$w_i$ of the assets
constituting the portfolio with different functional forms. It is thus
important
to determine the relative variation of the cumulants when the weights $w_i$ are
modified. In particular, we show now that the portfolio which minimize the
variance,
i.e. the relatively ``small'' risks, often increases larger risks as
measured by higher
normalized cumulants.

\subsection{Minimization of the variance}

For simplicity of notation, we restrict to the case where all assets have
the same
exponent $c=2/q$. It is straightforward to reintroduce the asset dependence
of the
$q$'s in the formulas.
In order to take into account the normalization constraint $\sum_i w_i =1$,
we solve
for the $N$-th weight $w_N= 1 - \sum_{i=1}^{N-1} w_i $ and we substitute it
in the explicit formula for the variance
\be
c_2(q) =  (2q -1)!! \left[\sum_{i=1}^{N-1} w_i^2 d_i^q
+ \left(1 - \sum_{i=1}^{N-1} w_i\right)^2 d_N^q\right]~.
\ee
The variance being quadratic in the weigths $w_i$, the solution to the
minimization
problem is the solution to the linear algebraic system of $N-1$ equations
in $N-1$ variables
\be
w_i d_i^q  =  \left(1 - \sum_{i=1}^{N-1} w_i\right) d_N^q,
~~~~~~~~~i=1,\ldots,N.
\ee
The solution can be written in a compact form by using the symmetric
polynomials
$\sigma_n^{(N)}(x_1,\ldots,x_N)$ generated by
$\prod_{i=1}^N (x + x_i) = \sum_{n=0}^N x^{N-n}\sigma_n^{(N)}(x_1,\ldots,x_N)$
such that $\sigma_0^{(N)} =1$, $\sigma_1^{(N)} = x_1+x_2+\ldots+x_N$,
... and $\sigma_N^{(N)}=x_1 x_2 \cdots x_N$.
The solution takes the form
\be
w_1 d_1^q = w_2 d_2^q = \ldots = w_N d_N^q =
\frac{\sigma_N^{(N)}(d_1^q,\ldots,d_N^q)}
{\sigma_{N-1}^{(N)}(d_1^q,\ldots,d_N^q)} = {1 \over \sum_i {1 \over d_i^q}}~.
\label{minvardiag}
\ee
The weights $w_i$ that minimize the portfolio variance are inversely
proportional
to the corresponding asset variance $d_i^q$.

Noting
\be
x_i = {1 \over d_i^q}~,
\label{qfqwxggw}
\ee
the cumulants corresponding
to the weights that minimize the variance are
\be
c_{2 r}^{V}(q) = C(r,q) ~{\sum_i x_i^r \over \biggl(\sum_i x_i\biggl)^{2r}}~.
\label{fjqoqoow}
\ee

\subsection{Minimizing the higher-order cumulants}

The weights $w_i$ that minimize the cumulants $c_{2 r}(q)$ given by
(\ref{gghJD/D})
are
\be
w_1 d_1^{qr \over 2r-1} = w_2 d_2^{qr \over 2r-1} = \ldots = w_N d_N^{qr
\over 2r-1} =
{1 \over \sum_i {1 \over d_i^{qr \over 2r-1}}}~.
\label{minvardiqfqag}
\ee
Note that these weights vary with the order $r$ of the cumulant that is
minimized.
It is thus not possible to
minimize simultaneously all the cumulants of order larger than two.
This is in contrast to the result for the normalized cumulants discussed below.
The weights that minimize the very large $r \to +\infty$ order
cumulants approach asymptotically the values
that minimize all the normalized cumulants of order larger than two, as
given by
(\ref{mincumdiag}) below. Thus, conclusions obtained for the normalized
cumulants
carry out for the cumulants in the limit $r \to \infty$.

The cumulants corresponding to (\ref{minvardiqfqag}) are
\be
c_{2 r}^{min}(q) = {C(r,q) \over \biggl( \sum_i x_i^{r \over
2r-1}\biggl)^{2r-1}}~,
\label{gghJD/Dqgqd}
\ee
where the $x_i$'s are given by (\ref{qfqwxggw}).
By definition, the two expressions (\ref{gghJD/Dqgqd}) and (\ref{fjqoqoow})
coincide
for $r=1$, i.e. give the same minimum variance.

It is interesting to compare the values taken by the
cumulants for the weights that minimize the variance
with the values taken by the cumulants for the weights that minimize a cumulant
of given order $2r$. For this, we study the ratio of the cumulants
which is also the ratio of the normalize cumulants since
the two expressions (\ref{gghJD/Dqgqd}) and (\ref{fjqoqoow}) coincide
for $r=1$\,:
\be
{c_{2 r}^{V}(q) \over c_{2 r}^{min}(q)} = {\lambda_{2 r}^{V}(q) \over
\lambda_{2 r}^{min}(q)} =
\biggl(\sum_i X_i^r \biggl)~\biggl( \sum_i X_i^{r \over 2r-1}\biggl)^{2r-1}~,
\label{rdfxffwffw}
\ee
where
\be
X_i = {x_i \over \sum_j x_j}~~~~~~~~{\rm~ are ~normalized}~~\sum_{j=1}^N
X_j = 1~,
\label{jsllamq}
\ee
and the $x_i$'s are defined by (\ref{qfqwxggw}).

Differentiating $c_{2 r}^{V}(q)/c_{2 r}^{min}(q)$
given by (\ref{rdfxffwffw}) with respect to one of the
$X_i$'s, we find that the derivative vanishes when all $X_j$'s are
equal to $1/N$. This corresponds to an extremum of all
$c_{2 r}^{V}(q)/c_{2 r}^{min}(q)$ equal to $1$. This extremum is in fact a
minimum. To see this, we parameterize
\be
X_i = {1 \over N} (1 + \epsilon_i)~,
\ee
where the $\epsilon_i$'s are small and sums up to zero ($\sum_{i=1}^N
\epsilon_i = 0$).
We then expand (\ref{rdfxffwffw})
in powers of $\epsilon_i$'s and find
\be
{c_{2 r}^{V}(q) \over c_{2 r}^{min}(q)} =
1 + {1 \over N}~{r (r-1)^2 \over 2r-1} ~\sum_{i=1}^N \epsilon_i^2 ~.
\ee

Thus, for unequal $X_i$'s, the ratios $c_{2 r}^{V}(q)/c_{2 r}^{min}(q)$
are generically larger than one. In words, the values of the
cumulants for the weights that minimize the variance are higher than
those for the weights that minimize a cumulant of given order $2r>2$\,:
minimizing ``small'' risk increases large risks.

\subsection{Minimization of the excess kurtosis and higher normalized
cumulants}

Normalized cumulants provide a better measure of large risks than the
(non-normalized) cumulants. The normalized cumulants are defined by
\be
\lambda_{2m} = {c_{2m} \over [c_2]^m} ~.
\label{hqkqkkl}
\ee
The fourth normalized cumulant $\lambda_4$ is often called the excess
kurtosis $\kappa$.
Recall that it is identically zero for Gaussian distributions.

The quantitative deviation of a distribution from normality is given by
normalized cumulants ${c_{2m} \over [c_{2}]^m}$. Indeed, the difference
${\cal P}_{>}(z)-g(z)$ between the cumulative distribution function of the
sum of $N$ random variables and its Gaussian asymptotic value is
given by [Gnedenko and Kolmogorov, 1954]
\be
{\cal P}_{>}(z)-g(z) \simeq
\frac{\exp(-z^2/2)}{\sqrt{2\pi}}\left( \frac{Q_1(z)}{N^{1/2}}+
\frac{Q_2(z)}{N}+\ldots+
\frac{Q_k(z)}{N^{k/2}}+\ldots\right)
\label{E_GD}
\ee
where $Q_k(x)$ are polynomials parameterized by the normalized cumulants
$\lambda_n$ of the distribution of the $N$ random variables. For instance,
the two first polynomials are
\be
Q_1(x)=\frac{1}{6}\lambda_3(1-x^2) ,
\ee
and
\be
Q_2(x)=\frac{1}{72}\lambda_3^2 x^5+
\frac{1}{8}(\frac{1}{3} \lambda_4-\frac{10}{9} \lambda_3^2)x^3+
(\frac{5}{24} \lambda_3^2-\frac{1}{8} \lambda_4)x  .
\ee

A more straightforward way to recognize the role of the normalized cumulants is
to work with the characteristic function defined in (\ref{fgjqkkkw}).
Since $k$ is the variable conjugate to $\delta x$ and the natural scale for
$\delta x$ is the standard deviation $\sqrt{c_2}$, this means that the natural
``dimensionless'' variable is $\delta x/\sqrt{c_2}$, i.e. in the conjugate
variable, it is
$k~\sqrt{c_2}$ that we define as ${\tilde k}$. In terms of ${\tilde k}$,
${\hat P}$
become
\be
{\hat P}({\tilde k}) = \exp \biggl(- {{\tilde k}^2 \over 2} +
\sum_{n=2}^{\infty} (-1)^n {\lambda_{2n} \over (2n)!} {\tilde k}^{2n}
\biggl) ~.
\label{cumuqfqfgqqdy}
\ee

Using (\ref{gghJD/D}) to construct the normalized cumulants $\lambda_{2m}$
defined by (\ref{hqkqkkl}), we find that the asset weights $w_i$ that minimize
$\lambda_{2m}$, {\it irrespective} of the order $2m$, are given by
\be
w_1 d_1^{q/2} = w_2 d_2^{q/2} = \ldots = w_N d_N^{q/2} =
\frac{\sigma_N^{(N)}(d_1^{q/2},\ldots,d_N^{q/2})}
{\sigma_{N-1}^{(N)}(d_1^{q/2},\ldots,d_N^{q/2})}
 = {1 \over \sum_i {1 \over d_i^{q \over 2}}}~.
\label{mincumdiag}
\ee

We recover (\ref{mincumdiag}) from the analysis of the tail
of the portfolio distribution for $c<1$ given in section (\ref{sectijj}).
The natural criterion is to find the weights $w_i$ that minimize
the characteristic decay rate ${\hat \chi}$ of the tail of the portfolio
distribution given by (\ref{jkqklql})\,:
\be
{\partial {\hat \chi} \over \partial w_j} = 0~, ~~~~~~~~~~~{\rm
for~all}~w_j's~.
\ee
After some calculation,  we find that the weights that
minimize ${\hat \chi}$ are exactly those given by (\ref{mincumdiag}), that
minimize the
high order normalized cumulants. This confirms that the normalized cumulants
are the relevant measures of the tail of the portfolio distribution.

The expressions of the cumulants corresponding to the weights
(\ref{mincumdiag}) are
\be
c_{2m}^{(K)} = N C(m,q) {1 \over \biggl(\sum_i {1 \over d_i^{q \over
2}}\biggl)^{2m}}~.
\label{rfkjlkm}
\ee

If all assets have the same nonlinear
variance $d_i=d$, then $c_{2m}^{(K)} = c_{2m}^{(V)}$ for all orders $2m$.
Thus, minimizing
the variance minimizes all normalized cumulants at the same time. This special
result is not true in general as we now show.

Using (\ref{jsllamq}) and the definition (\ref{hqkqkkl}) of the normalized
cumulants, we get
\be
{\lambda_{2m}^{(K)} \over \lambda_{2m}^{(V)}} = {1 \over N^{m-1}} {
\biggl(\sum_i X_i^2 \biggl)^{m} \over \biggl(\sum_j X_j^{2m}\biggl)}.
\label{mlfgggglm}
\ee
Using the Lagrange multiplier method to impose the normalization condition
on the
$X_i$'s, we find that the values $X_i$'s that {\it maximize}
the ${\lambda_{2m}^{(K)} \over \lambda_{2m}^{(V)}}$ for $m > 1$ are all
equal to $1/N$,
for which ${\lambda_{2m}^{(K)} \over \lambda_{2m}^{(V)}} = 1$. For any other
set of ``nonlinear variances'', the ratio is less than one. Thus,
the weights that minimize the variance {\it increase} the higher normalized
cumulants
such as the excess kurtosis.

We now compare the cumulants of the portfolio with minimum variance to the
cumulants of the portfolio with minimum normalized excess kurtosis, for
forming the ratio
\be
{c_{2m}^{(K)} \over c_{2m}^{(V)}} = N {
\biggl(\sum_i {1 \over d_i^{q}}\biggl)^{2m} \over
\biggl(\sum_j {1 \over d_j^{qm}}\biggl)~\biggl(\sum_k {1 \over d_k^{q \over
2}}\biggl)^{2m}
}~.
\label{glzhgmq}
\ee
With the definition (\ref{jsllamq},\ref{qfqwxggw}), we get
\be
{c_{2m}^{(K)} \over c_{2m}^{(V)}} = N {
\biggl(\sum_i X_i^2 \biggl)^{2m} \over \biggl(\sum_j X_j^{2m}\biggl)}.
\label{mlfjqmvllm}
\ee
In particular,
\be
{c_{2}^{(K)} \over c_{2}^{(V)}} = N  \sum_i X_i^2~,
\label{lsfkll}
\ee
and
\be
{c_{4}^{(K)} \over c_{4}^{(V)}} = N {
\biggl(\sum_i X_i^2 \biggl)^{4} \over \biggl(\sum_j X_j^{4}\biggl) }.
\label{rsgdghx}
\ee

We find that the ratio ${c_{2}^{(K)} \over c_{2}^{(V)}}$ is a
minimum equal to $1$ when all $X_i$'s are equal to $1/N$. For any other
set of nonlinear covariance matrix, the variance of the portfolio
which minimize the excess kurtosis is always larger or equal to the minimum
variance. This is expected by construction.

The ratio ${c_{4}^{(K)} \over c_{4}^{(V)}}$ is also found to be a minimum
equal to $1$ when
$X_i$'s being equal to $1/N$. This result generalizes to higher orders.
For any other
set of nonlinear covariance matrix, we find that the higher order cumulants
of the portfolio
which minimize the excess kurtosis are always larger or equal to the
cumulants of
the portfolio that minimizes the variance.

\subsection{Comparison between the excess kurtosis of the minimum-variance
portfolio
and the benchmark $w_i = 1/N$}

The benchmark with $w_i = 1/N$ has the following cumulants
\be
c_{2m}^{(1/N)}  = {C(m,q) \over N^{2m}}~  \sum_i d_i^{qm}~.
\label{lsfkfqfdll}
\ee
We construct the ratio of the excess kurtosis of the portfolio with minimum
variance
to the excess kurtosis of the benchmark\,:
\be
R \equiv {\lambda_{4}^{(V)} \over \lambda_{4}^{(1/N)}} = {
\biggl(\sum_i x_i^2 \biggl) ~\biggl(\sum_j {1 \over x_j}\biggl)^2 \over
\biggl(\sum_k x_k \biggl)^2 ~ \biggl(\sum_l {1 \over x_l^2}\biggl)} ~,
\label{mlfgggglqgqdsm}
\ee
where the $x_i$'s are defined by (\ref{qfqwxggw}).

For all $x_i$'s equal, $R=1$ as it should.
Changing all $x_i$'s into their inverse change the ratio $R$ into
its inverse $1/R$. This implies that the two situations are equally probable\,:
either the excess kurtosis of the benchmark is smaller/larger than the
excess kurtosis
of the portfolio with minimum variance. Indeed, for each set
of assets with $x_i$'s for which
$R < 1$, then the set of assets with inverse $1/x_i$'s gives $R>1$.
Finding the portfolio with minimum variance may thus
either increase or decrease its excess kurtosis compared to that of
the benchmark.

Notice that for $N=2$, $R$ is identically equal to unity for any possible
choice of $x_1$ and
$x_2$. A portfolio constituted of two uncorrelated assets is thus
such that the excess kurtosis of the benchmark and of the optimized
variance portfolio are the same.
This results is not true anymore for $N > 2$ for
which $R$ is usually different from one.

It is also interesting to investigate the ratio of the cumulants
\be
{c_{2m}^{(V)} \over c_{2m}^{(1/N)}} = N^{2m} {\sum_i X_i^m \over \sum_j {1
\over X_j^m}}~.
\ee
Expanding the $X_i$'s around the equal nonlinear variance case
$X_i = {1 \over N} (1 + \epsilon_i)$ where the $\epsilon_i$'s are small and
sums up to zero  $\sum_{i=1}^N \epsilon_i = 0$, we get
\be
{c_{2m}^{(V)} \over c_{2m}^{(1/N)}} = 1 - {m \over N} ~\sum_{i=1}^N
\epsilon_i^2 ~,
\ee
up to second-order in $\epsilon_i$'s.
The higher-order cumulants of the portfolio with minimum variance are
smaller than those of
the benchmark.

\subsection{Synthesis}

We have obtained the following results.
\begin{enumerate}
\item {\it Minimizing the variance versus minimizing the normalized
higher-order cumulants
$\lambda_{2m}$}\,:
the portfolio with minimum variance may have smaller cumulants than
the portfolio with minimum normalized cumulants but may have larger
normalized cumulants (which characterize the large risks). Thus, minimizing
small risks (the variance) may increase the large risks.

\item {\it Minimizing the variance versus minimizing the non-normalized
higher-order cumulants
$c_{2m}$}\,:
the portfolio with minimum variance may have larger non-normalized and
normalized cumulants of order larger than two
than the portfolio with minimum non-normalized cumulants.
Thus, minimizing small risks (the variance) may increase the large risks.

\item {\it Minimizing the variance versus the benchmark portfolio}\,:
the portfolio with minimum variance has smaller non-normalized higher-order
cumulants than the benchmarck portfolio, even if it
may have smaller or larger higher-order
normalized higher-order cumulants. This situation is
similar to the first case, i.e.
it is preferable to minimize the variance than taking the benchmark, even
if the resulting portfolio distribution is less Gaussian.

\end{enumerate}

\subsection{The expected utility approach}

The investor, starting the period with initial capital $W_0>0$, is assumed
to have preferences that are rational in the von-Neumann-Morgenstern
[1944] ~sense
with respect to the end-of-period distribution of wealth $W_0 + \delta S$.
His preferences are
therefore representable by a utility function $u(W_0 + \delta S)$
determined by the wealth
variation $\delta S$ at the end-of-period. The expected utility theorem
states that the investor's problem is to maximize
$\E[u(W_0 + \delta S)]$, where $\E[x]$ denotes the expectation operator\,:
\be
\E[u(W_0 + \delta S)] = \int_{-W_0}^{+\infty} d\delta S~ u(W_0 + \delta S)
~P_S(\delta S)~.
\label{hqlql}
\ee
$u(W)$ has a positive first derivative (wealth is prefered) and a negative
second derivative (risk aversion).

Consider first the case of a constant absolute measure of
risk aversion $-U''/U' = a$, for which $U(W) = -\exp(-a W)$. In the limit of
an initial wealth $W_0$ that is sufficiently large such that
the probability of ruin occuring in a one-period
(i.e. $\delta S < - W_0$) is negligible, expression (\ref{hqlql})
transforms into
\be
\E[u(W_0 + \delta S)] = - e^{-a W_0} ~{\hat P}(k=ia) =
- e^{-a W_0} \exp \biggl( \sum_{n=1}^{+\infty} {c_{2n} \over (2n)!}~a^{2n}
\biggl)~.
\ee
Due to the negative sign in front of the expression,
maximizing $\E[u(W_0 + \delta S)]$ is equivalent to minimizing
$\sum_{n=1}^{+\infty} {c_{2n} \over (2n)!}~a^{2n}$, i.e. a weighted sum
over the cumulants,
each of them function of the asset weights $w_i$ given by (\ref{gghJD/D}).
For a small
risk aversion $a<<1$, the sum in the exponential is essentially given by the
first term $c_2 a^2/2$. In this limit of small risk aversion, maximizing
the expected
utility retrieved the standard procedure of
minimizing the portfolio variance. However, for larger
risk aversions, the higher-order cumulants bring in non-negligible
contributions. It
is straightforward to solve for the optimization with explicit analytical
formulas.

Consider as another illustration the case where the utility is a member of
the power
(isoelastic) functions
\be
u(W) = W^{\gamma}~~~~~~~~{\rm with}~~0< \gamma <1~.
\label{hqlkql}
\ee
Then, assuming again that $|\delta S| < W_0$ for a single period (which is
not at
all restrictive an assumption in practice), we expand
\be
u(W_0 + \delta S) = (W_0 + \delta S)^{\gamma} = W_0^{\gamma} (1 + \delta
S/W_0)^{\gamma}
= W_0^{\gamma} \sum_{j=0}^{+\infty} {\gamma! \over (\gamma-j)! ~j!}~(\delta
S/W_0)^j~.
\label{fjqjM}
\ee
Putting (\ref{fjqjM}) in (\ref{hqlql}) gives
$$
\E[u(W_0 + \delta S)] =
W_0^{\gamma} \sum_{r=0}^{+\infty} {\gamma! \over (\gamma-2r)! ~(2r)!}~
\E[(\delta S/W_0)^{2r}]
$$
\be
= W_0^{\gamma} \biggl( 1 - {\gamma (1-\gamma) \over 2} ~{c_2 \over W_0^2}
- {\gamma(1-\gamma)(2-\gamma)(3-\gamma) \over 4!} {c_4 + 3 c_2^2 \over
W_0^4} - ...\biggl)
\label{hfqjqmm}
\ee
where we have replaced the lower bound $-W_0$ by $-\infty$ in the integral and
$\E[(\delta S/W_0)^{2}]$ by $c_2/W_0^2$ and $\E[(\delta S/W_0)^{4}]$ by
$(c_4 + 3 c_2^2)/W_0^4$.
Recall that the cumulants $c_{2n}$ are given by (\ref{gghJD/D}) with
(\ref{jfzjmg}) and contain
the dependence on the asset weights $w_i$.
The parameters controlling the behavior of $\E[u(W_0 + \delta S)]$
are the exponent $\gamma$ and the $N$ ratios $d_i^{q_i/2}/W_0$
of the asset price standard deviations normalized by the initial wealth.

For large initial wealth $W_0$ (compared to the one-period price standard
deviations),
only the first few cumulants need to be considered in the sum as the others are
weighted by a negligible factor. In this limit, the best portfolio
retrieves the
optimal variance portfolio, since maximizing $\E[u(W_0 + \delta S)]$
corresponds to
minimizing the first term $c_2/W_0^2$ in the expansion (\ref{hfqjqmm}).

In the other limit where the initial wealth $W_0$ is not
large compared to the one-period price standard deviations, $\E[u(W_0 +
\delta S)]$
receives non-negligible contributions from higher-order cumulants. The best
portfolio is a weighted compromise between the different dimensions of
risks provided by
the different cumulants.

An important insight obtained by this analysis is
that the optimal portfolio depends on the initial wealth $W_0$, everything else
being equal. This results from the existence of the many different
dimensions of risks provided by
the different cumulants which are each weighted by the appropriate power of
the initial
wealth. In other words, the initial wealth quantifies the relative
importance of the
high-order cumulants.

Notice the parallel between the analysis of the two cases with
constant absolute measure $a$ of risk aversion and with a power
(isoelastic) utility function\,: the relevant relative measure of the
impact of the
various cumulants (i.e. the importance of the tail of the distributions)
is $a$ for the first case and $1/W_0$ in the second case. It is interesting to
retrieve the standard variance minimization approach in the limit of small
risk aversion
or large initial wealth. In other words, when you use the standard variance
minimization
method, you implicitely express (perhaps unwillingly) a small risk aversion.
This is logical since you drop all information on the large-order cumulants
that
control the large risks.
In contrast, the full treatment incorporating the
higher cumulants as relevant measures of risks allows us to respond to a much
larger spectrum of trader's sensitivity and risk aversions.


\section{Empirical tests}

\subsection{$r \to y$ transformation by Eq.(\ref{EQ:transform})\,: examples
and statistical tests}

\subsubsection{Multivariate and marginal return distributions of six
currencies}

In Figures
\ref{JPYCHF(r)},\ref{JPYUKP(r)},\ref{JPYCAN(r)},\ref{JPYRUB(r)},\ref{JPYMYR(r)}
are shown the empirical bivariate distributions ($N=2$) for pairs made
of the Japanese Yen (JPY) and one of five other currencies, all quoted in US
dollars. Each point $(r_1(t),r_2(t))$ is defined as the daily annualized return
\be
r_i(t) = 250 ~\ln {s_i(t+1) \over s_i(t)}
\label{return}
\ee
where $s_i(t)$ is the price of currency $i$ at day $t$. We use the index
values $i=2$
for the Japanese Yen and $i=1$ for the other currencies which are respectively
the Swiss Franc (CHF), the British Pound (UKP),
the Russian Ruble (RUR), the Canadian Dollar (CAD) and the Malaysian Ringgit
(MYR). The time interval is from Jan. 1971 to Oct. 1998 except for the data
with RUR where the time interval is from Jun. 1993 to Oct. 1998.

The contour lines are obtained from Eq.(\ref{EQ:pca}) and corresponds to
different probability levels. A given level gives the probability that
an event $(r_1,r_2)$ falls inside the domain limited by the corresponding
contour line. Three
probability levels corresponding to $90\%$, $50\%$ and $10\%$ are shown
for each pair except for the pair RUR-JPY where only the $50\%$ and $10\%$
levels are shown. In order to distinguish the contour lines from the data
points, only one fourth of the data points (randomly chosen) are shown.

In these figures, we also present the price time series $s_i(t)$ (top and
right panels)
and the corresponding monovariate distributions $P(r_i(t))$. The scales of the
axis for the
bivariate and the abcissa of the monovariate distribution plots are chosen
identical
in order to highlight that monovariate distributions are
obtained as projections of the multivariate distribution.
The continuous lines correspond to the best fits of the Weibull pdf
defined by (\ref{aeqgdfbbzre}) to the tail for large returns {\bf r}
of the empirical distributions shown as open circles.
It can be compared to the best Gaussian fit shown as an inverted parabola
in these semi-log
plots. The values for the exponent  $c$ and the
constant $A$ are given in the figure captions.
 The fat tail nature of these pdf's is strongly apparent to the eye
and is also reflected quantitatively by the fact that the exponents $c$'s
of the
Weibull pdfs are all found of the order or less than one, while a Gaussian pdf
corresponds to $c=2$.
These fits show that the distribution Eq.(\ref{aeqgdfbbzre})
provides an excellent representation for all six currencies.

It is quite clear that the bivariate distributions
are very non-Gaussian, as can be seen from
the shape of the countour lines at the $90\%$, $50\%$ and $10\%$ confidence
levels.
For Gaussian distributions, the contour lines would be given by
the equation ${\bf r}'{\cal V}^{-1} {\bf r}= C$, where $C$ is a constant
and ${\cal V}$
is the covariance matrix for the ${\bf r}$ variables.
Depending on ${\cal V}$, the contour lines would take
the shape of
\begin{itemize}
\item circles for ${\cal V}_{11}={\cal V}_{22}$ and ${\cal V}_{21}\equiv
{\cal V}_{12}=0$;
\item ellipses with principal axis along the coordinates
for ${\cal V}_{11} \neq {\cal V}_{22}$ and ${\cal V}_{21} =0$;
\item ellipses with principal axis tilted with respect to the coordinates for
${\cal V}_{21} \neq 0$ with the tilt and the
eccentricity of the ellipses depending on ${\cal V}_{11},{\cal V}_{22},
{\cal V}_{21}$.
\item In particular, if ${\cal V}_{11}={\cal V}_{22}$ and ${\cal V}_{21}
\neq 0$,
the ellipses are always tilted at $\pm 45$ degrees with the ratio
of the small over large principal axis given by
$\sqrt{(1-|V_{12}|)/(1+|V_{12}|)}$.
For perfect correlations $|V_{12}| = 1$, the ellipse becomes a segment of
straight line
as expected for the dependence of two perfectly correlated variables.
\end{itemize}

Consider first the $90\%$ contour level for the CHF-JPY data
Fig.\ref{JPYCHF(r)}.
It is apparent that the data is not described by an ellipse as
prescribed by a Gaussian distribution, but rather the contour line takes
the shape of a ``bean''. For the fairly large {\bf r} events that fall outside
the $90\%$ level,
events with different signs of $r_1,r_2$ are
less likely to occur, or equivalently events which have same signs of
$r_1,r_2$
are more likely to occur, compared to an ellipse described by the multivariate
Gaussian distribution.  At the
$50\%$ level, the same ``bean''-like structure is found. Similar comments
apply to the UKP-JPY pair. This ``bean'' structure is thus the signature of a
stronger correlation of the sign of the returns, i.e. in the trend of the
US \$ against
these other currencies, for relatively small returns compared to the large
returns
for which this correlation weakens.

For the RUR-JPY data, the limited statistics
makes it hard to state anything decisive concerning
the large {\bf r} bivariate distribution, while the marginal distributions
are clearly strongly non-Gaussian. However, the small
${\bf r}$ data seems not incompatible
with Gaussian behavior, both for the marginal and bivariate distributions.
This illustrated in figure (\ref{figsmallr1}) that shows a close-up of the
countour
lines of the bivariate distributions.  For small ${|\bf r|} (< 0.5 \%)$,
the contour lines in all five cases appear either as approximate circles or
ellipses with principal axis along the coordinates,
thereby suggesting {\em uncorrelated} multivariate Gaussian
behavior for small changes of exchange rate.

The CAD-JPY contour lines shown in Fig.\ref{JPYCAN(r)} have a diamond like
shape with
principal axis along the coordinates. This signals that correlations are
very weak and
that, compared to an ellipse, the CAD-JPY data shows a higher probability
for having
events in which one of the returns is small. In other words, either one or
the other
exhibits a large move but rarely the two together.

The MYR-JPY data looks similar
except that the diamond principal axis appear tilted and ``twisted''. The
tilt reflects the
fact that the MYR-JPY pairs are correlated, significantly more so than the
CAD-JPY pair.

\subsubsection{Multivariate Gaussian distributions of the transformed
variables $y_i$}

In Figures
\ref{JPYCHF(y)},\ref{JPYUKP(y)},\ref{JPYRUB(y)},\ref{JPYCAN(y)},\ref{JPYMYR(y)},
we plot the bivariate distributions ${\hat P}({\bf y})$ defined by
Eq.(\ref{EQ:ngaus}), obtained from Figures
\ref{JPYCHF(r)},\ref{JPYUKP(r)},\ref{JPYCAN(r)},\ref{JPYRUB(r)},\ref{JPYMYR(r)}
using the transformation
Eq.(\ref{EQ:transform}). In each figure, we also construct the marginal
distributions
obtained by the transformation Eq.(\ref{EQ:transform}) performed on the
empirical
data points. The continuous lines represent the parabola (in this semi-log
representation)
corresponding to a Gaussian of unit variance. It is not surprising to find an
excellent agreement between the empirical marginal distributions of the $y$
variables
and the Gaussian pdf, as the mapping (\ref{EQ:transform}) ensures this
correspondence.
We note however some discrepancy near $y=0$, due to the
large number of quotes with zero or very small absolute values of
the returns and the limited accuracy of the data (the quotes are measured
with a finite number of digits, in units of points).

The contour lines for the
bivariate distributions of the $y$ variables are defined as in Figures
\ref{JPYCHF(r)},\ref{JPYUKP(r)},\ref{JPYCAN(r)},\ref{JPYRUB(r)},\ref{JPYMYR(r)}.
Let us mention that,
due to the phenomenon just mentioned of the abnormal behavior of $P(r)$
near $r=0$, there is
a high degeneracy of the points near $y=0$, i.e. one single dot may
correspond to many events near $y=0$.

Since the transformation
Eq.(\ref{EQ:transform}) ensures that $V_{11} \approx V_{22} \approx 1$, it
is easy
to show that the contour lines should be ellipses tilted
at an angle of $\pm 45$ degrees (for correlated/anticorrelated $V_{12}$
respectively) with a ratio of the small over large pricipal axis
of the ellipse given by $\sqrt{(1-|V_{12}|)(1+|V_{12}|)}$. Visual inspection
of the Figures
\ref{JPYCHF(y)},\ref{JPYUKP(y)},\ref{JPYCAN(y)},\ref{JPYRUB(y)},\ref{JPYMYR(y)}
confirms that the transformation (\ref{EQ:transform}), which is exact for
each of the
marginal distribution, provides an excellent representation of the
bivariate distributions.

\subsubsection{Goodness of the nonlinear transformation and statistical tests}

The transformation Eq.(\ref{EQ:transform}) offers a
new way for characterizing marginal distributions and their deviations
from normality.

In Fig.\ref{y(r)data}, we represent
the transformation Eq.(\ref{EQ:transform}) for all six currencies. The
$r$-variables
are calculated from the empirical data. From each data point, the
transformation
Eq.(\ref{EQ:transform})  provides the corresponding $y$ value.
The negative returns have been folded back to the positive quadrant (by taking
their absolute values).
The double logarithmic representation (log-log plot) is useful to characterize
a power law (\ref{rtgbn,jj}), corresponding to marginal distributions in $r$
given by the Weibull pdf (\ref{aeqgdfbbzre}).

Two straight lines are drawn on the figures. The $y=r$ line
corresponds to the exponent $c=2$ in
Eq.(\ref{aeqgdfbbzre}) and Eq.(\ref{rtgbn,jj}), which would qualify
a Gaussian pdf.
The other straight line shown for each currency describes the
transformation law
(\ref{rtgbn,jj}), where the exponents $c$
 have been determined independently by fitting the
tails of the marginal distributions reported in Figures
\ref{JPYCHF(r)},\ref{JPYUKP(r)},\ref{JPYCAN(r)},\ref{JPYRUB(r)},\ref{JPYMYR(r)}.
Thus, the Weibull distributions
Eq.(\ref{aeqgdfbbzre}) correspond to exact power laws Eq.(\ref{rtgbn,jj}),
translating into straight lines of slope $c/2$ in the log-log representation of
Fig.\ref{y(r)data}.
 Without any adjustment of the exponent, we find an excellent consistency
of the
description in terms of the modified Weibull distributions for large returns.

For smaller returns, we observe in Figure \ref{y(r)data} that the $y(r)$ curve
bends down to become approximately parallel to the line $y=r$ suggesting again,
in agreement what was found above from the contour lines
of the bivariate distributions,
that the distributions are approximately Gaussian for small absolute returns.

We now present a statistical test for the reliability and
``goodness of fit'' of the representation
Eq.(\ref{EQ:ngaus}) of empirical multivariate distributions.
We plot in thick line in Fig.\ref{figchiy}a-e the $\chi^2$ cumulative
distribution for
$N=2$ degrees of freedom versus the fraction of events as shown
in Figures
\ref{JPYCHF(y)},\ref{JPYUKP(y)},\ref{JPYRUB(y)},\ref{JPYCAN(y)},\ref{JPYMYR(y)}
within an  ellipse of equation $\chi^2 = {\bf y}' V^{-1} {\bf y}$. If the
empirical distributions
strictly followed the pdf Eq.(\ref{EQ:ngaus}), one should have a straight
line.
The thin lines correspond to the $90\%$ confidence
levels obtained from 100 Monte Carlo simulations\,: 100 random
realizations with the same number of points as the empirical data,
taken from a Gaussian multivariate
distribution, are transformed into the representation Eq.(\ref{EQ:ngaus}).
The thick line is found close to the diagonal, implying that the empirical
${\hat P}({\bf y})$ is very well approximated by a multivariate Gaussian.
There is
however a small but statistically significant departure from the diagonal,
mostly in the upper
parts of the figures, corresponding to the small returns. This seems to
tell us that
the correlation has a different structure for small and moderate returns
compared to the
larger returns, again in agreement with the Weibull (resp. Gaussian)
representation
of large (resp. small) returns. The largest
deviation can be seen for the RUR-JPY data, which is to be expected since
the statistics
is the poorest for this set.

In Fig.\ref{figchiy}f, we construct a similar plot for the multivariate
distributions with the $N=6$ currencies taken all together.
Since the empirical determination of the multivariate distribution
requires us to keep only the quotes for all six
currencies occurring simultaneously on the same day,
this statistics for the multidimensional plot of Fig.\ref{figchiy}f is
necessarily
the poorest and is controlled by the poorest of all bivariate distributions
which is
the RUR-JPY set. Fig.\ref{figchiy}f shows that the
method works well also with more than $N=2$ assets.

We have checked for the stationarity of these results by dividing
the total time window into three sub-windows of equal lengths
and have performed the same analysis using Eq.(\ref{EQ:transform}).
We find that the representation
Eq.(\ref{EQ:ngaus}) tested in Fig.\ref{figchiy} is extremely robust\,: no
statistically significant differences are found between the three sub-windows.
The stationarity of the nonlinear covariance matrix $V$ is quite
remarkable.

The same cannot be said with a represention in terms of the return $r$
variables.
Fig.\ref{figchir} illustrates this fact
by constructing the same tests as shown in Fig.(\ref{figchiy}) but for the
returns $r$ instead
of the transformed variables $y$. One can observe the dramatic departure
from the
diagonal, with an extremely large statistical significance, confirming the
highly non-gaussian
structure of the empirical multivariate distribution.
This figure is presented to stress (1) how far from Gaussian
are the empirical distributions and (2) how good is our transformation
Eq.(\ref{EQ:transform}) in comparison. The transformed variable $y$ is
indeed the
correct one to work with to get a robust and stationary representation.

Notice that, in Fig.\ref{figchir}, the thick line does not
approach the diagonal near abcissa values close to $1$, as one
could naively expect from our argument that the empirical distribution are
not far from Gaussian for small returns. The reason is that the distribution
may be close to Gaussian for small returns, but with a {\it different}
covariance matrix. This interpretation is born out by the observed deviations
from the diagonal in Fig.\ref{figchir} occurring for small $y$'s.

The fact that the thick line in Fig.\ref{figchir} is below the diagonal for
most of the range of
$\chi^2$ can be explained as follows. The correlation matrix of the returns
is estimated over all returns, including the large realizations.
Therefore, for a given value of $\chi^2$, the corresponding ellipse
is much larger than the ellipse that would be estimated from the covariance
matrix
(let us call it $R$)
of the small returns. For small
values of $r$ (close to $1$ on the abcissa of Fig.\ref{figchir}), there are
therefore
more points inside, i.e. fewer points outside, than one should
expect for an ellipse evaluated from the small returns.
Another way to see this result is to recall that $y \approx r$ for small $r$
as shown in Fig.\ref{y(r)data}. Since
$R^{-1}$ is much smaller than $V^{-1}$ (think of a matrix with only
diagonal elements), $\chi^2$ is therefore much
smaller for small $r$-values than for small $y$-values. This implies that,
for small r, the integral on the abcissa stays approximately constant as one
decreases the fraction of events outside the ellipse on the ordinate.
In comparison in Fig.\ref{figchir}, since $V^{-1}$
is much larger than $R^{-1}$, $\chi^2$ is larger and the integral
on the abscissa decreases with the same rate as the fraction
of events outside the ellipse on the ordinate.

\subsection{Variance and excess kurtosis}

Consider one of the five portfolios with two currencies investing a fixed
fraction $p$ of
its wealth $W$ calculated in US\$ in currency 1 and the remaining fraction
$1-p$ in
currency 2, where 1 and 2 refers to the five pairs of currencies analyzed
in the previous
figures. Using the historical time series,
we construct numerically the time series for the portfolio value $W(t)$
from the recursion
\be
W(t+1) = p W(t) s_1(t) + (1-p) W(t) s_2(t)
\label{hgjjjjjd}
\ee
with $p$ fixed. This expression ensures that the fraction of the wealth
invested
in a given currency is constant. We do not address here the issue of
friction and
transaction costs that would be involved in this dynamical reallocation but
rather
focus on the tests of the theory.

The annualized daily return $r_W$ of $W(t)$ is
defined by $r_W(t) = 250~ \ln {W(t+1) \over W(t)}$.
Fig.~\ref{figVARKURT} shows the dependence as a function of $p$ of the
variance
\be
c_2 \equiv \langle (r_W - \langle r_W \rangle)^2 \rangle~,
\ee
of the excess kurtosis
\be
\kappa \equiv {c_4 \over c_2^2} = {\langle (r_W - \langle r_W \rangle)^4
\rangle
\over \langle (r_W - \langle r_W \rangle)^2 \rangle^2} - 3~,
\ee
and of the sixth-order normalized cumulant
\be
\lambda_6 \equiv {c_6 \over c_2^3}~,
\ee
of the daily portfolio returns. The excess kurtosis and
the sixth-order normalized cumulant quantify the deviation from a Gaussian
distribution and provide measures for the degree of ``fatness'' of the
tails, i.e.
a measure of the ``large'' risks. Taking into account only the variance and
the excess kurtosis
and neglecting all higher order cumulants, a distribution can be
approximated by the
following expression valid for small excess kurtosis [Sornette, 1998]
\be
P_(r_W) \simeq \exp \biggl[-{(r_W-\langle r_W \rangle)^2 \over 2 c_2}
\biggl(1 - {5 \kappa \over 12} {(r_W-\langle r_W \rangle)^2 \over
c_2}\biggl)\biggl] ~ .
\label{gaussddd}
\ee
The negative sign
of the correction proportional to $\kappa$ means that large deviations are
more probable
than extrapolated from the Gaussian approximation.
For a typical fluctuation $|r_W - \langle r_W \rangle| \sim \sqrt{c_2}$,
the relative size of the
correction in the exponential is ${5 \kappa \over 12}$. For large values of
$\kappa$, this approximation (\ref{gaussddd}) break down and the deviation
from a Gaussian is much more dramatic.

The most interesting feature of Fig.~\ref{figVARKURT} is that the weight
that minimizes
the variance does not correspond to the minimum of $\kappa$ or of $\lambda_6$.
This illustrates one of the main message of our work, derived
in section 4 analytically for non-Gaussian distributions\,: minimizing the
portfolio variance
is not optimal with respect to the large risks. The strength of this effect
depends on the relative shape of the distributions of the assets
constituting the portfolio and their correlations.
The strongest effect in Fig.~\ref{figVARKURT} is found for the
Malaysian-Yen pair, for
which $p \approx 0.5$ minimizes the variance but gives an almost four-fold
increase
of the excess kurtosis compared to its minimum. Fig.~\ref{figVARKURT} shows
that it
is possible to do much better and construct a portfolio which
has not much more ``small'' risks (as measured by the variance) while having
significantly smaller ``large'' risks (as measured by the excess kurtosis
and six-order
normalized cumulant). This is due to the fact that the excess kurtosis and the
six-order normalized cumulant have in general a direct or inverted
S-shape with a rather steep dependence or narrow well as a function of the
asset weight $p$, while
the variance exhibits comparatively a smoother and rather slower
dependence. The investor
will be well-inspired in using the additional information provided by the
higher-order
cumulants to control for the portfolio's risks.

It is also very interesting to observe that
$\kappa$ and $\lambda_6$ exhibit similar behaviors with minima occurring
essentially
for the same weight. This confirms a prediction of our theory, according to
which
the asset weights that minimize
$\lambda_{2m}$, {\it irrespective} of the order $2m$ for $m>1$, are given
by Eq.(\ref{mincumdiag}).
Notice that this prediction holds in principle only for uncorrelated
assets, while the
correlation coefficients are $\rho(y_1,y_2)= 0.57$ (CHF-JPY); $0.43$
(UKP-JPY);
$-0.06$ (RUR-JPY); $0.07$ (CAN-JPY); $0.32$ (MYR-JPY), where
$\rho$ is defined by (\ref{jqjfnmqnqmq}). The case where the correlations
are weak
(RUR-JPY and CAN-JPY) exhibit the best agreement with our prediction. The
agreement is
still reasonable for the three other cases with larger correlations.
The main lesson learned here is that
such correlations are not very important for ``fat tail'' distributions ($c
\leq 1.5$) in
the determination of cumulants of large orders (in practice larger than two).

Fig.\ref{figkurtests}a-b and d-e compare the empirical excess kurtosis (fat
solid line)
shown in Fig.~\ref{figVARKURT} for the five portfolios
to our theoretical prediction (\ref{gghJD/D}) with (\ref{jfzjmg}) (solid line).
We use the result for uncorrelated assets as the coefficient of
correlations are small
for two of the five portfolios. In addition, we find that the empirical
exponents $c$
are much less than $2$ for which the calibration of the
exponents $c$ plays a much more important role in the determination of the
high-order
cumulants than the correlation. We leave for a future paper the extension
of our theory
to the case of correlated assets with different exponents $c$.
The exponents $c_1$ and $c_2$ are those determined in the fits of the pdf's
tail,
as given in Fig.\ref{y(r)data}. There is thus {\it no} adjustable
parameters in
the comparison shown in Fig.\ref{figkurtests}a-b and d-e. The
thin solid lines and dashed lines plot the theoretical formula
(\ref{gghJD/D}) for values of the exponents $\alpha_i \equiv c_i/2 \pm
0.05$, so as
to provide uncertainty brackets for the comparison.

Fig.\ref{figkurtests}c (RUR) shows the empirical excess kurtosis (fat solid
line) and the
theoretical prediction (\ref{gghJD/D})  with the fixed exponents
$c_1$ and $c_2$ given in Fig.\ref{y(r)data}. The thin solid line gives the
predicted excess kurtosis for exponents $c_1+0.05, c_2 \pm 0.05$. A better
agreement
is observed for an exponent $c_1+0.05$ for the Russian currency slightly
larger than determined
from fitting the tail of its pdf. We interpret this result by the fact that
this
fat tail pdf embodies the effect of much higher-order cumulants, while
the excess kurtosis is relatively a still rather low-order cumulant.

Fig.\ref{figkurtests}f compares the empirical excess kurtosis (fat solid line)
of the portfolio CHF-JPY (Fig.\ref{figkurtests}a) to the prediction
(\ref{kjlskskl})
with (\ref{hqkqk}) for correlated assets with the fixed exponents
$c_1=c_2=2/3$. The thin solid line correspond to the empirical value
$\rho(y1,y2)= 0.57$
while the dashed line is obtained from the same formula with $\rho(y1,y2)=0$.
The lesson we learn here by comparing the dashed and thin solid line
together with the
solid line of Fig.\ref{figkurtests}a is that the existence or absence of
a correlation for ``fat tail'' distributions
is not very important for the determination of the excess kurtosis. Much
more important is the correct determination of the tail exponents $c$ of
the Weibull
distributions.

In summary, we find a good agreement between the empirical excess kurtosis
and our
prediction with (\ref{gghJD/D}) with fixed exponents
$c_1$ and $c_2$ determined in Fig.\ref{y(r)data}. The other theoretical curves
provide the range of uncertainty in the determination of
the excess kurtosis  coming from measurement
errors in the exponents $c$. In fact, the fits with the predicted excess
kurtosis look amazingly good, because even
minor effects like the double well structure in Fig.\ref{figkurtests}c
near $p=0$ and in Fig.\ref{figkurtests}d near $p=1$ are in accordance with
theory.
These results are very consistent\,: a bad choice of $c$ leads to a bad fit
of the pdf tails
of each asset constituting the portfolio
and change completely the kurtosis of the portfolio pdf,
as can be seen from its large sensitivity under
relatively small variations of $\pm 0.1$ of the exponent $c$.

The main point here is that the theory adequately identifies the
set of portfolios which have small excess kurtosis and thus small `large
risks' and still reasonable
variance (`small risk'). We stress the importance of such precise
analytical quantification to
increase the robustess of risk estimators\,: historical data becomes
notoriously
unreliable for medium and large risks for lack of suitable statistics.

\section{Conclusion}

We have presented a novel and general methodology to deal with multivariate
distributions with non-Gaussian fat tails and non-linear correlations.
In a nutshell, our approach consists in projecting the marginal
distributions onto
Gaussian distributions, through highly nonlinear changes of variables. In turn,
the covariance matrix of these nonlinear variables allows us to define a novel
measure of dependence between assets, coined the ``nonlinear covariance
matrix'',
which is specifically adapted to remain stable in the presence of non-gaussian
structures. We have then presented the formulation of the corresponding
portfolio theory which requires to perform non-Gaussian integrals in order
to obtain the full distribution of portfolio returns. We have developed a
systematic perturbution theory using the technology borrowed from particle
physics of
Feynmann diagrams to calculate the cumulants of the portfolio
distributions, in the
case where the marginal distributions are of the Weibull class. The main
prediction
is that minimizing the portfolio variance may in general increase the large
risks
quantified by the higher-order cumulants. Our detailed empirical tests on
a panel of six currencies confirm the relevance of the Weibull description
and allows
us to make precise comparisons with our theoretical predictions. For ``fat
tail''
distributions, we find in particular
that the valid determination of
large risks, as quantified by the excess kurtosis, are much more sensitive to
the correct measurement of the Weibull exponent of each asset than to their
correlation, which
appears almost negligible.

Plenty of works remain to be done to explore further this approach.
\begin{itemize}
\item The case of assets with different exponents $c$ have been treated only
for uncorrelated assets and the corresponding problem of heterogeneous
$c$'s in the
correlated case is relevant for a precise comparison with empirical data.
Furthermore, we have not studied assets with large exponents $c \geq 1.5$. The
relevance of correlations increases with increasing $c$ and we expect a precise
determination of the correlation matrix to become more important as $c \to 2$.
\item We have focused our analysis on the risk dimension of the problems by
studying symmetric distributions, i.e. assets which are not expected to exhibit
long-term trends. A natural and relevant extension of our theory is to
treat the case
where the mean return is non-zero and different from asset to asset.
\item The next level of complexity is to have non-symmetric distributions,
with variable
Weibull exponents $c$ and with correlations.
\item The perturbation theory in terms of Feynmann diagrams can be used for
other classes
of distributions and it would be interesting to explore in details other
potentially
useful classes.
\item We have assumed and found to be reasonably verified that the nonlinear
covariance matrices are stationary. There is however no conceptual
difficulty in generalizing and adapting the
ARCH [Engle, 1982] and GARCH [Bollerslev et al., 1992]
models of time-varying covariance to this formulation in terms of effective
$y$ variables.
\item Our empirical tests have been performed on small portfolios with two
and six
assets, with the purpose of a pedagogical exposition and easier tests of
our theory. It
is worthwhile to extend our work to larger and more heterogeneous portfolios.
\end{itemize}

These goals all seem within reach and we intend to address these questions
in future works.

\pagebreak

\section{APPENDIX A: Consistency condition for Elliptic distributions}

Elliptic multivariate distributions $P(X)$ are defined by
\be
P(X) = F \biggl( (X-X^0)^T V^{-1} (X-X^0) \biggl)  ~.
\label{quasiggaus}
\ee
$X^0$ is the unit column vector of the average returns and $V$ is a dependence
matrix proportional to the covariance matrix when it exists.
The function $F$ is kept a priori arbitrary (non-negative and normalized).
If $F$ is an exponential, (\ref{quasiggaus})
retrieves the normal distribution and $V$ becomes the covariance matrix.

The proposition that the results of the
CAPM extends to  elliptical distributions [Owen and Rabinovitch, 1983;
Ingersoll, 1987]
relies on the ``consistency'' condition, according to which
any unconditional marginal distribution is of the form $F(V_{ii}^{-1}
(X_i-X_i^0)^2)$ with the
same function $F$. In particular, this leads to the fact that the
distribution of the
portfolio is a function solely of $W' V W$ (where $W$
is the column vector of the asset weights) with a functional form {\it
independent} of
the weights $W$ and the number $N$ of assets in the portfolio.

Kano [1994] has shown that this is not true for most marginal distributions.
For instance, for $F(x) = C \exp [-\sqrt{|x|}]$ where $C$ is a normalizing
constant,
the unconditional variance of a single
variable calculated using (\ref{quasiggaus}) is equal to $(N+1)/2$ times
the variance
obtained for a single variable using the same functional form. The result
is thus
not independent of $N$. This inconsistency implies
that the overall shape of the distribution will change as a function of the
asset weights in the
portfolio. As a consequence, $W' \Sigma W$ is not the sole measure of the
portfolio risks.
The conclusions of Owen and Rabinovitch [1983] and Ingersoll [1987] thus
apply only to
a restricted class of elliptic distributions for which the consistency
condition apply.

Let us show now explicity that
\begin{enumerate}
\item the dependence of the portfolio distribution on its wealth variation
$\delta S$ can be
expressed solely in terms of the ratio $(\delta S)^2/ W' V W$, but
\item the distribution itself has a functional form which, in general, is
still dependent on
the asset weights $W$ constituting the portfolio.
\end{enumerate}

For this, we write its variation
during a unit time step as
\be
\delta S(t) = \sum_{a=1}^N w_a \delta x^a(t) = W' X~.
\ee
The density distribution $P(\delta S)$ can be written as
\be
P(\delta S) = \int dX F \biggl( (X-X^0)^T V^{-1} (X-X^0) \biggl)  \delta
(\delta S-W' X) ~,
\label{azetpw}
\ee
where $\delta(x)$ is the Dirac distribution.
To estimate this integral, we isolate one of the assets $x_1 = x^0_1+ y_1$
and write,
using $Y=  X-X^0$,
\be
Y^T V^{-1} Y = V^{-1}_{11} y_1^2 + 2 (v^T y) y_1 + y^T {\cal V}^{-1} y ~,
\label{froemru}
\ee
where $V^{-1}_{ij}$ is the element $ij$ of the matrix $V^{-1}$, $y$ is the
unit column vector
$(y_2, y_3, ...., y_N)'$ of dimension $N-1$, $v$ is the unit column vector
$(V^{-1}_{21},
V^{-1}_{31}, ..., V^{-1}_{N1})^T$ of dimension $N-1$, and ${\cal V}^{-1}$
is the square matrix of
dimension $N-1$ by $N-1$ derived from $V^{-1}$ by removing the first row
and first column.
The factor $2$ in $2 (v^T y) y_1$ comes from the symmetric structure of the
matrix $V^{-1}$.

We can now express the condition $\delta(\delta S - W' X)$ in the integral
(\ref{azetpw})\,:
\be
{1 \over w_1}  \delta (y_1 - {1 \over w_1}(\delta S - W' X_0 - {\cal W}' y)) ~,
\ee
where  ${\cal W}$ is the unit column vector $(w_2, w_3, ..., w_N)'$ of
dimension $N-1$.
The integration over the variable $y_1$ cancels out the Dirac function and
we obtain the
argument of the function $F$ under a quadratic form in the variables $S$
and $y$. Using the
identity
\be
X' V^{-1} X +  X' Y = {\hat X}' V^{-1} {\hat X} - {1 \over 4} Y' V Y ~,
\ee
where ${\hat X} = X + VY$, we obtain
\be
P(\delta S) = \int d{\hat y} F\biggl( {\hat y}' M^{-1} {\hat y} + {\delta
S^2 \over W^T V W}
\biggl) ~,
\label{qusrdfgscy}
\ee
where the integral is carried out over the space of vectors ${\hat y}$ of
dimension $N-1$ and
\be
M^{-1} \equiv {\cal V}^{-1} - {2 \over w_1} (v - {V^{-1}_{11} \over 2w_1}
{\cal W}) {\cal W}' ~.
\ee
We can finally write
\be
P(\delta S) = {\cal F} \biggl({\delta S^2 \over W' V W}\biggl) ~,
\label{qusrcy}
\ee
where ${\cal F}(x)$ is defined by (\ref{qusrdfgscy}).

This confirms that the typical volatility of the portfolio is
controlled by the quasi-variance $W' V W$ as for the normal case. It is
then natural
to optimize the portfolio using this measure of the risk. However, it is
clear that,
for arbitrary functional forms of $F$, the
specific functional form of the distribution $P(\delta S)$ is in general a
function of the
asset weights constituting the portfolio.
Minimizing only $p^T V p$ may thus be insufficient because it may be linked
to a
dangerous deformation of ${\cal F}(x)$ in the tail.

\pagebreak


\section{APPENDIX B: Derivation of the approximate stability in family of
the Weibull distributions}

Consider an asset with
daily returns distributed according to (\ref{aeqgdfbbzre}) with $c<1$. What
is the
distribution of returns over $T$ days? It is convenient to use the
representation of the
distribution in terms of its cumulants defined by the derivatives of the
logarithm of its characteristic function\,:
\be
c_n=(-i)^n\left.\frac{d^n}{dk^n}\log {\hat P}(k)\right|_{k=0}~.
\label{cumulant}
\ee
The characteristic function thus reads
\be
{\hat P}(k) = \exp\{ \sum_n^{\infty} {c_n \over n!} (ik)^n \} .
\label{cumudy}
\ee

Assuming the absence of correlations between successive daily returns, the
distribution
of returns over $T$ days is the distribution of the sum of $T$ independent
random variables. Its cumulants are thus $T$ times the cumulants of the
distribution of
the random variables \footnote{This stems from the fact that the
distribution of the
sum is a convolution integral and its Fourier transform is then the product
of the
Fourier transforms of each individual pdf's. From the definition
(\ref{cumudy}), the
result follows.}. Adapting the results derived in Appendix C to the case of
a single
asset with ``nonlinear'' variance $d=\chi^{2 \over q}$ according to
(\ref{fjqkmqmkqmnqm})
with $c=2/q$, we obtain
the expression of the cumulants of the returns over $T$ days as
\be
c_{2 r}(T) = T c_{2 r}(1) = T~C(r,q)~d^{rq}~,
\label{hqkkq,bfa}
\ee
where $C(r,q)$ is a numerical factor given by (\ref{jfzjmg}) in the main
text and in Appendix C.
This expression (\ref{hqkkq,bfa}) is valid for any real value $q$, i.e.
describe the case of a
Weibull exponential with arbitrary real exponent $c=2/q$.

The deviation from the normal distribution is quantified by the normalized
cumulants
\be
\lambda_{2r}(T) = {c_{2r}(T) \over [c_2(T)]^r} = {C(r,q) \over
[C(1,q)]^r}~{1 \over T^{r-1}}~.
\ee
The cumulants of order $2r$ larger than $2$ decay to zero
as $T$ increases to infinity. This constitutes a signature of the central limit
theory according to which the distribution of returns for the sum of $T \to
\infty$
i.i.d. random variables tends to the normal law. In practice, the
distribution of returns
over a finite $T$ is not exactly normal because the convergence to the
normal law is rather slow. For instance, the excess kurtosis decays to zero
only as $1/T$, starting
from a rather large value $\kappa (T=1) = 7.2$.

For finite $T>1$, the distribution of
returns over the time scale $T$ may be approximated by a Weibull pdf
with an apparent exponent $c_T = 2/q_T$ larger than the exponent $c$
determined at the daily time scale. For this, we approximate the first
cumulants
by their expression for a Weibull distribution with adjustable
nonlinear variance $d_T$ and different exponent $c_T=2/q_T$.
This amounts to look for an approximate representation of
$T~C(r,q)~d^{qr}$ by $C(r,q_T)~d_T^{q_T r}$. Since there are only two variables
$d_T$ and $q_T$ to optimize, they can be determined from two conditions, that
we take to be the correspondence of
the variance and excess kurtosis
\be
T~C(1,q)~d^{q} = C(1,q_T)~d_T^{q_T}~,
\label{jfqjmqqqmqmq}
\ee
\be
T~C(2,q)~d^{2q} = C(2,q_T)~d_T^{2 q_T}~.
\ee
Eliminating $d_T$ between the two equations gives $q_T$ as the solution of
\be
T = {C(1,q_T) \over C(1,q)} ~{C(2,q) \over C(2,q_T)}~.
\label{fjquv}
\ee
For a distribution of daily returns given by a Weibull distribution with
exponent $c_1=2/3$ ($q=3$), we find that the monthly ($T\approx 25$)
returns are
distributed according to an exponential ($c_{25} \approx 1$).
This approximation illustrates the very slow convergence to the
normal law since the value $c_{25} \approx 1$ is still very far from the
asymptotic
normal value $c_{\infty}=2$.

The effective variables $d_T$ and $q_T$ can be determined by more global
conditions
such that the weighted sum of the square of the
differences $T~C(r,q)~d^{qr} - C(r,q_T)~d_T^{q_T r}$ be minimum over a certain
set of $r$'s, this set controlling how far in the tail the approximation is
valid.

We test this idea by the following synthetic tests. Let us call
$c_1=2/q_1=2/3$ the
exponent of the Weibull pdf $P_1$ of the returns $r_1$ at the daily time
scale. We construct
the pdf $P_T$ of the returns $r_T$
over $T$ days by taking the characteristic function of $P_1$ to the
$T$-th power and then taking the inverse Fourier transform \footnote{We use
the theorem
that the Fourier transform of the convolution of two functions is the
product of the
Fourier transforms.}. Let us now test whether $P_T$ can be approximated by
a Weibull
distribution with an effective exponent $c_T=2/q_T$ and determine its value
as a function of $T$.

For this, we perform the change of variable $r_T \to y_T(r_T)$ given by
(\ref{rtgeejj}) with
(\ref{qkjqklkqqq}), with a given choice for the
 exponent $c_T$ and using (\ref{jfqjmqqqmqmq},\ref{fjquv}) to
get $\chi_T=d_T^q$. If the  $T$-fold convolution distribution $P_T$
of the Weibull distribution $P_1$
is approximately a Weibull, this change of variable should lead to an
approximate
gaussian with unit variance for the correct choice of $c_T$.
We check the consistency of this program for $T=1$ for which
we do retrieve, as expected, an exact Gaussian with unit variance
independent of $c$ and $\chi$.

Figs.\ref{figweibullstable},\ref{figweibullstable1},\ref{figweibullstable2},\ref
{figweibullstable3}
plots the pdf's $P_T(y)$ as a function of $z \equiv y^2$ so that
a Gaussian (in the $y$ variable) is qualified as a straight line (dashed
line on the plots).
Thus, from the series of transformations, a straight line
qualifies a Weibull distribution. We show the cases $T=2, 4, 8$ and $20$
for which the best
$c_T$ are respectively $c_2=0.73, c_4=0.80, c_8=0.90$ and $c_{20}\approx
1.05$. The other
curves allow one to estimate the sensitivity of the representation of $P_T$
in terms of
a Weibull as a function of the choice of the exponent $c_T$. These
simulations confirm
convincingly our proposal that a Weibull distribution remains quasi-stable
for many orders
of convolutions, once the exponent $c_T$ is correspondingly ajusted. We
observe on
Fig.\ref{figweibullstable},\ref{figweibullstable1},\ref{figweibullstable2},\ref{
figweibullstable3}
 that the Weibull representation is accurate over more than
five orders of magnitude of the pdf $P_T$. Only for the largest time scale
$T=20$, we observe
significant departure from the Weibull representation.

\pagebreak


\section{APPENDIX C: Calculation of the cumulants of $P(\delta S)$ in the
diagonal case}

The integral
\be
I_i = \int_{-\infty}^{+\infty} du ~ e^{-\frac{u^2}{2 d_i}+ i k w_i u^q}~.
\ee
can be perturbatively expanded as
\begin{eqnarray}
I_i &=& 2 \sum_{m=0}^{\infty} \frac{(-k^2  w_i^2)^m}{(2m!)} \int_0^{+\infty}
du ~e^{-\frac{u^2}{2 d_i}} u^{2 q m} \\
&=& (2 d_i)^{\frac{1}{2}}
\sum_{m=0}^{\infty} \frac{(-2^q k^2  w_i^2 d_i^q)^m}{(2m!)} \int_0^{+\infty}
dt ~e^{-t} t^{qm-\frac{1}{2}}\nonumber \\
&=& (2 d_i)^{\frac{1}{2}}
\sum_{m=0}^{\infty} \frac{(-2^q k^2  w_i^2 d_i^q)^m}{(2m!)}
\Gamma\left(q m +\frac{1}{2}\right) \nonumber \\
&=&\sqrt{2 \pi d_i} \sum_{m=0}^{\infty} \frac{(2 q m -1)!!}{(2 m)!}
(w_i^2 d_i^q)^m (- k^2)^m \nonumber.
\eea
We recall the definition $\Gamma(x) = \int_0^{\infty} dt\,e^{-t} t^{x-1}$ and
the property $\Gamma(n+ 1/2)=\pi^{1/2} \frac{(2n-1)!!}{2^n}$ when $n$ is an integer
and $(2n-1)!!= (2n-1)(2n-3)(2n-5)...5.3.1$.

The density $\hat{P}_S$ the
refore becomes
\bea
\hat{P}_S(k) &=&  1 - k^2 \frac{(2 q-1)!!}{2!} \sum_i w_i^2 d_i^q \\
&& + k^4 \left[ \frac{(4 q -1)!!}{4!} \sum_i (w_i^2 d_i^q)^2 +
\left(\frac{(2 q -1)!!}{2!}\right)^2 \sum_{i<j} (w_i^2 d_i^q) (w_j^2 d_j^q)
\right]
+ \cdots
\eea
that can be exponentiated to extract the cumulants $c_m(q)$
\begin{eqnarray}
\hat{P}_S (k) &=&  \exp\biggl[ \sum_m \frac{c_m}{m!} (i k)^m \biggl]\\
&=&
\exp\biggl[ - k^2 \frac{(2 q -1)!!}{2!} \sum_i (w_i^2 d_i^q) +
k^4 \left[\frac{(4 q -1)!!}{4!} -\frac{1}{2} \left(\frac{(2
q-1)!!}{2!}\right)^2\right]
\sum_i (w_i^2 d_i^q)^2 +\cdots \biggl] ~.\nonumber
\eea

Pushing the calculation to the next orders, we get the sixth and eigth
cumulants
\begin{eqnarray}
\frac{c_6(q)}{6!} &=&
\left\{ \frac{(6q-1)!!}{6!} -\frac{(4q-1)!!}{4!}\frac{(2q -1)!!}{2!}
+\frac{1}{3}\left[\frac{(2q-1)!!}{2!}\right]^3\right\}\sum_i (p_i^2 d_i^q)^3 \\
\frac{c_8(q)}{8!} &=&
\left\{ \frac{(8q-1)!!}{8!} -\frac{(6q-1)!!}{6!}\frac{(2q-1)!!}{2!}
+\frac{(4q-1)!!}{4!}\left[\frac{(2q-1)!!}{2!}\right]^2
-\frac{1}{4}\left[\frac{(2q-1)!!}{2!}\right]^4\right\}\times\nonumber\\
&&\,\,\,\,\,\,\sum_i (w_i^2 d_i^q)^4 .
\nonumber
\eea
The general cumulant is found by recurrence\,:
\be
\frac{c_{2 r}(q)}{(2 r)!} = \left\{\sum_{n=0}^{r-2} (-1)^n
\frac{[2 (r- n)q -1]!!}{(2r-2n)!}\left[\frac{(2q-1)!!}{2!}\right]^n\,\,\,
- \frac{(-1)^r}{r} \left[\frac{(2q-1)!!}{2!}\right]^r\right\}
\sum_i (w_i^2 d_i^q)^{r}.
\ee
Note that the above computation  is valid even when $q$ is real and the
interaction
term is proportional to ${\rm sign}(u_i) |u_i|^q$. In this case the
interaction
is still an odd function of $u$ and the derivation goes through exactly
with the same combinatorics as above. The result is
\be
\frac{c_{2 r}(q)}{(2 r)!} = 2^{q r} \left\{\sum_{n=0}^{r-2} (-1)^n
\frac{\Gamma\left((r-n)q+\frac{1}{2}\right)}{(2r-2n)! \pi^{1/2}}
\left[\frac{\Gamma\left(q+\frac{1}{2}\right)}{2!\pi^{1/2}}\right]^n\,\,\,
- \frac{(-1)^r}{r}
\left[\frac{\Gamma\left(q+\frac{1}{2}\right)}{2!\pi^{1/2}}\right]^r\right\}
\sum_i (w_i^2 d_i^q)^{r}.
\ee

\pagebreak
\section{APPENDIX D: Generalization of the extreme deviation theorem of
Frisch and
Sornette [1997] to obtain the tail structure of $P(\delta S)$ in the
diagonal case for $c>1$}

We start from the definition $\delta S = \sum_{i=1}^N w_i \delta x_i$ and
the corresponding
equation for its probability density function\,:
\be
P_N(\delta S) = \underbrace{\int \cdots\int}_N
e^{-\sum_{i=1}^N f_i(\delta x_i)}\,\delta\left(\delta S - \sum_{i=1}^N
w_i \delta x_i\right)\,d\delta x_1\cdots d\delta x_N~,
\label{solut}
\ee
where we have used the parameterization
\be
P_i(\delta x_i) \equiv e^{-f_i(\delta x_i)}~,
\ee
with
\be
f_i(\delta x_i) = ({\delta x_i \over \chi_i})^c~, ~~~{\rm with}~~c > 1~.
\label{gshswhjwhj}
\ee

All integrals in (\ref{solut}) are from $-\infty$ to $+\infty$. The delta
function expresses
the constraint on the sum.

We need the following conditions on the functions $f_i$ (see [Frisch and
Sornette, 1997] for
precisions)\,:
\begin{itemize}
\item[(i)] $f_i(\delta x_i) \to +\infty$ sufficiently fast to ensure the
normalization of the
pdf's.
\item[(ii)] $f_i''(\delta x_i)>0$ (convexity), where $f''$ is the second
derivative
of $f$\,.
\item[(iii)] $\lim_{x \to \infty} x^2f''(x)=+\infty$.
\end{itemize}

Under these assumptions, the leading-order
expansion of $P_N(\delta S)$ for large $\delta S$ and finite $N \ge 1$ is
obtained
by a generalization of the Laplace's method which here amounts to remark
that the set of
$\delta x_i^*$'s that maximize the integrant in (\ref{solut}) are solution of
\be
f_i(\delta x_i^*) = C_N(\delta S)~,
\label{wdwdv}
\ee
where $C_N(\delta S)$ is independent of $i$. In words, the leading behavior of
$P_N(\delta S)$ is obtained by the set of $\delta x_i$'s that occurs with
the same
probability. The $\delta x_i^*$ obey
\be
\sum_{i=1}^N  w_i \delta x_i^* = \delta S~.
\label{qgqjcqjxn}
\ee
Expanding $f_i(\delta x_i)$ around $\delta x_i^*$ yields
\be
f_i(\delta x_i) = f_i(\delta x_i^*) + a_i h_i + b_i h_i^2 + ...
\ee
where $a_i \equiv f_i'(\delta x_i^*)$, $b_i = {1 \over 2} f_i''(\delta
x_i^*)$ and
$h_i \equiv \delta x_i - \delta x_i^*$ obey the condition
\be
\sum_{i=1}^N  w_i h_i = 0~.
\ee
We ignore the terms of order higher than two as they do not contribute to
the leading order.
We rewrite
\be
a_i h_i + b_i h_i^2 = b_i (h_i +{a_i \over 2b_i})^2 -{a_i^2 \over 4b_i}
= {b_i \over w_i^2} (H_i +{a_i w_i \over 2b_i})^2 -{a_i^2 \over 4b_i}~,
\ee
where the $H_i = w_i h_i$ verify $\sum_{i=1}^N  H_i = 0$.
Expression (\ref{solut}) then becomes
\be
P_N(\delta S) = e^{-N ~C_N(\delta S)}~e^{\sum_{i=1}^N {a_i^2 \over 4b_i}}~
\underbrace{\int \cdots\int}_{N-1}
e^{-\sum_{i=1}^N {b_i \over w_i^2} (H_i +{a_i w_i \over 2b_i})^2}
\,dH_1 \cdots d H_{N-1}~.
\label{soluqfvqt}
\ee
The integral in (\ref{soluqfvqt}) is evaluated by setting $y=\sum_{j=1}^N
{a_j w_j \over 2b_j}$ and
$\lambda_j = {b_j \over w_j^2}$ in the identity
\be
\underbrace{\int\cdots\int}_{N-1}
e^{- \sum_{j=1}^{N-1} \lambda_j h_j^2 - \lambda_N (y- h_1-\cdots
-h_{N-1})^2 } ~dh_1\cdots dh_{N-1} = \pi^{N-1 \over 2} ~\sqrt{\Lambda \over
\prod_{j=1}^N \lambda_j} ~ e^{-\Lambda y^2}~,
\label{gaussident}
\ee
where $\Lambda$ is defined by
\be
{1 \over \Lambda} = \sum_{j=1}^N {1 \over \lambda_j}~.
\ee
This identity is obtained by viewing $y$ as the sum of the $N$ Gaussian
variables $y_i$'s.

For the case (\ref{gshswhjwhj}) with $c > 1$, the condition (\ref{wdwdv})
together with (\ref{qgqjcqjxn}) yields
\be
f_i(\delta x_i^*) = ({\delta x_i^* \over \chi_i})^c = ({\delta S \over
\chi})^c~, ~~~~{\rm i.e.}~~
~~~~~x_i^* =  \delta S ~{\chi_i \over \chi}~,
\label{wdwdrgqv}
\ee
where
\be
\chi = \sum_{j=1}^N w_j \chi_j = \sum_{j=1}^N w_j d_j^{q \over 2}
\ee
(with $c=2/q$) and
\be
{a_i^2 \over 4b_i} = {c \over 2(c-1)} ~({\delta S \over \chi})^c~.
\ee
We also have
\be
y \equiv \sum_{j=1}^N {a_j w_j \over 2b_j} = \sum_{j=1}^N { w_j f'(\delta
x_j^*)
 \over f''(\delta x_j^*)} = {\chi \delta S \over c-1}~.
\ee
\be
{1 \over \Lambda} = {c(c-1) \over 2 X^2}~\biggl({\delta S \over
\chi}\biggl)^{c-2}~,
\ee
where
\be
X^2 \equiv \sum_{j=1}^N w_j^2 \chi_j^2 ~.
\ee
This yields the contribution
\be
e^{-\Lambda y^2} = \exp \biggl( - {c \over 2(c-1)} {\chi^2 \over X^2}
({\delta S \over \chi})^c
\biggl)~.
\ee
Regrouping all terms in (\ref{soluqfvqt}) leads to
\be
P_N(\delta S) = \pi^{N-1 \over 2}~{1 \over X ~\prod_{j=1}^N w_j \chi_j}~
\biggl[{2 \over c(c-1)} ({\delta S \over \chi})^{2-c} \biggl]^{N-1 \over 2}~
\exp \biggl( -{N \over 2(c-1)} ({\delta S \over {\hat \chi}})^c \biggl)~,
\label{eezesdfwx}
\ee
where
\be
{\hat \chi}^c \equiv {\chi^c \over c{\chi^2 \over N X^2} - (2-c)}
= {(\sum_{j=1}^N w_j \chi_j )^c \over c {(\sum_{j=1}^N w_j \chi_j )^2
\over N \sum_{j=1}^N w_j^2 \chi_j^2} -(2-c)}~.
\ee
These results are valid for $c>1$. The extreme tail of the portfolio wealth
distribution is thus controlled completely by ${\hat \chi}^c$ which is its
characteristic decay value.

Note that ${\chi^2 \over N X^2} = 1$ for identical assets $\chi_i = \chi$
when all
weights $w_i$ are equal to $1/N$. In this case, the exponential term in
(\ref{eezesdfwx}) simplifies into
\be
P_N(\delta S) \sim \exp \biggl( -N ({\delta S \over \chi})^c \biggl)~.
\label{eezeqqsdfwx}
\ee

\pagebreak


\section{APPENDIX E: Computation of the characteristic function defined by
eq.(\ref{spopojklrft})}

The characteristic function of the distribution
$P_S(\delta S)$ of portfolio wealth variations for correlated assets with
Weibull
distributions is given by
\be
{\hat P}_S(k) = {1 \over (2\pi)^{N/2} \det{V}^{1/2} }
 \prod_{i=1}^N \biggl( \int du_i \biggl) ~ e^{-{1 \over 2} ~U'V^{-1} U +
ik~\sum_{i=1}^N ~w_i ~u_i^q} ~ .
\label{spopojklrft}
\ee

We recall the definition of the functional generator [Sornette, 1998]
\be
\hat{P}^q_S(k,J_i) = {1 \over (2\pi)^{N/2} \det{V}^{1/2} }
\int\left(\prod_i^N du_i\right)
e^{-\frac{1}{2} u V^{-1} u  + i k \sum_i w_i u_i^q + \sum_i J_i u_i}~.
\ee
When the integral is a Gaussian ($k=0$), we get
\bea
\hat{P}^q_S(0,0) &=& 1\\
\hat{P}^q_S(0,J_i) &=&  e^{\frac{1}{2} J V J}~.
\eea
With the property
\be
f\left(\frac{\delta}{\delta J_i}\right) \int\left(\prod_i^N du_i\right)
e^{-\frac{1}{2} U V^{-1} U + \sum_i J_i u_i} =
\int\left(\prod_i^N du_i\right)
e^{-\frac{1}{2} U V^{-1} U  + \sum_i J_i u_i} f(u_i)~,
\ee
we can formally  express the characteristic function as
\be
\hat{P}^q_S(k) =
\left. e^{i k \sum_i w_i \frac{\delta^q}{\delta J_i^q}} e^{\frac{1}{2} J V J}
\right|_{J_i=0}~.
\label{kldqmqn}
\ee

We first consider the case $q=3$.
The first non-vanishing perturbative contribution for this case is obtained by
expanding the formal expression above up to second order in $k$
\be
\hat{P}^3_S(k) =
\left. \left[1+ i k \sum_i w_i \frac{\delta^3}{\delta J_i^3} -
\frac{k^2}{2} \sum_{i,j} w_i w_j \frac{\delta^3}{\delta J_j^3}
\frac{\delta^3}{\delta J_i^3}
\right] e^{\frac{1}{2} J V J} \right|_{J=0}~.
\ee
The first order vanishes because
\be
\frac{\delta^3}{\delta J_i^3} e^{\frac{1}{2} J V J} = e^{\frac{1}{2} J V J}
\left\{ (V J)_i^3 + 3 V_{i i} (V J)_i\right\}
\ee
is zero when $J=0$.
\footnote{We adopt here the compact notation $J V J = \sum_{i j} J_i V_{i
j} J_j$
and $(V J)_i = \sum_l V_{i l} J_l$.}
The second order contribution comes from
\begin{eqnarray}
\frac{\delta^3}{\delta J_j^3} \frac{\delta^3}{\delta J_i^3}
 e^{\frac{1}{2} J V J} &=&
e^{\frac{1}{2} J V J} \times \\
& & \,\,\,\,\,\,
\left\{
(V J)_j^3 (V J)_i^3 + 9 V_{i j} (V J)_j^2 (V J)_i^2 + 3 V_{i i} (V J)_j^3
(V J)_i +
3 V_{j j} (V J)_j (V J)_i^3 + \right.\nonumber\\
& &\,\,\,\,\,\,  9 V_{i j} V_{i i} (V J)_j^2 + 9 (2 V_{i j}^2 +
V_{i i} V_{jj} ) (V J)_j (V J)_j +
9 V_{i j} V_{j j} (V J)_i^2 + \nonumber\\
& &\,\,\,\,\,\, \left. 6 V_{i j}^3 + 9 V_{ii} V_{i j} V_{jj}
\right\}\nonumber ~.
\eea
By putting to zero the source term $J$, this expression leads to
\be
\hat{P}^3_S(k) =
1 - \frac{k^2}{2} \sum_{ij}\left(6 w_i \left(V_{i j}\right)^3 w_j +
9 w_i V_{ii} V_{ij} V_{jj} w_j \right)~.
\label{qml,klSD}
\ee
\vskip 0.5cm

This result can be usefully represented with diagrams in the following way.
Let us associate to each factor $V_{ij}$ the propagator diagram and to each
factor $i g_3 w_j$ the vertex diagram as shown in Fig.\ref{D1}, where we
have defined
the coupling constant
\be
g_3 = 3! k~.
\label{hfqklqmkqm}
\ee

The two contributions in (\ref{qml,klSD}) can thus be represented
by propagators connecting vertices as in Fig.\ref{D2}.
Having defined the coupling constant $g_3$ in (\ref{hfqklqmkqm})
allows us to interpret easily the coefficient in front of each diagram
corresponding to the two terms in the expression
\be
\hat{P}^3_S(k) =
1 - g_3^2 \left(\frac{1}{3! 2}\sum_{ij} w_i
\left(V_{i j}\right)^3 w_j + \frac{1}{2^3} \sum_{ij} w_i V_{ii} V_{ij}
V_{jj} w_j \right)~.
\ee
The coefficient in front of each term (diagram)
is equal to $1/S$ where $S$ is the symmetry factor of each diagram with
respect to permutation of lines and vertices.
The first diagram has a symmetry under exchange of $3$ propagators and $2$
vertices,
therefore we have $S=3!\times 2!$.
The second diagram has a symmetry under exchange of the $2$ lines forming
the loop of
each vertex, giving a contribution of $2!\times 2!$. It also has the
symmetry under permutation
of the two vertices. The total symmetry factor is therefore $S=2!\times
2!\times 2!$.

A systematic way to  keep under control the symmetry factor is to compute
it diagramatically as follows. The second-order derivative operator with
respect to
$J_i$ and $J_j$ is  represented by the two vertices in the
left hand side of Fig.\ref{D3}.

The $J$-independent term is given by pairwise combining each leg of each vertex
in all the possible topologically inequivalent way, taking  into account
the multiplicity of each configuration.
Fig.\ref{D3} shows how to proceed. The coefficient in front of  the
vertices of the
left hand side comes from the perturbative expansion.

As a first step, let us consider the first leg of the first  vertex.
We can either contract it with another leg (two possibile
contractions) of the same vertex or with a leg
of the second vertex (three possible contractions). This is summarized in
the first equality of Fig.\ref{D3}.
The first contribution of the second equality is the result
of the contraction with multiplicity $3$ of the residual
leg of the first vertex of the first diagram of the line above.
Considering the possible contractions of the second contribution of the first
equality generate the  other two diagrams.
At the end of the combining procedure, we have two inequivalent diagrams,
with multiplicities $3!$ and $3^2$. The resulting coefficient in front
of each diagram can now be interpreted in terms of the symmetries of the
diagram as anticipated.

These rules can be generalized to higher orders and other ``interaction''
terms,
i.e. $\sum_i w_i u_i^q$ for general value of $q$ or even generic functions
$f(u_i)$ which admit expansions in power series of $u_i$.
As an example, we give the first order correction to the characteristic
function for
generic value of $q$ odd.

The perturbative expansion is
\be
\hat{P}^q_S(k) =
\left. \left[1+ i \frac{g_q}{q!}\sum_i w_i \frac{\delta^q}{\delta J_i^q} +
\frac{1}{2} \left(\frac{i g_q}{q!}\right)^2
\sum_{i,j} w_i w_j \frac{\delta^q}{\delta J_j^q} \frac{\delta^q}{\delta J_i^q}
\right] e^{\frac{1}{2} J V J}
\right|_{J=0}
\ee
where we have defined the auxiliary coupling constant $g_q= q! k$.
Now, instead of taking explicitely the derivatives and keep the term
independent on $J$
as the result, let us use the diagrammatic formalism.
At this order, we have two vertices each with $q$ legs, representing the
partial derivatives
with respect to $J_i$ and $J_j$ and to which we associate the value $i g_q
w_i$
and $i g_q w_j$ respectively.
Now, let us combine pairwise each leg of the two vertices to form all the
possible
topologically inequivalent diagrams. We have collected in Fig.\ref{D4} the
sequence of all
the diagrams.
Each diagram is characterized by the number $l$ of loops on each vertex and
the number
$q-2 l$ of lines connecting the two vertices giving therefore a contribution
\be
(i g_q)^2 \frac{1}{S_l} \sum_{i,j} w_i \left(V_{ii}\right)^l
\left(V_{ij}\right)^{q-2 l} \left(V_{jj}\right)^l w_j~,
\label{yhdjqmqmmq}
\ee
where each loop around vertex $i$ contributes to a factor $V_{ii}$ and each
propagator
connecting the vertices $i$ and $j$ gives a factor $V_{ij}$.
It is now easy to determine the symmetry factor: we have $l$ symmetries
under the exchange
of the two lines of each loop at each vertex ($(2!)^l \times (2!)^l$),
the symmetry of the $l$ loops at each vertex ($(l!)\times (l!)$),
one symmetry under the exchange of the $q-2 l$ internal propagators
($(2q-l)!$),
and the reflection symmetry under the exchange of the two vertices ($2!$).
The total factor is therefore
$S_n = (q-2l)! (2!)^{2l+1} (l!)^2 $ and the characteristic function up to
second
order in $k$ reads
\be
\hat{P}^q_S(k) =
1-g_q^2 \sum_{l=0}^{(q-1)/2} \frac{1}{(q-2l)!}
\frac{1}{(2!)^{2l +1}} \frac{1}{(l!)^2}
\sum_{i,j} w_i \left(V_{ii}\right)^l
\left(V_{ij}\right)^{q-2l} \left(V_{jj}\right)^l w_j ~.
\ee

The diagrammatic expansion becomes very useful for calculating the higher
cumulants.
A well known result of diagrammatic perturbation theory tells us that
neglecting
the disconnected diagrams, which do not occur at second order, corresponds
to compute
the logarithm of characteristic function. Therefore, the set of connected
diagrams
at $m$-th order give us directly the $m$-th cumulant coefficient $c_m$.

Let us give as an explicit example the computation of the
fourth  cumulant   $c_4(3)$.
In fig.\ref{D5}, the  result of the contraction procedure is shown, where
we keep
only the  connected diagrams. It is explicitely
\begin{eqnarray}
c_4(3) &=& 4! (3!)^4 \sum_{i_1,i_2,i_3,i_4} w_{i_1} w_{i_2} w_{i_3} w_{i_4}
\left\{ \frac{1}{2^4}
V_{i_1 i_2}^2 V_{i_1 i_3} V_{i_2 i_4} V_{i_3 i_3} V_{i_4 i_4} +
\right. \nonumber\\
& & \,\,\,\, \frac{1}{2^3}
V_{i_1 i_2}^2 V_{i_1 i_3} V_{i_2 i_3} V_{i_3 i_4} V_{i_4 i_4} +
\frac{1}{2^4}
V_{i_1 i_2}^2 V_{i_1 i_3} V_{i_2 i_4} V_{i_3 i_4}^2 + \nonumber\\
& &\,\,\,\,\left.
\frac{1}{3! 2^3}
V_{i_1 i_2} V_{i_1 i_3} V_{i_1 i_4} V_{i_2 i_2} V_{i_3 i_3} V_{i_4 i_4} +
\frac{1}{4!}
V_{i_1 i_2} V_{i_1 i_3} V_{i_1 i_4} V_{i_2 i_3} V_{i_2 i_4} V_{i_3 i_4}
\right\}~.
\label{fjqmmqkqqmm}
\eea

We can thus formally generalize the result for the $m$-th cumulant
as
\be
c_m(q)= m! (q!)^m \sum_{i_1,\ldots,i_m} w_{i_1}\cdots w_{i_m}
\sum_{{\cal G}_m (q)}
\frac{1}{S(\{l_r\},\{n_{rs}\})} \prod_{r=1}^m \left(V_{i_r i_r}\right)^{l_r}
\prod_{r<s=1}^m \left(V_{i_r i_s}\right)^{n_{rs}}
\ee
where
${\cal G}_m(q)$ is the set of  all the topologically inequivalent connected
diagrams with $m$ vertices of $q$ legs. Each diagram is characterized
by the number of loops at  each  vertex, $\{l_r\}$, and the number
of lines connecting each couple of vertices, $\{n_{rs}=n_{sr} \}$.
These numbers have to satisfy the constraints
\be
2 l_r + \frac{1}{2} \sum_{s\neq r} n_{rs} =  q~,
\ee
which embody the fact that each vertex has $q$ legs.
The symmetry factor $S$ is obviously the most difficult part to determine.
The safest procedure is to compute it with the contraction rule case by case.
For each diagram, it is of the form
\be
S(\{l_r\},\{n_{rs}\})= (2!)^{\sum_{r=1}^m l_r}
\prod_{r=1}^m l_r! \prod_{r<s=1}^m n_{rs}! S_v(\{l_r\},\{n_{rs}\})\,\,.
\ee
Each factorial comes from the various symmetries under the exchange of the
propagators, and
we have  isolated the contribution of the residual symmetries
of the diagram under exchange of the vertices.
Fig.\ref{D6} summarizes the vertex symmetry factors for the diagrams
contributing to
$c_4(3)$.

\pagebreak

\vfill\eject

REFERENCES\,:

R. Arad Wiener, The implications of a long-tailed distribution structure
to portfolio selection and capital asset pricing, PhD thesis, Princeton
University (1975).

V. Bawa, E.J. Elton and M.J. Gruber, Journal of Finance, vol. XXXIV,
1041-1047 (1979).

T. Bollerslev, R.Y. Chous and K.F. Kroner, ARCH modeling in Finance - A review
of the theory and empirical evidence,
J. Econometrics 52, 5-59 (1992).

J.-P. Bouchaud, D. Sornette, C. Walter and J.-P. Aguilar,
Taming large events\,: Optimal portfolio theory for strongly fluctuating
assets,
International Journal of Theoretical and Applied Finance 1, 25-41 (1998).

S. Cambanis, S. Huang and G. Simons, On the theory of elliptically
contoured distributions,
Journal of Multivariate Analysis 11, 368-385 (1981).

P. Embrechts, C. Kl\"uppelberg and T. Mikosh, Modelling extremal events
(Springer-Verlag, Applications of Mathematics 33, 1997).

R.F. Engle, Econometrica 50, 987 (1982).

E. Fama, Management Science 11, 404-419 (1965).

U. Frisch and D. Sornette, Extreme deviations and applications,
J. Phys. I France 7, 1155-1171 (1997).

C.C. G\'eczy, Some generalized tests of mean-variance efficiency and
performance,
working paper

B.V. Gnedenko and Kolmogorov, A.N., Limit distributions
for sum of independent random variables, Addison Wesley, Reading MA (1954).

E.J. Gumbel, {\it Statistics of extremes} (Columbia University
Press, New York, 1960).

M.J.R. Healy, Matrices for statistics (Clarendon Press, Oxford, 1986).

J.E. Ingersoll, Jr., Theory of financial decision making (Totowa,
N.J. : Rowman \& Littlefield, 1987).

B. Jorgensen, Exponential dispersion models, J. R. Statist. Soc. B 49 (2),
127-162 (1987).

Y. Kano, Consistency property of elliptical probability density functions,
Journal of Multivariate Analysis 51, 139-147 (1994).

D. Karlen, Using projections and correlations to approximate probability
distributions, Computer in Physics 12, 380-384 (1998).

J. Laherr\`ere and D. Sornette,
Stretched exponential distributions in Nature and Economy: ``Fat tails''
with characteristic scales, European Physical Journal B 2, 525-539 (1998)

P. L\'evy, Th\'eorie de l'addition des variables al\'eatoires (Gauthier
Villars, Paris, 1937-1954).

R. Litterman and K. Winkelmann, Estimating covariance matrices, Risk Management
Series, Goldman Sachs (1998).

B.B. Mandelbrot, Journal of Business, 36, 394 (1963)

B.B. Mandelbrot, Fractals and scaling in finance : discontinuity,
concentration, risk :
   selecta volume E (New York : Springer, 1997).

H. Markovitz, Portfolio selection : Efficient diversification of
investments (John Wiley and Sons, New York, 1959).

R. C. Merton, Continuous-time finance, (Blackwell, Cambridge,1990).

S. Mittnik and S.T. Rachev, Reply to comments on 'Modelling asset returns
with alternative stable distributions' and some extensions, Econometric
Reviews 12, 347-389 (1993).

J. Owen and R. Rabinovitch, On the class of elliptical distributions and their
applications to the theory of portfolio choice, The Journal of Finance 38,
745-752 (1983).

C. R. Rao, Linear statistical inference and its applications, 2d ed.
(New York, Wiley, 1973).

P.A. Samuelson, J. Financial and Quantitative Analysis Vol.2, pp. 107-122
(1967).

D. Sornette, Large deviations and portfolio optimization, Physica A 256,
251-283 (1998).

J. von-Neumann and O. Morgenstern, Theory of games and economic behavior
(Princeton, Princeton University Press, 1944)

M. Veltman, Diagrammatica, the path to Feynman diagrams (Cambridge
University Press, 1995).

\pagebreak

\newpage
\begin{figure}
\begin{center}
\epsfig{file=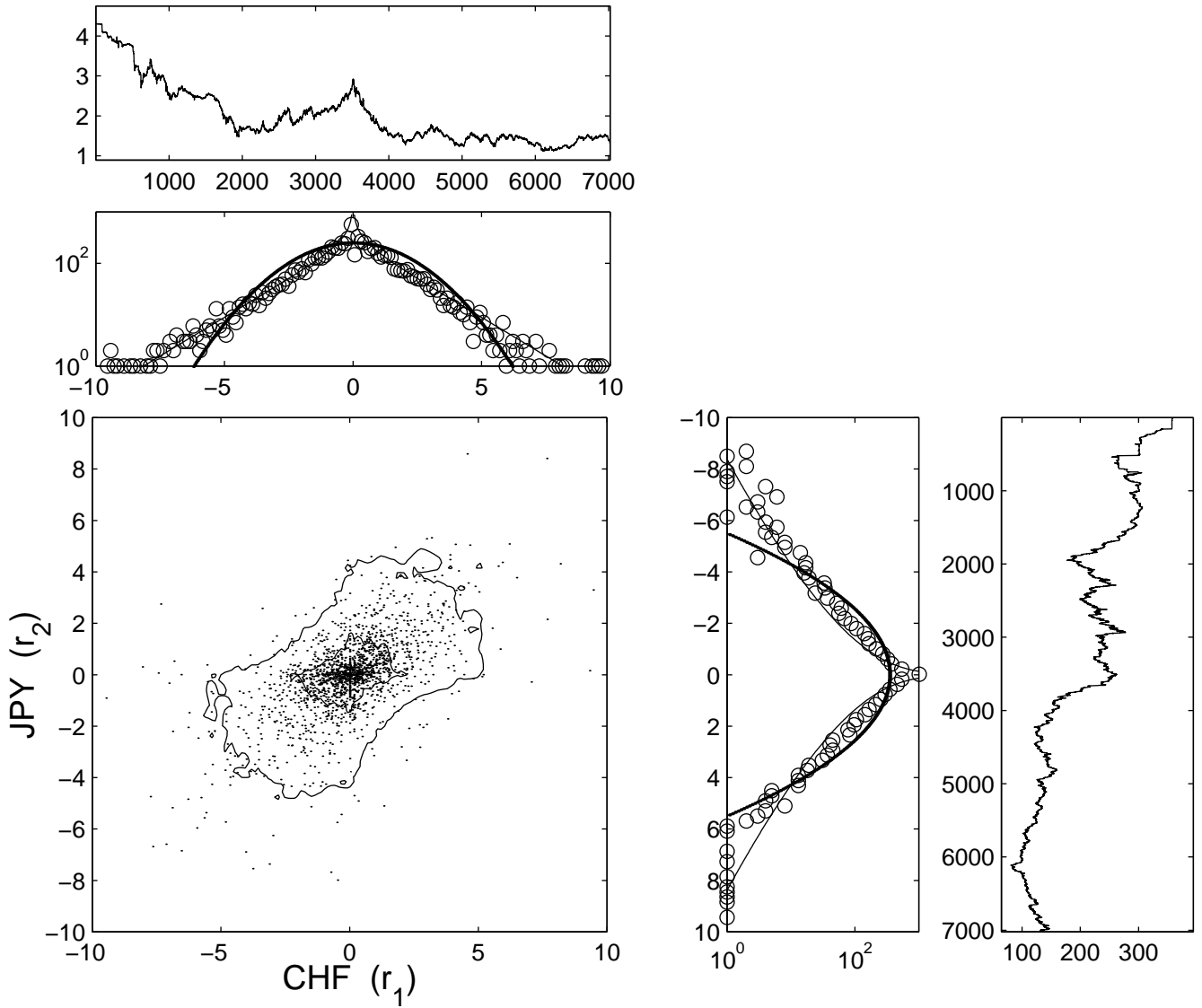}
\caption{\protect\label{JPYCHF(r)} Bivariate distribution of
the daily annualized returns of the
CHF in US \$ ($i=1$) and of JPY in US \$ ($i=2$) for the time interval from
Jan. 1971 to Oct. 1998. One fourth of the data points are represented for
clarity of the
figure. The contour lines define the probability confidence level of 90\%
(outer line), 50\% and 10\%. Also shown are the time series and the marginal
distributions in the panels at the top and on the side.
The parameters for the fit of the marginal pdf's are: CHF in US \$: $A_1=250,
c_1=1.14, r_{01}=2.13$ and JPY in US \$: $A_2=350, c_2=0.8, r_{02}=1.25$.}
\end{center}
\end{figure}

\newpage

\begin{figure}
\begin{center}
\epsfig{file=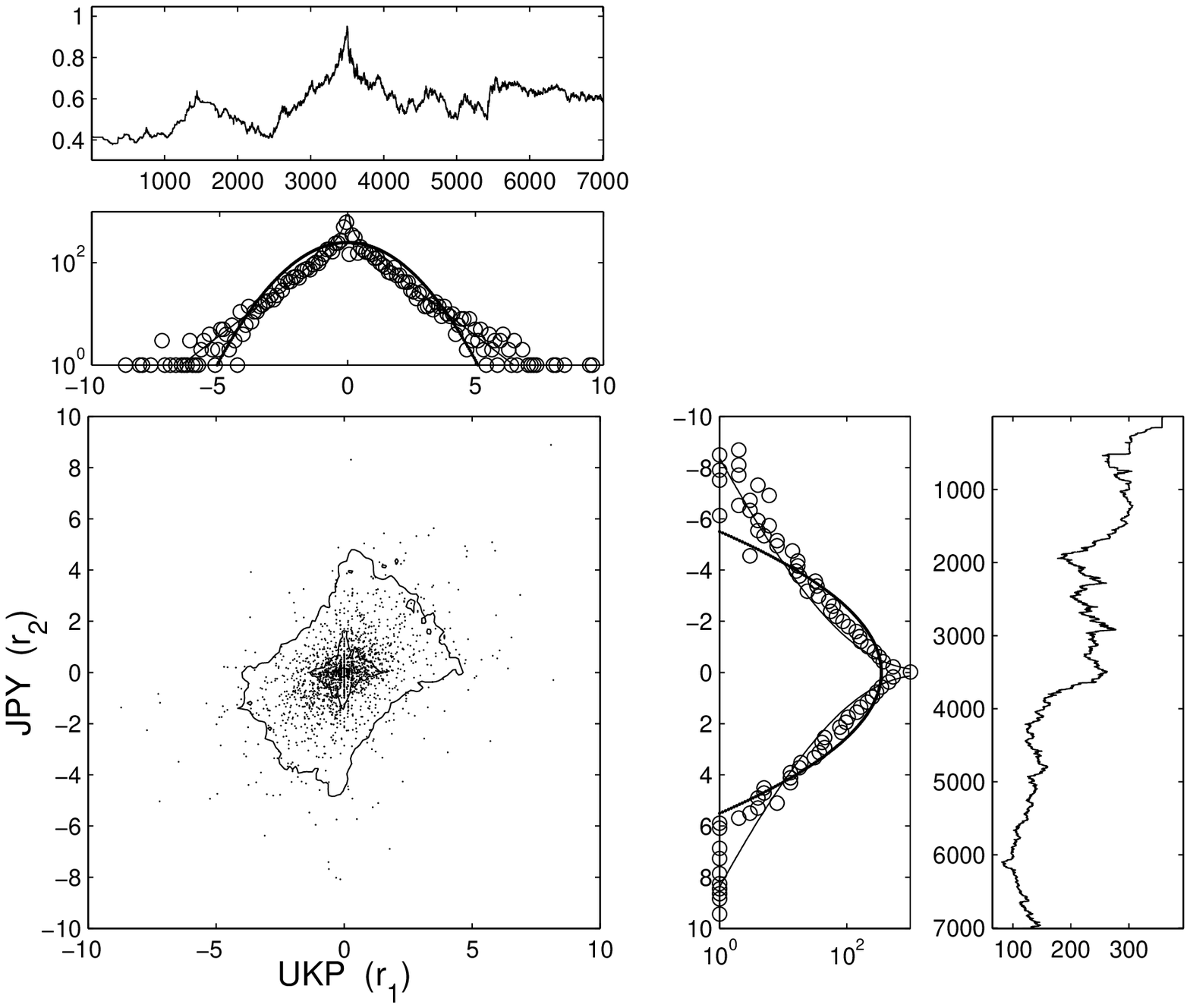}
\caption{\protect\label{JPYUKP(r)}
Bivariate distribution of the daily annualized returns of the
UKP in US \$ ($i=1$) and of JPY in US \$ ($i=2$) for the time interval from
Jan. 1971 to Oct. 1998. One fourth of the data points are represented for
clarity of the
figure. The contour lines define the probability confidence level of 90\%
(outer line), 50\% and 10\%. Also shown are the time series and the marginal
distributions in the panels at the top and on the side.
The parameters for the fit of the marginal pdf's are: UKP in US \$: $A_1=250,
c_1=1.14, r_{01}=1.67$ and JPY in US \$: $A_2=350, c_2=0.8, r_{02}=1.25$.
}
\end{center}
\end{figure}

\newpage

\begin{figure}
\begin{center}
\epsfig{file=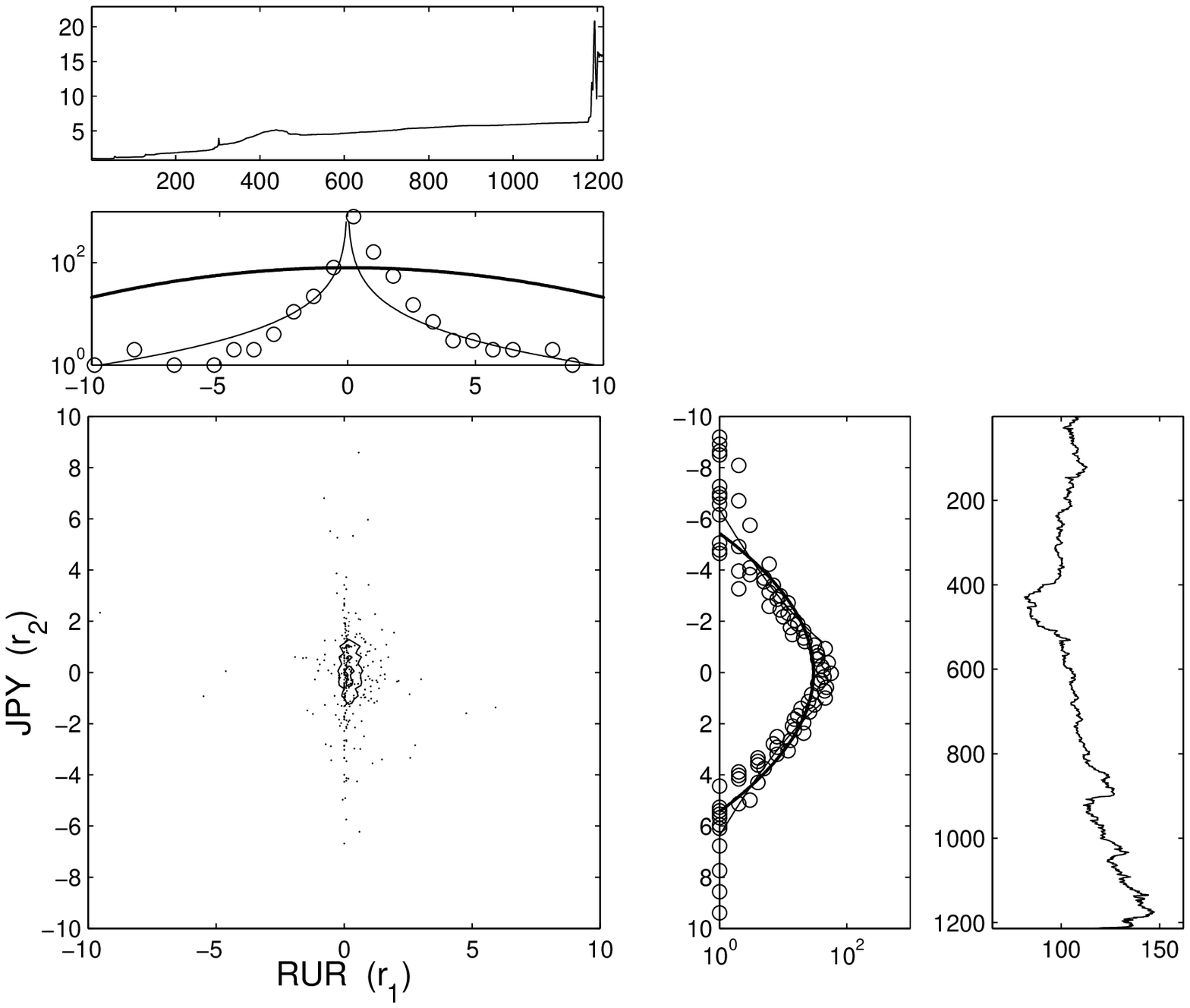}
\caption{\protect\label{JPYRUB(r)}
Bivariate distribution of the daily annualized returns of the
RUR in US \$ ($i=1$) and of JPY in US \$ ($i=2$) for the time interval from
Jun. 1993 to Oct. 1998. One fourth of the data points are represented for
clarity of the
figure. The contour lines define the probability confidence level of 50\%
(outer line) and 10\%. Also shown are the time series and the marginal
distributions in the panels at the top and on the side.
The parameters for the fit of the marginal pdf's are: RUR in US \$: $A_1=80,
c_1=0.38, r_{01}=0.83$ and JPY in US \$ $A_2=120, c_2=0.8, r_{02}=1.25$.
}
\end{center}
\end{figure}

\newpage

\begin{figure}
\begin{center}
\epsfig{file=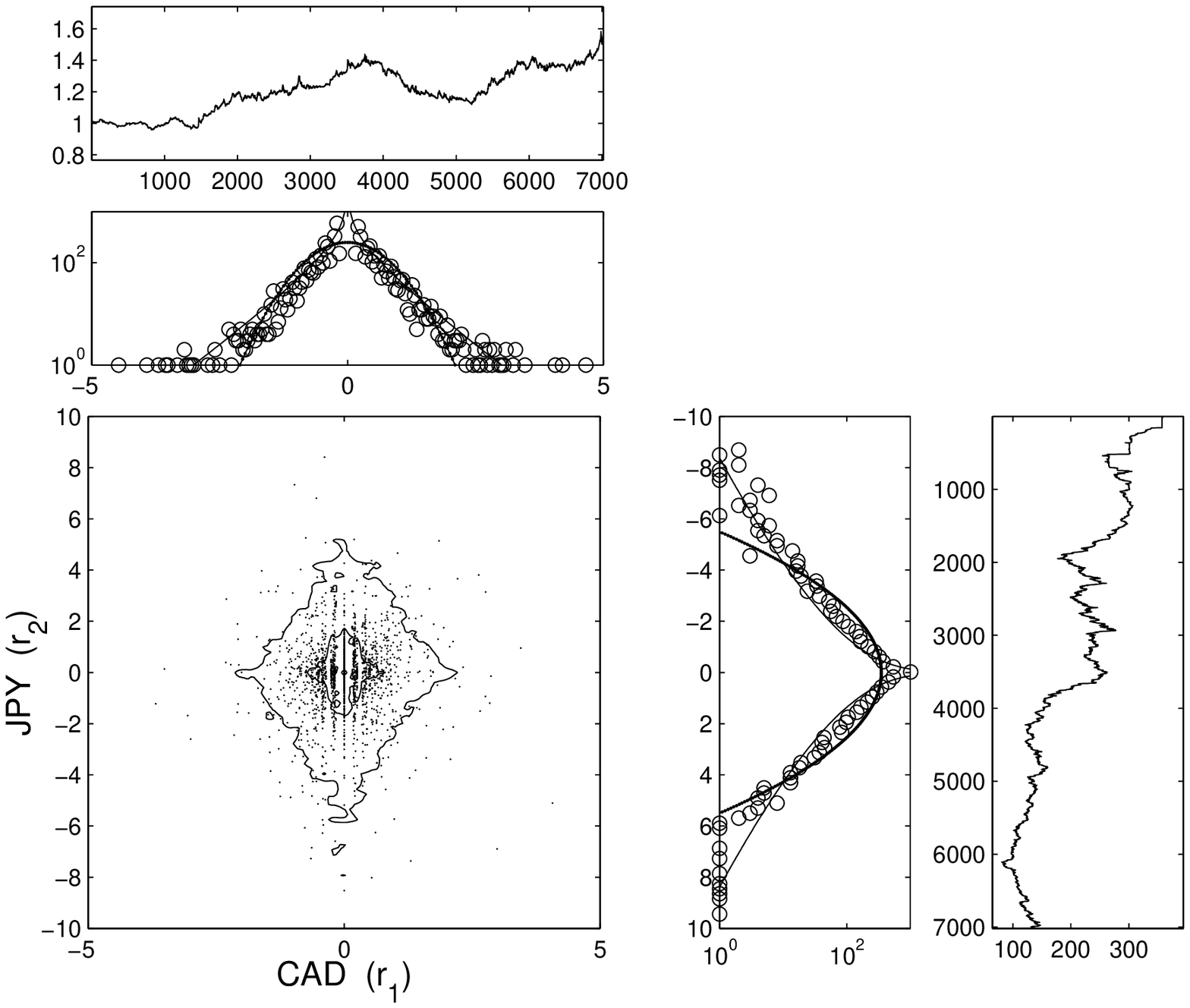}
\caption{\protect\label{JPYCAN(r)}
Bivariate distribution of the daily annualized returns of the
CAD in US \$ ($i=1$) and of JPY in US \$ ($i=2$) for the time interval from
Jan. 1971 to Oct. 1998. One fourth of the data points are represented for
clarity of the
figure. The contour lines define the probability confidence level of 90\%
(outer line), 50\% and 10\%. Also shown are the time series and the marginal
distributions in the panels at the top and on the side.
The parameters for the fit of the marginal pdf's are: CAD in US \$: $A_1=250,
c_1=0.98, r_{01}=0.59$ and JPY in US \$: $A_2=350, c_2=0.8, r_{02}=1.25$.
}
\end{center}
\end{figure}

\newpage

\begin{figure}
\begin{center}
\epsfig{file=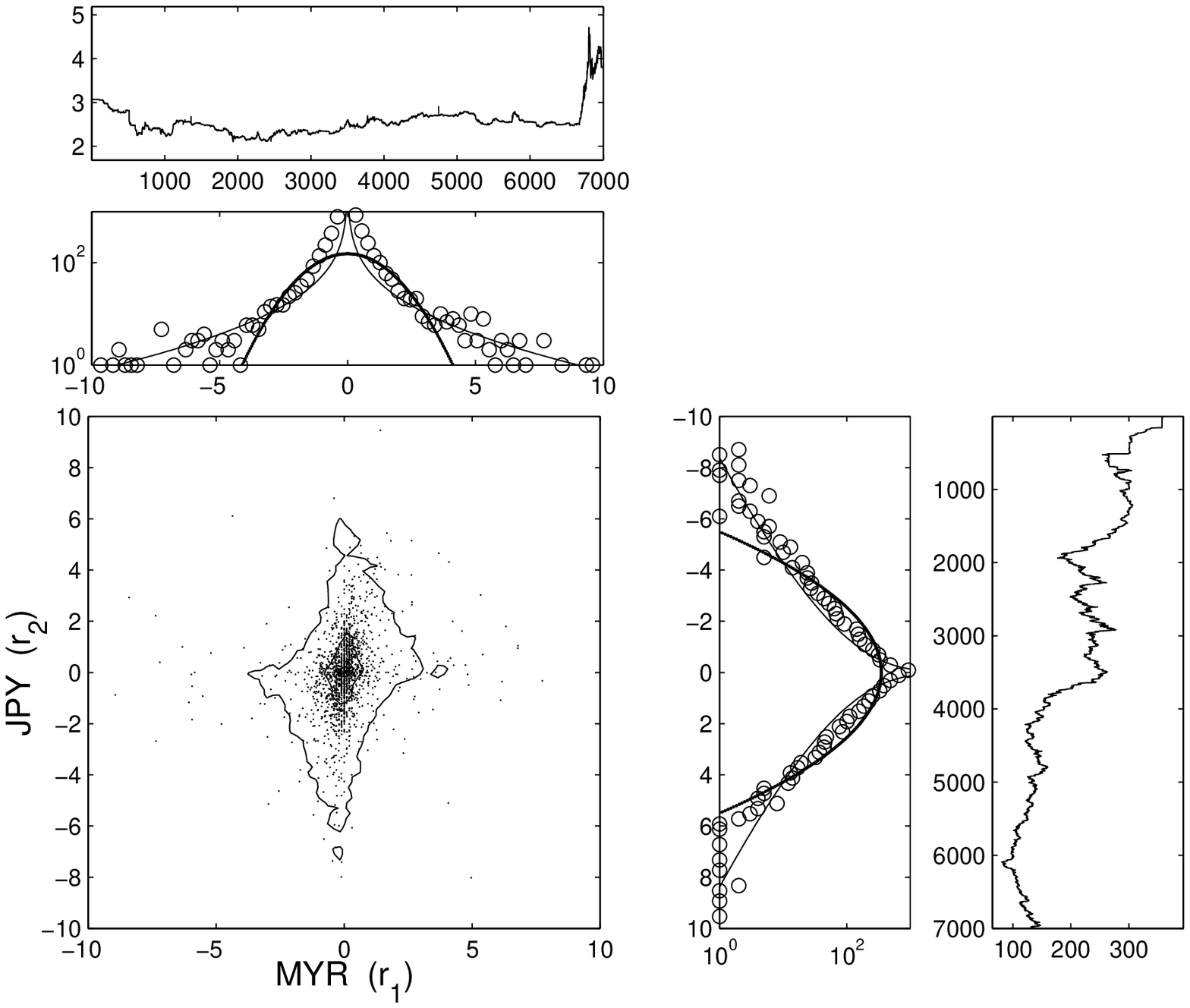}
\caption{\protect\label{JPYMYR(r)}
Bivariate distribution of the daily annualized returns of the
MYR in US \$ ($i=1$) and of JPY in US \$ ($i=2$) for the time interval from
Jan. 1971 to Oct. 1998. One fourth of the data points are represented for
clarity of the
figure. The contour lines define the probability confidence level of 90\%
(outer line), 50\% and 10\%. Also shown are the time series and the marginal
distributions in the panels at the top and on the side.
The parameters for the fit of the marginal pdf's are: MYR in US \$: $A_1=150,
c_1=0.56, r_{01}=1.00$ and JPY in US \$: $A_2=350, c_2=0.8, r_{02}=1.25$.
}
\end{center}
\end{figure}

\newpage

\begin{figure}
\begin{center}
\epsfig{file=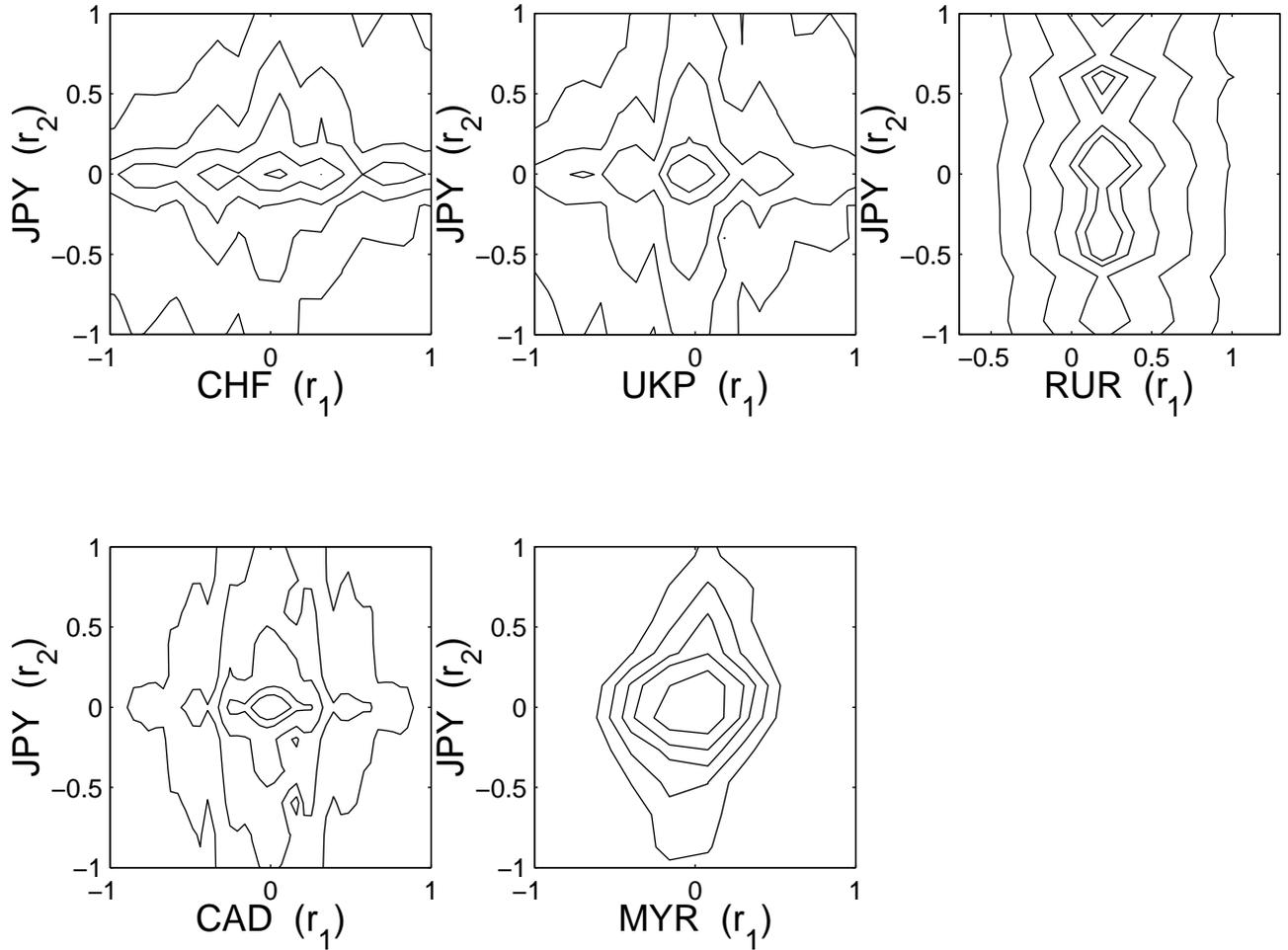}
\caption{\protect\label{figsmallr1}
Close-up of the countour
lines of the bivariate distributions. One can observe that the countour
lines at
the center are not far from elliptical while they depart more and more from
ellipses for larger levels. The corresponding probability levels are:
CHF: 0.44, 0.24, 0.13, 0.07, 0.04
UKP: 0.48, 0.30, 0.17, 0.09, 0.07
RUR: 0.66, 0.56, 0.43, 0.36, 0.34
CAD: 0.60, 0.37, 0.22, 0.12, 0.10
MYR: 0.42, 0.32, 0.25, 0.20, 0.17}
\end{center}
\end{figure}

\newpage

\begin{figure}
\begin{center}
\epsfig{file=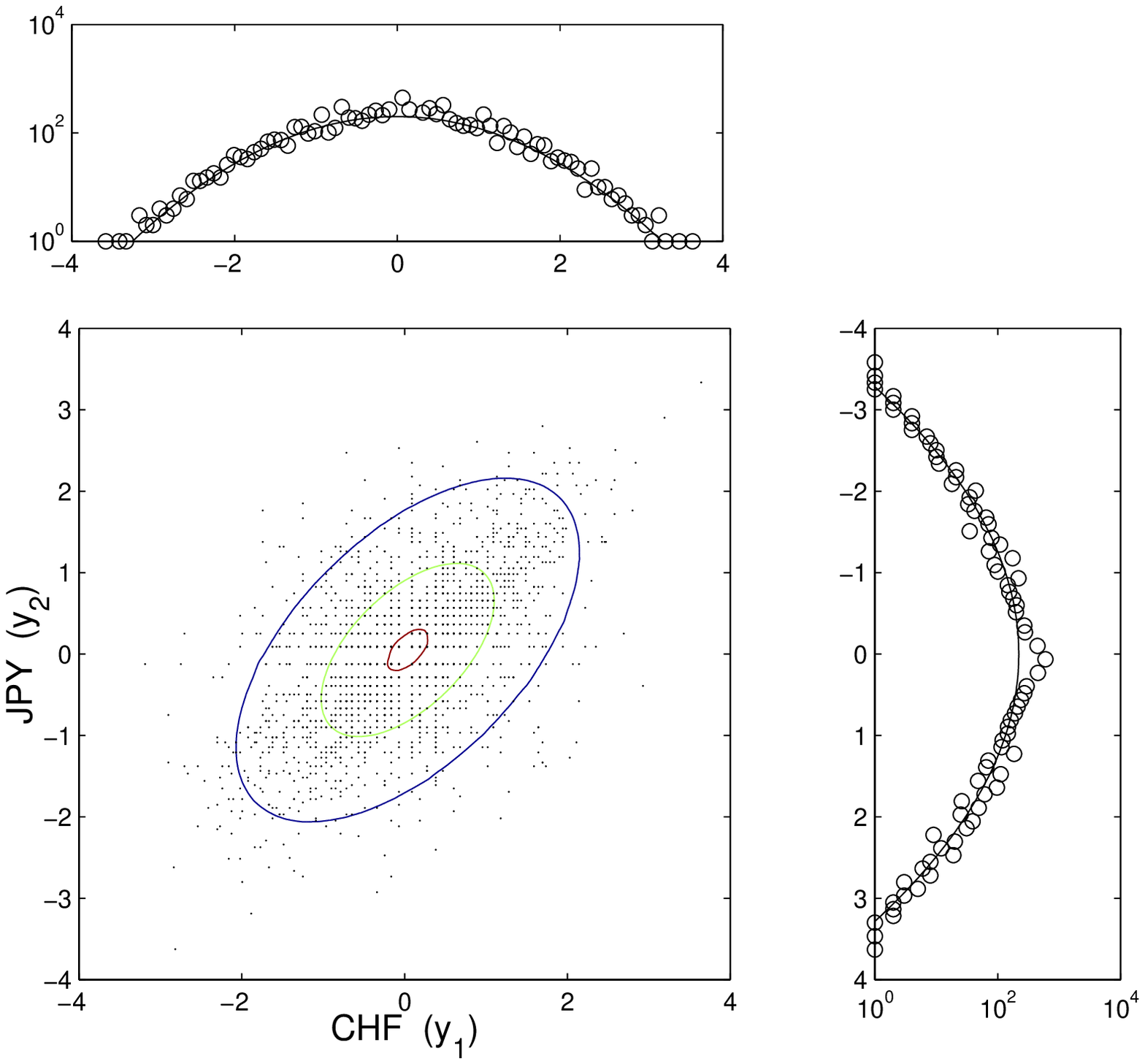}
\caption{\protect\label{JPYCHF(y)}
Bivariate distribution ${\hat P}({\bf y})$ obtained from
Fig.{JPYCHF(r)} using the transformation Eq.(\ref{EQ:transform}). The
contour lines
are defined as in Fig.{JPYCHF(r)}. The upper and right diagrams show the
corresponding
projected marginal distributions, which are Gaussian by construction of the
change
of variable Eq.(\ref{EQ:transform}). The solid lines are fits of the form
$A \exp (-|y|^2/2)$ with $A_1=200, A_2=220$.}
\end{center}
\end{figure}

\newpage

\begin{figure}
\begin{center}
\epsfig{file=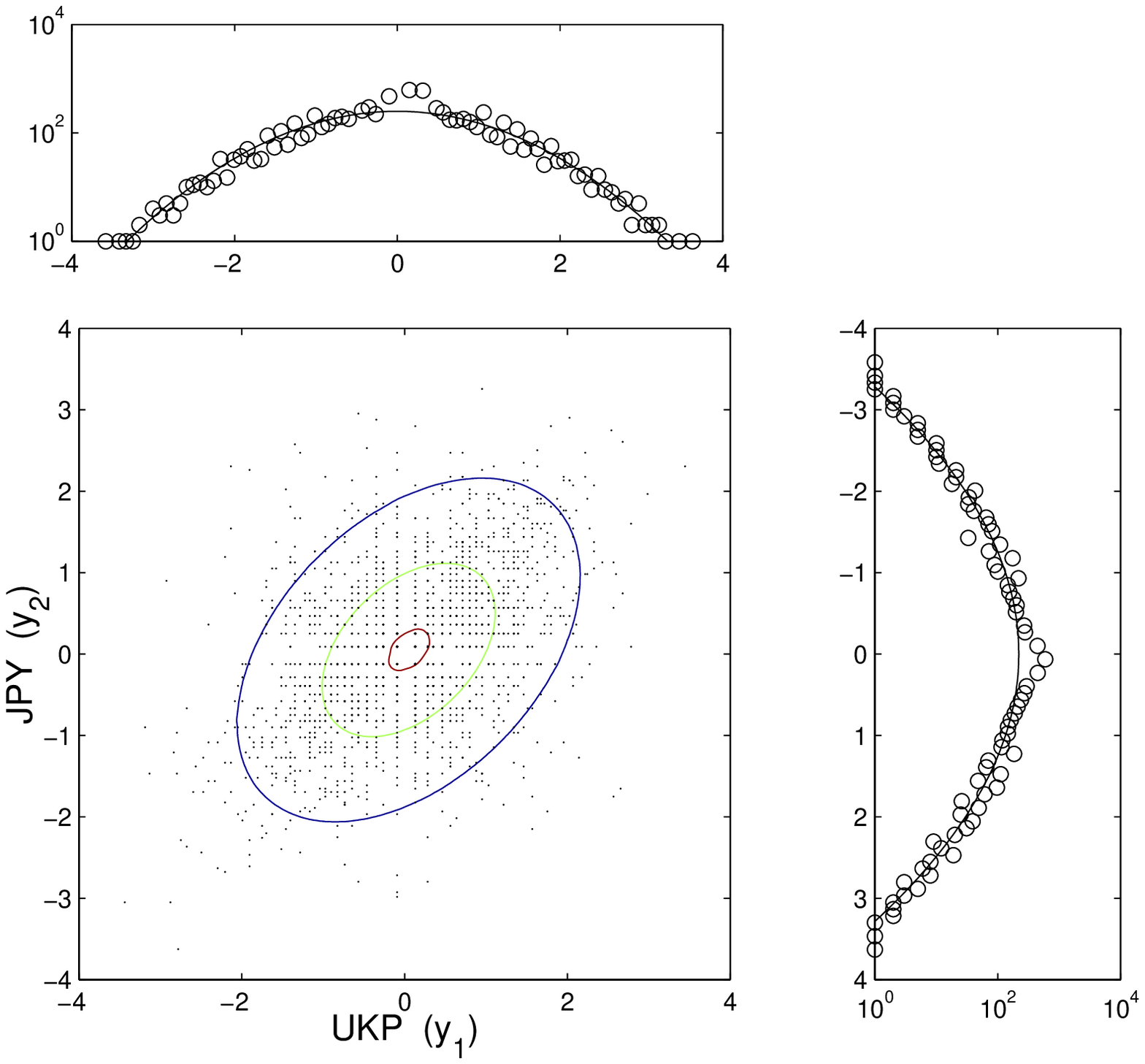}
\caption{\protect\label{JPYUKP(y)}
 Bivariate distribution ${\hat P}({\bf y})$ obtained from
Fig.\ref{JPYUKP(r)} using the transformation Eq.(\ref{EQ:transform}). The
contour lines
are defined as in Fig.\ref{JPYUKP(r)}. The upper and right diagrams show
the corresponding
projected marginal distributions, which are Gaussian by construction of the
change
of variable Eq.(\ref{EQ:transform}). The solid lines are fits of the form
$A \exp (-|y|^2/2)$ with $A_1=250, A_2=220$.}
\end{center}
\end{figure}

\newpage

\begin{figure}
\begin{center}
\epsfig{file=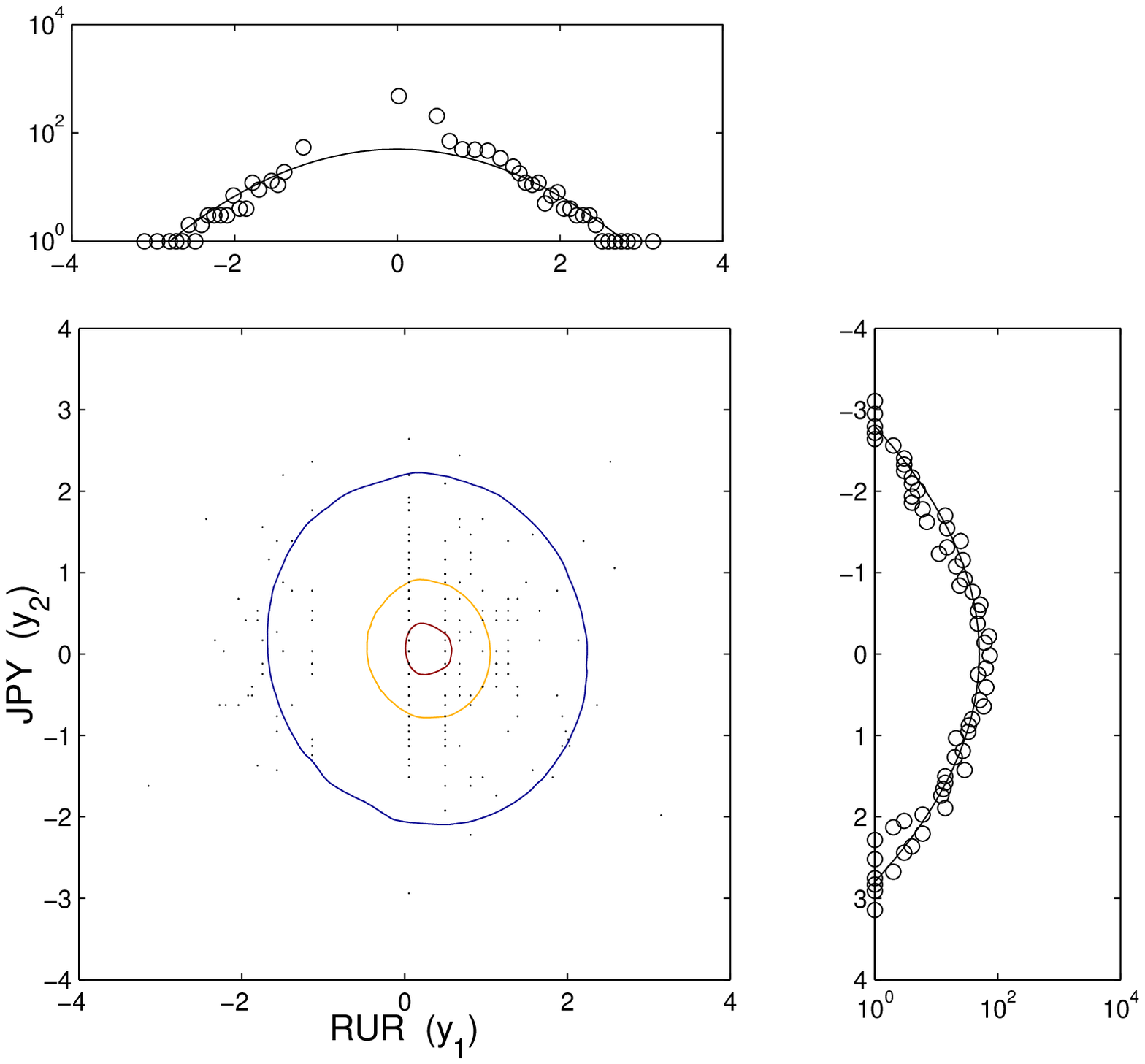}
\caption{\protect\label{JPYRUB(y)}
 Bivariate distribution ${\hat P}({\bf y})$ obtained from
Fig.\ref{JPYRUB(r)} using the transformation Eq.(\ref{EQ:transform}). The
contour lines
are defined as in Fig.\ref{JPYRUB(r)}. The upper and right diagrams show
the corresponding
projected marginal distributions, which are Gaussian by construction of the
change
of variable Eq.(\ref{EQ:transform}). The solid lines are fits of the form
$A \exp (-|y|^2/2)$ with $A_1=50, A_2=50$.}
\end{center}
\end{figure}

\newpage

\begin{figure}
\begin{center}
\epsfig{file=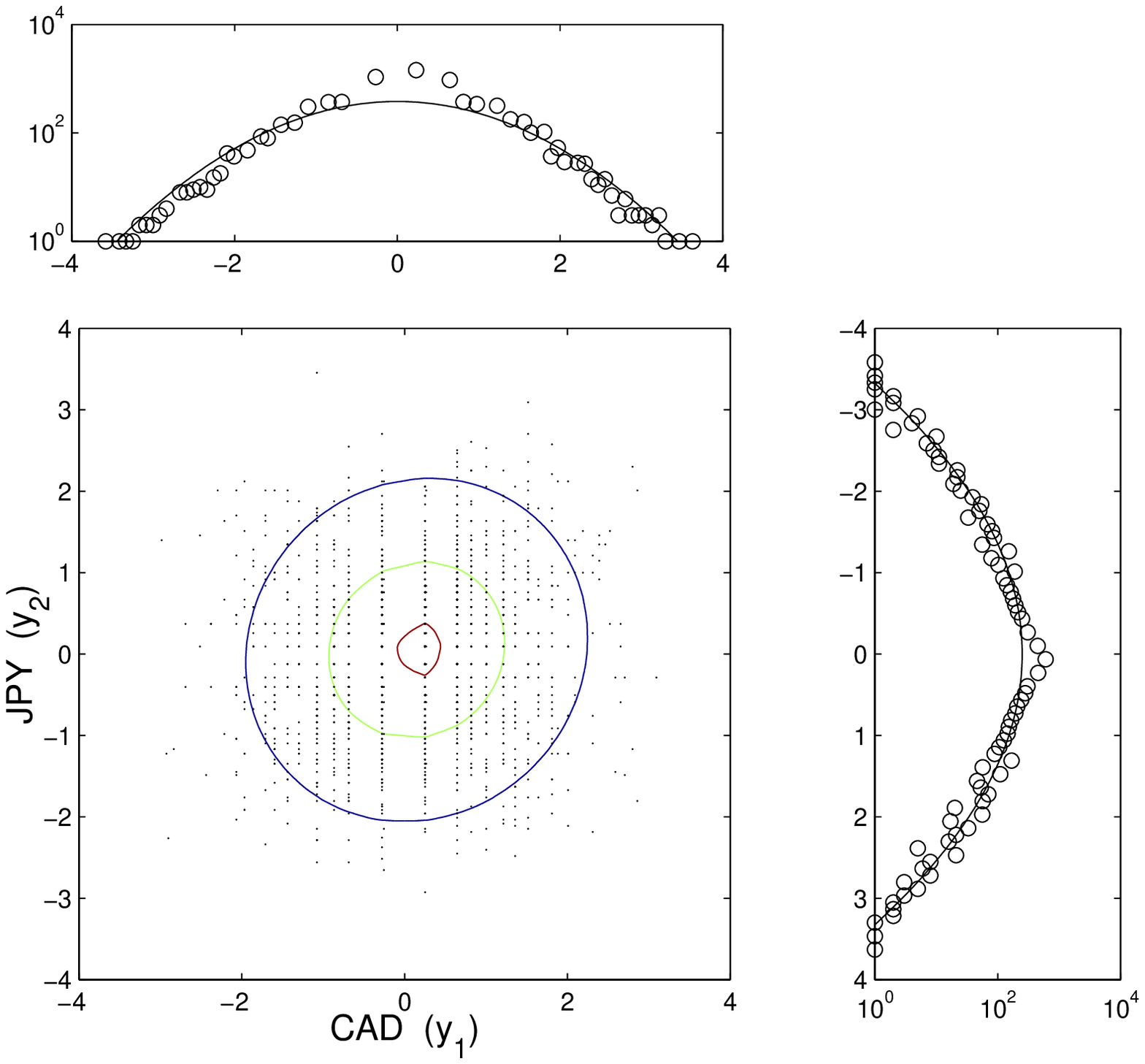}
\caption{\protect\label{JPYCAN(y)}
Bivariate distribution ${\hat P}({\bf y})$ obtained from
Fig.\ref{JPYCAN(r)} using the transformation Eq.(\ref{EQ:transform}). The
contour lines
are defined as in Fig.\ref{JPYCAN(r)}. The upper and right diagrams show
the corresponding
projected marginal distributions, which are Gaussian by construction of the
change
of variable Eq.(\ref{EQ:transform}). The solid lines are fits of the form
$A \exp (-|y|^2/2)$ with $A_1=380, A_2=220$.}
\end{center}
\end{figure}

\newpage

\begin{figure}
\begin{center}
\epsfig{file=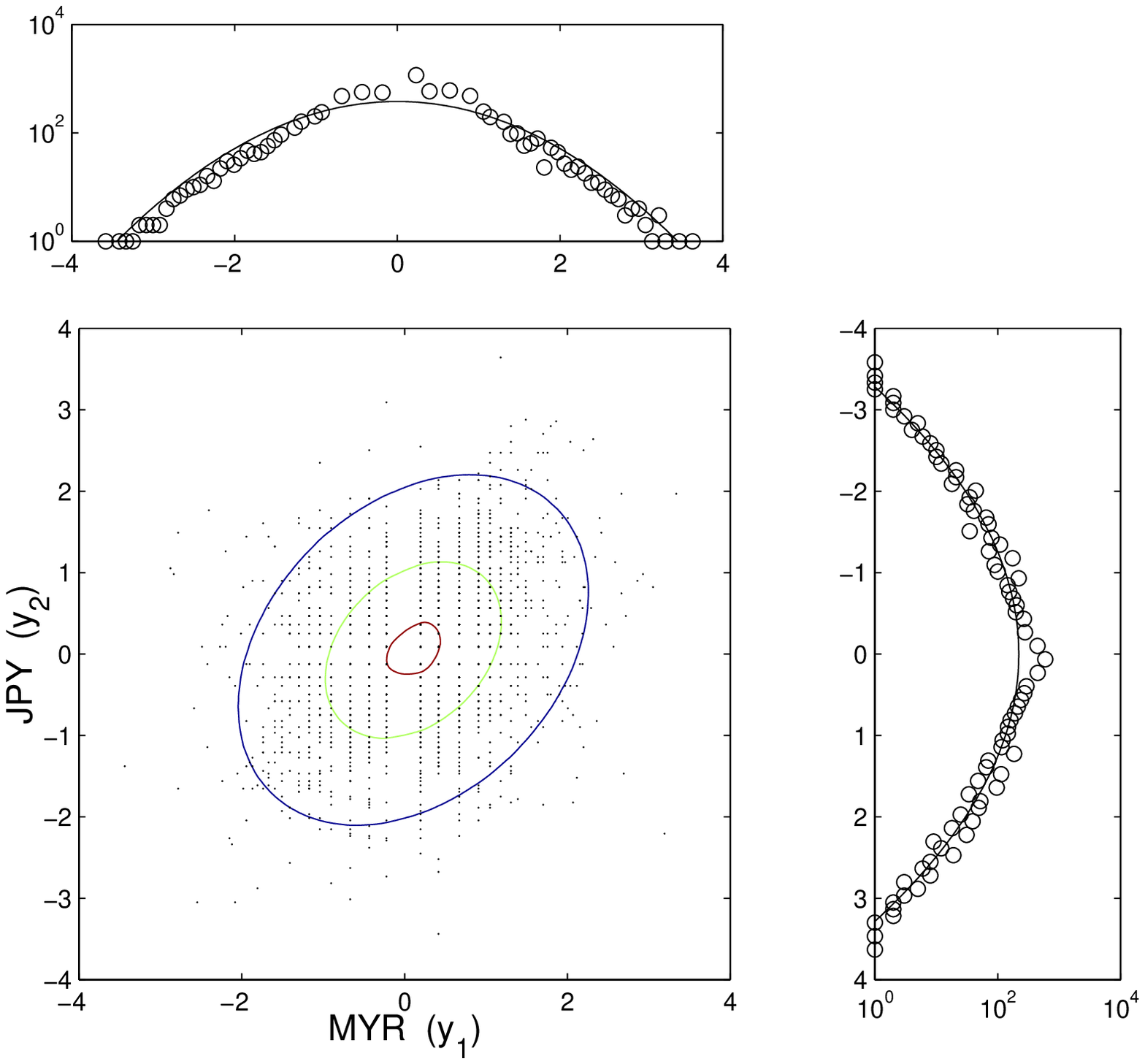}
\caption{\protect\label{JPYMYR(y)}
 Bivariate distribution ${\hat P}({\bf y})$ obtained from
Fig.\ref{JPYMYR(r)} using the transformation Eq.(\ref{EQ:transform}). The
contour lines
are defined as in Fig.\ref{JPYMYR(r)}. The upper and right diagrams show
the corresponding
projected marginal distributions, which are Gaussian by construction of the
change
of variable Eq.(\ref{EQ:transform}). The solid lines are fits of the form
$A \exp (-|y|^2/2)$ with $A_1=380, A_2=220$.}
\end{center}
\end{figure}

\newpage

\begin{figure}
\begin{center}
\epsfig{file=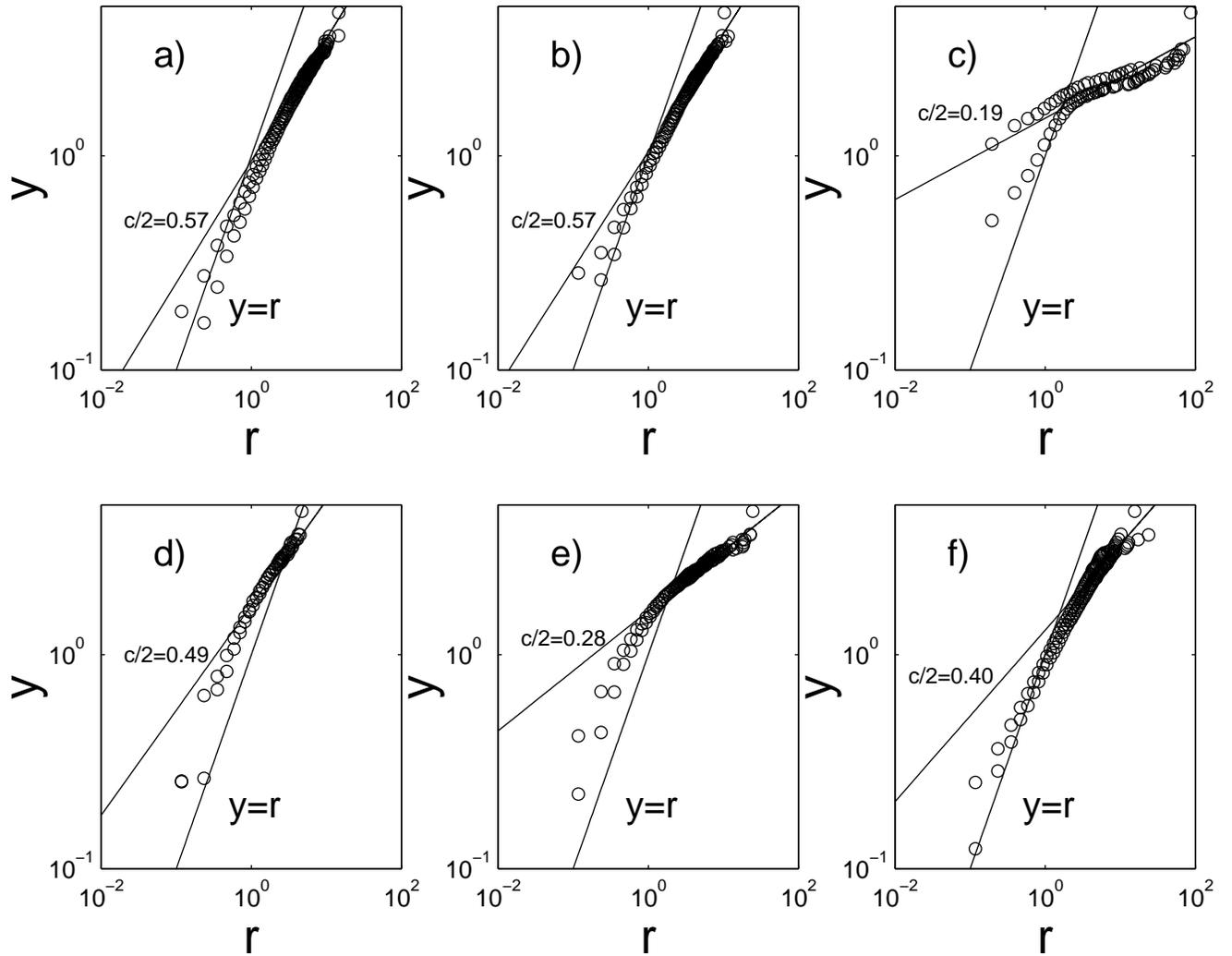}
\caption{\protect\label{y(r)data}
 $y(r)$-transformation defined by Eq.(\ref{EQ:transform}) for a) CHF,
b) UKP, c) RUR, d) CAD, e) MYR and f) JPY.
The negative returns have been folded back to the positive quadrant.
}
\end{center}
\end{figure}

\newpage

\begin{figure}
\begin{center}
\epsfig{file=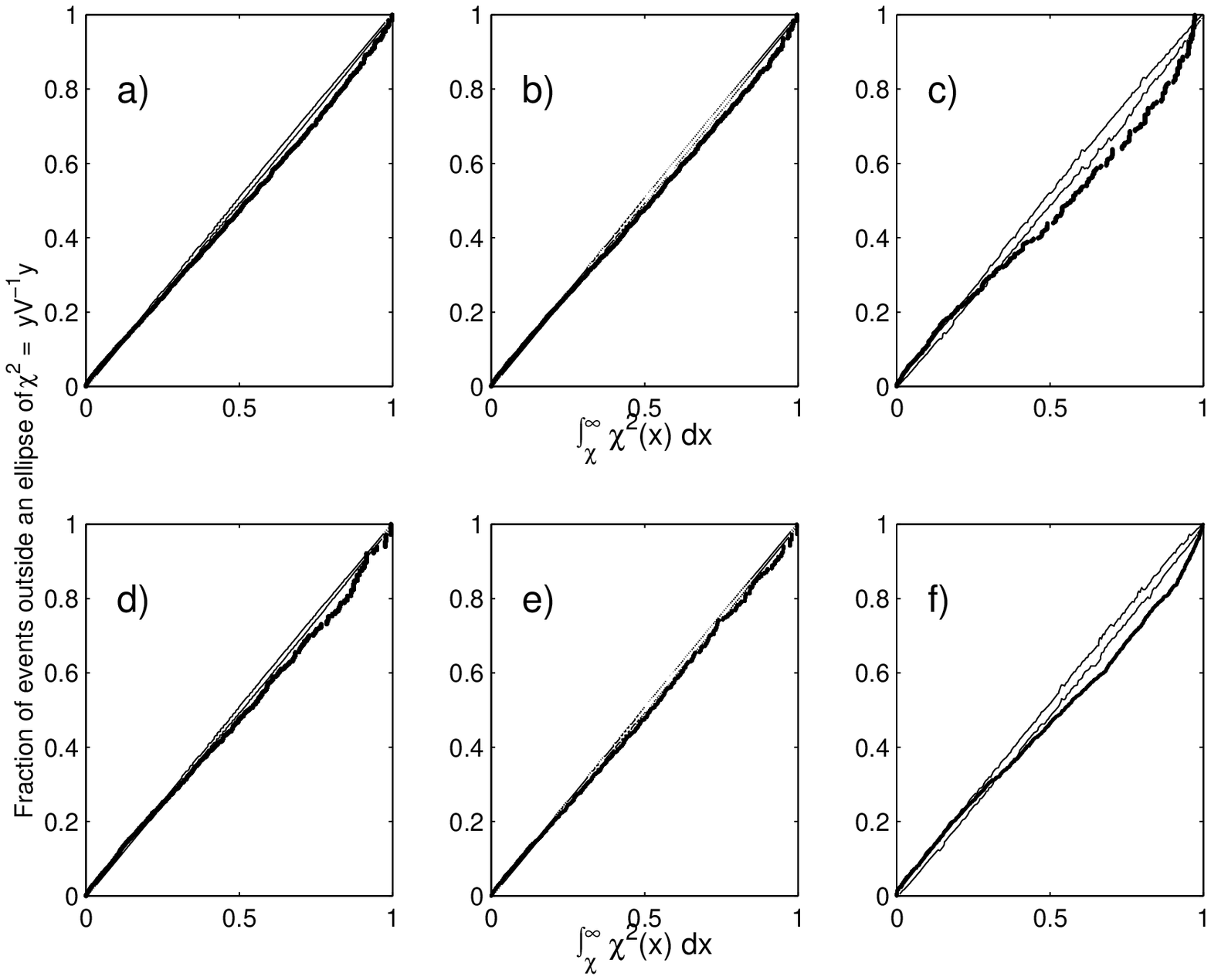}
\caption{\protect\label{figchiy}
a-e: $\chi^2$ cumulative distribution for
$N=2$ degrees of freedom versus the fraction of events
shown in Figures
\ref{JPYCHF(y)},\ref{JPYUKP(y)},\ref{JPYCAN(y)},\ref{JPYRUB(y)},
\ref{JPYMYR(y)} outside an
ellipse of equation $\chi^2 = {\bf y}'V^{-1}) {\bf y}$.
a) CHF-JPY,
b) UKP-JPY, c) RUR-JPY, d) CAD-JPY, e) MYR-JPY. f) same plot as a)-e) but for
$N=6$ degrees of freedom for the data set CHF-UKP-RUR-CAD-MYR-JPY.}
\end{center}
\end{figure}

\newpage

\begin{figure}
\begin{center}
\epsfig{file=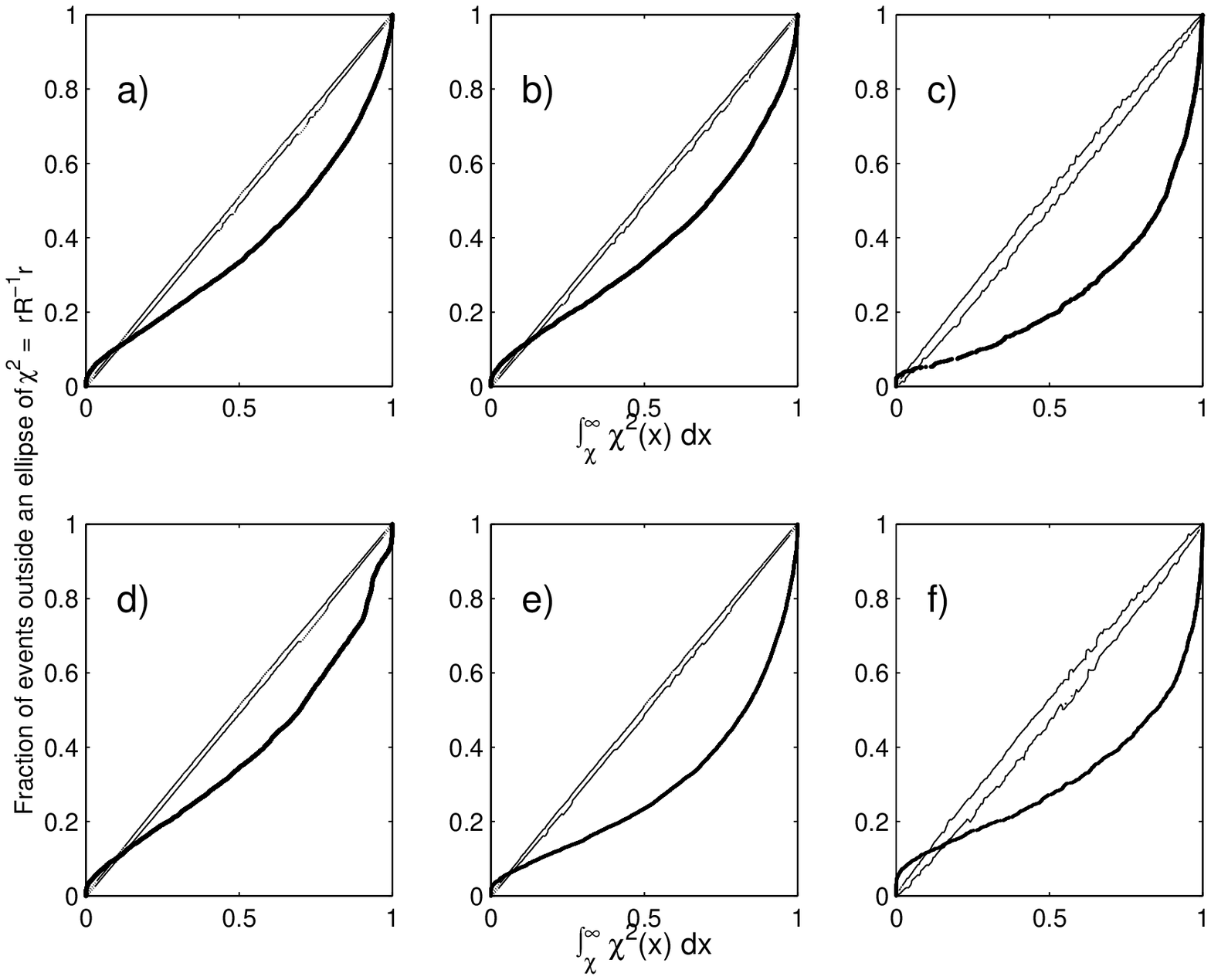}
\caption{\protect\label{figchir}
Same as Fig.(\ref{figchiy}) but for the returns $r$.
a-e: $\chi^2$ cumulative distribution for
$N=2$ degrees of freedom versus the fraction of events
shown in Figures
\ref{JPYCHF(r)},\ref{JPYUKP(r)},\ref{JPYCAN(r)},\ref{JPYRUB(r)},
\ref{JPYMYR(r)} outside an
ellipse of equation $\chi^2 = {\bf r}'{\cal V}^{-1}) {\bf r}$.
a) CHF-JPY,
b) UKP-JPY, c) RUR-JPY, d) CAD-JPY, e) MYR-JPY. f) same plot as a)-e) but for
$N=6$ degrees of freedom for the data set CHF-UKP-RUR-CAD-MYR-JPY.}
\end{center}
\end{figure}

\clearpage
\newpage

\begin{figure}
\begin{center}
\epsfig{file=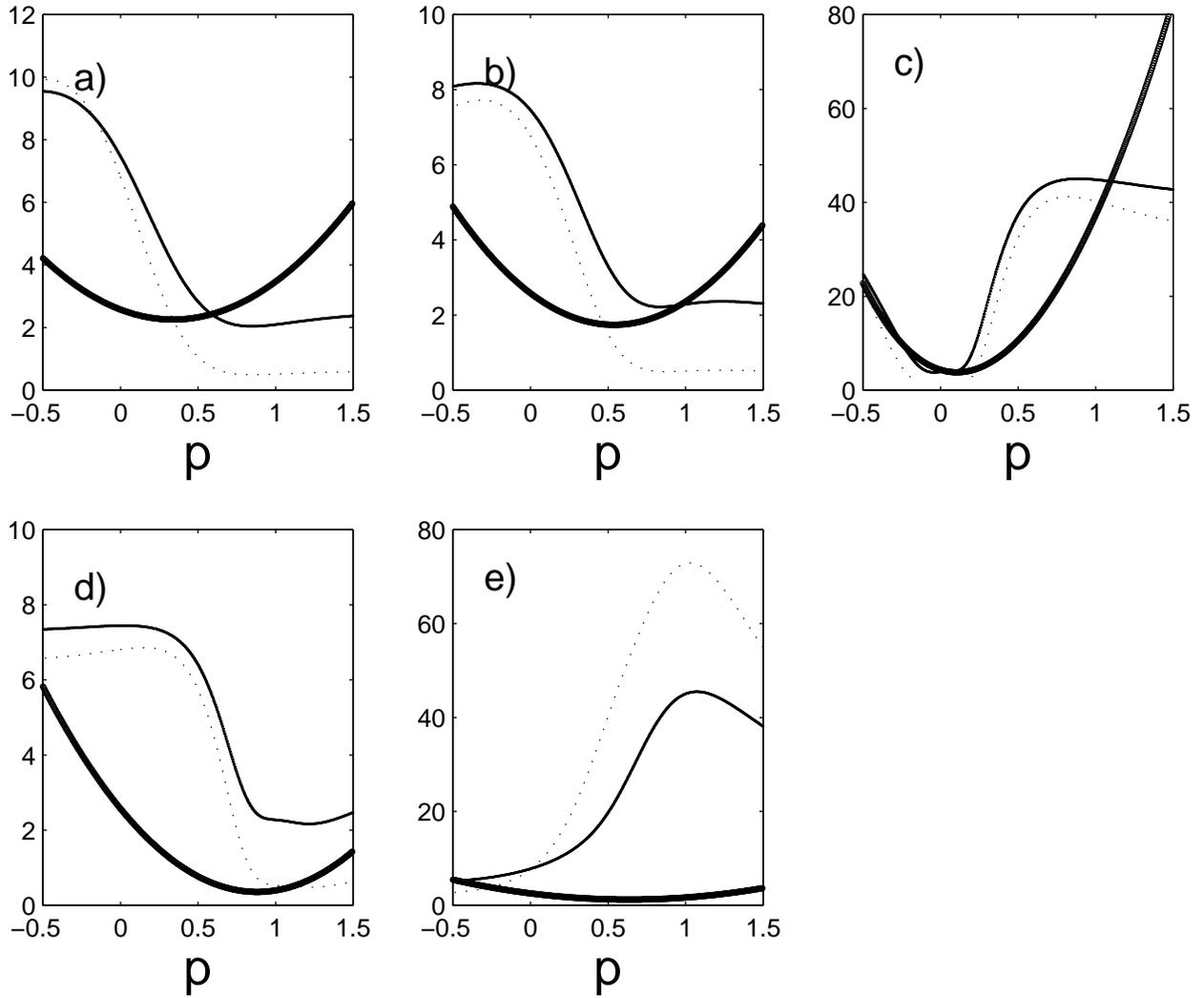}
\caption{\protect\label{figVARKURT}
Variance (thick solid line), excess kurtosis $\kappa$ (thin solid line)
and sixth-order normalized cumulant $\lambda_6$ as a function of the weight
$p$ of currency
$1$ (the weight of currency $2$ is $1-p$).
for the data sets:
a) CHF-JPY,
b) UKP-JPY, c) RUR-JPY, d) CAD-JPY and e) MYR-JPY.
$\kappa$ is divided by $2$ and $\lambda_6$ is divided by $300$.
}
\end{center}
\end{figure}

\clearpage
\newpage

\begin{figure}
\begin{center}
\epsfig{file=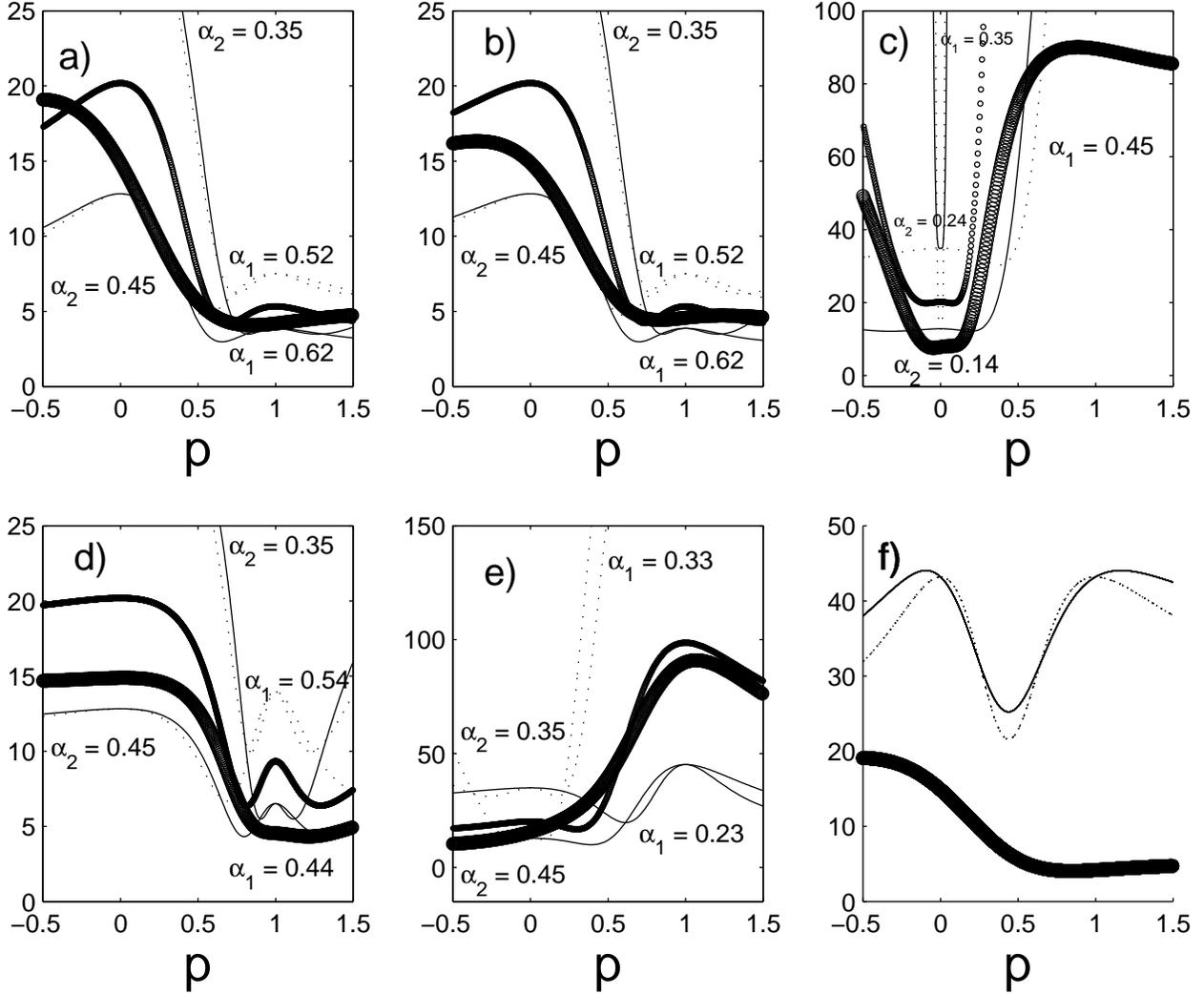}
\caption{\protect\label{figkurtests} Comparison of
the empirical excess kurtosis (fat solid line)
shown in Fig.~\ref{figVARKURT} for the five portfolios
to our theoretical prediction (\ref{gghJD/D}) with (\ref{jfzjmg}) for
uncorrelated assets (solid line).
The exponents $c_1$ and $c_2$ are those determined in the fits of the pdf's
tail,
as given in Fig.\ref{y(r)data}. The
thin solid lines and dashed lines plot the theoretical formula
(\ref{gghJD/D}) for values of the exponents $\alpha_i \equiv c_i/2 \pm 0.05$.
Figure c (RUR) is the same as a-b and d-e but the thin solid line gives the
predicted excess kurtosis for exponents $c_1 + 0.05, c_2 \pm 0.05$.
Figure f compares the empirical excess kurtosis (fat solid line)
of the portfolio CHF-JPY (Figure a) to the prediction (\ref{kjlskskl})
with (\ref{hqkqk}) for correlated assets with the fixed exponents
$c_1=c_2=2/3$. The thin solid line correspond to the empirical value $\rho
(y1,y2)= 0.57$
while the dashed line is obtained from the same formula with $\rho (y1,y2)=0$.
}
\end{center}
\end{figure}

\clearpage
\newpage

\begin{figure}
\begin{center}
\epsfig{file=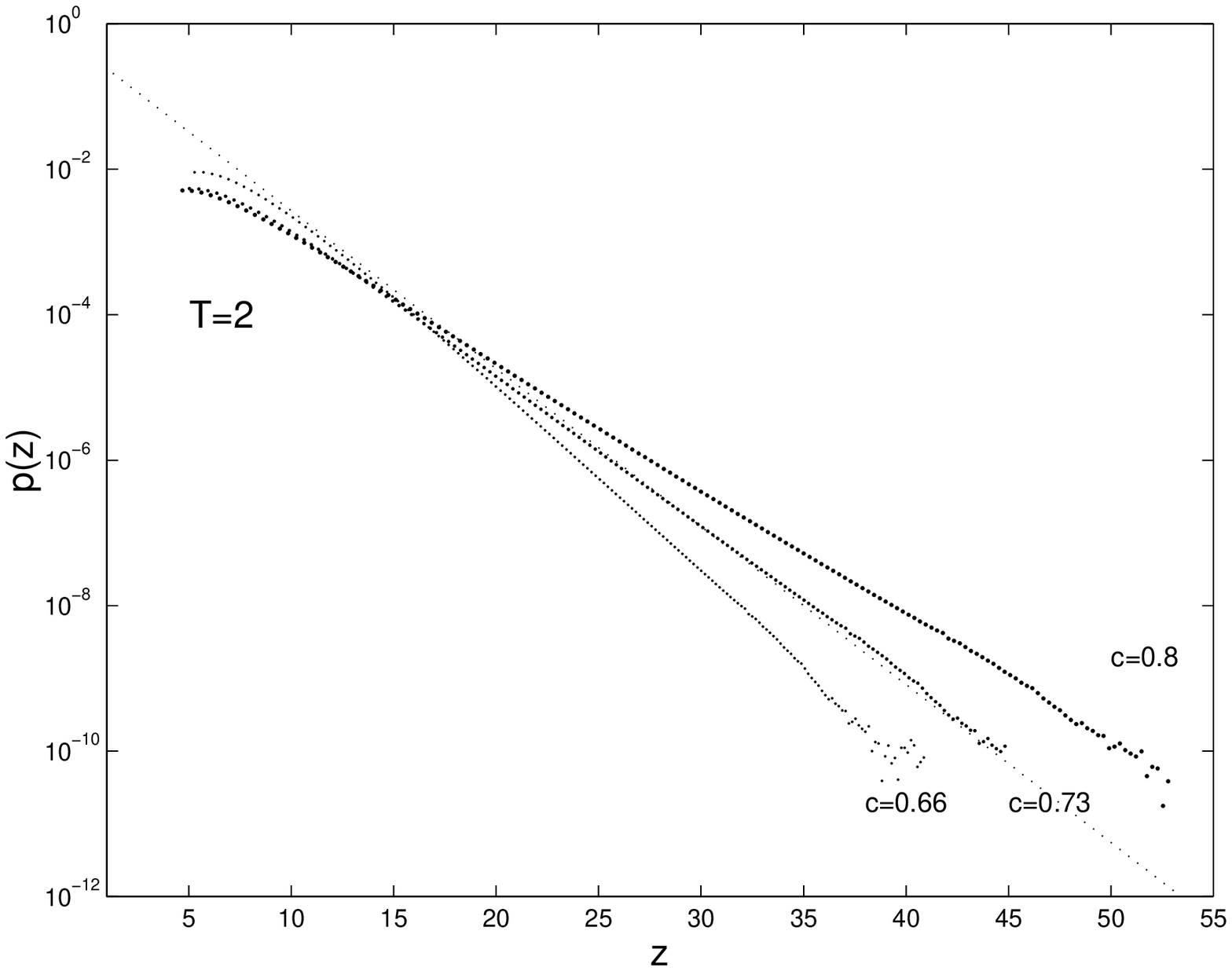}
\caption{\protect\label{figweibullstable}
Plots of the pdf's $P_T(y)$ as a function of $z \equiv y^2$ so that
a Gaussian (in the $y$ variable) is qualified as a straight line (dashed
line). In turn, by the
construction explained in the text, a straight line
qualifies a Weibull distribution. Here is shown the case $T=2$ for which
the best
$c_T$ is $c_2=0.73$. The other
curves allow one to estimate the sensitivity of the representation of $P_T$
in terms of
a Weibull as a function of the choice of the exponent $c_T$. The curves
have been normalized
by a coefficient $A_T^T$, with
$A_2=10$ for $c_2=0.66$ and $A_2=8$ for $c_2=0.73$ and $0.8$.
}
\end{center}
\end{figure}

\clearpage
\newpage

\begin{figure}
\begin{center}
\epsfig{file=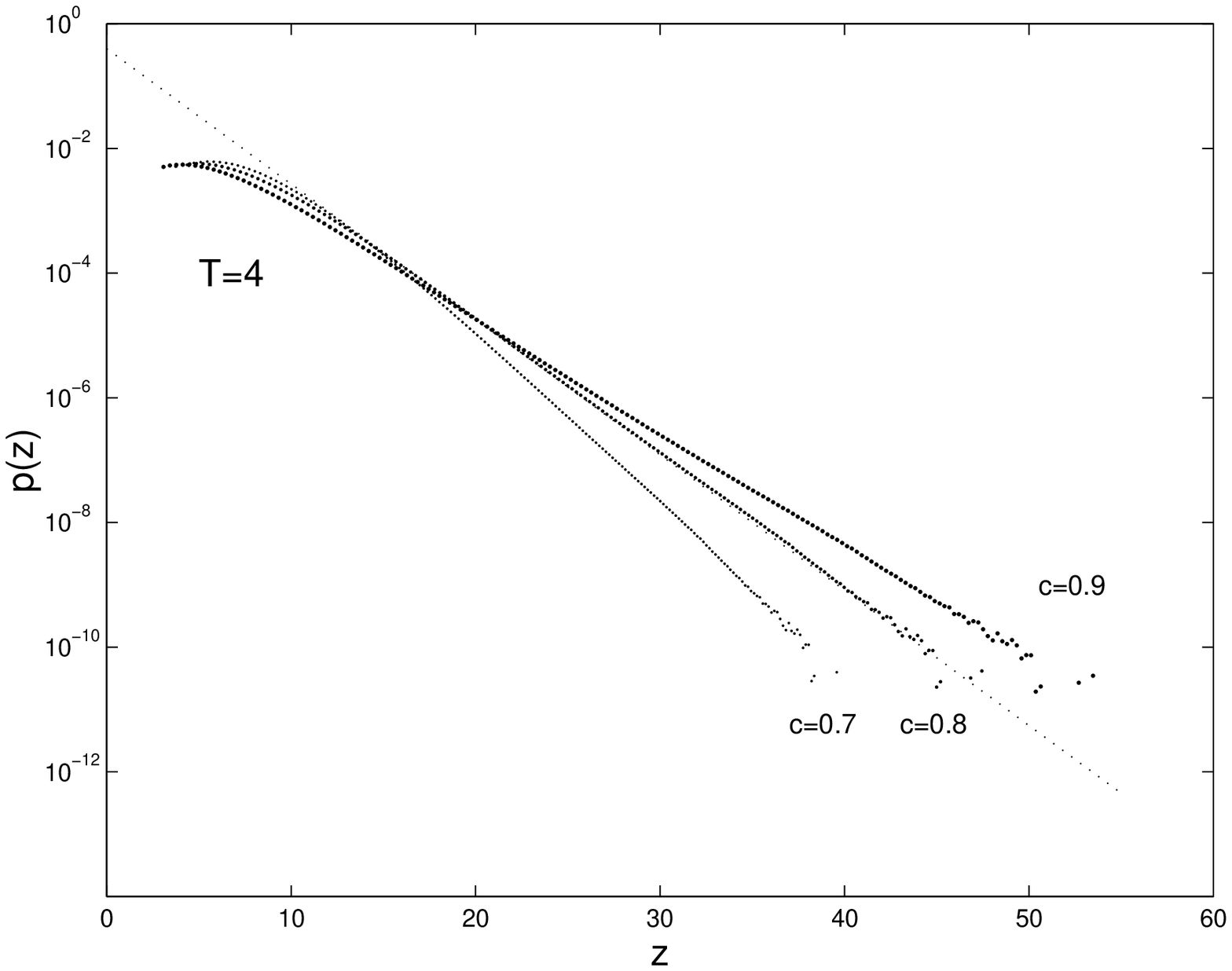}
\caption{\protect\label{figweibullstable1}
Plots of the pdf's $P_T(y)$ as a function of $z \equiv y^2$ so that
a Gaussian (in the $y$ variable) is qualified as a straight line (dashed
line). In turn, by the
construction explained in the text, a straight line
qualifies a Weibull distribution. Here is shown the case $T=4$ for which
the best
$c_T$ is $c_4=0.80$. The other
curves allow one to estimate the sensitivity of the representation of $P_T$
in terms of
a Weibull as a function of the choice of the exponent $c_T$. The curves
have been normalized
by a coefficient $A_T^T$, with
$A_4=25$ for all $c_4$'s.
}
\end{center}
\end{figure}

\clearpage
\newpage

\begin{figure}
\begin{center}
\epsfig{file=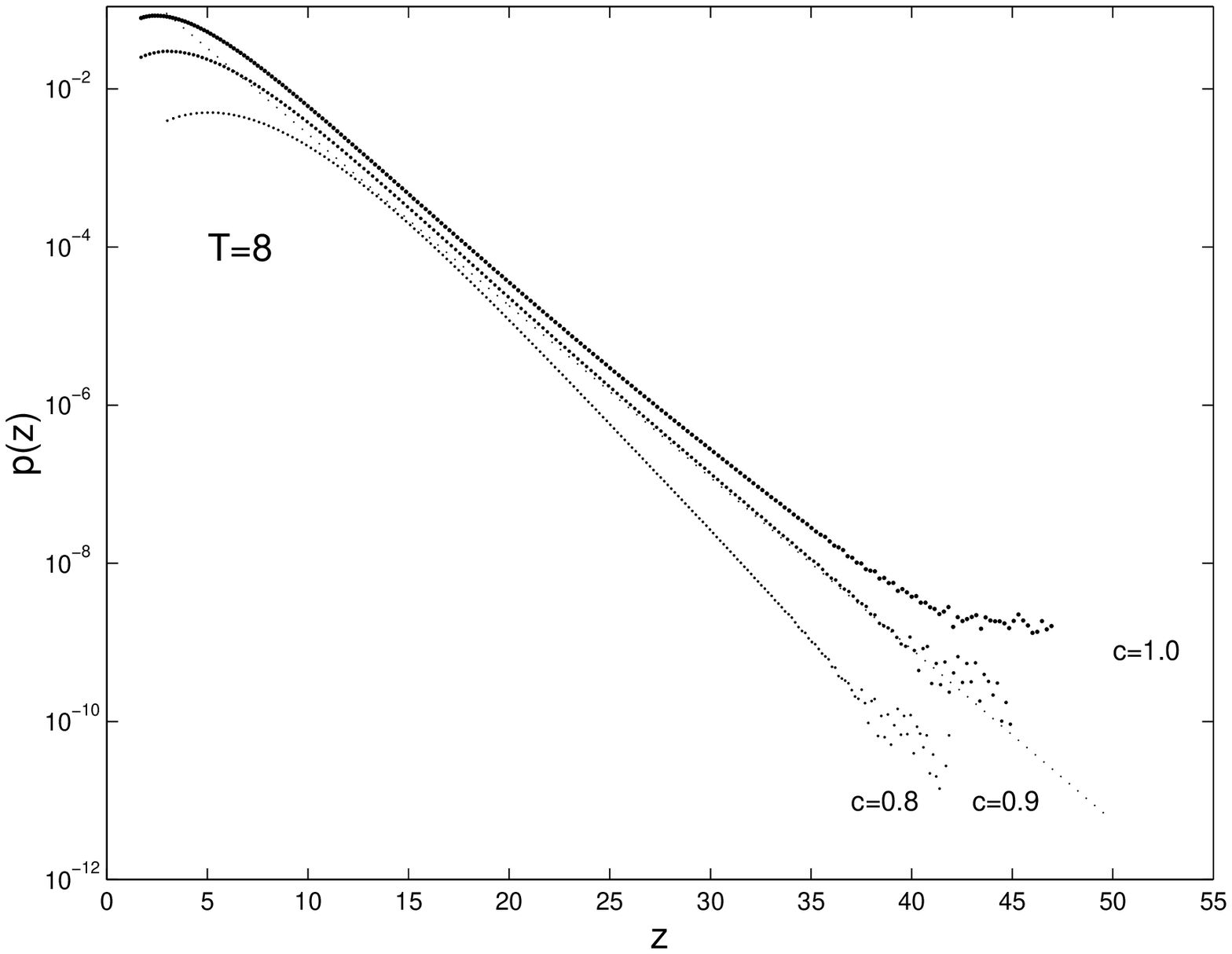}
\caption{\protect\label{figweibullstable2}
Plots of the pdf's $P_T(y)$ as a function of $z \equiv y^2$ so that
a Gaussian (in the $y$ variable) is qualified as a straight line (dashed
line). In turn, by the
construction explained in the text, a straight line
qualifies a Weibull distribution. We show here the case $T=8$ for which the
best
$c_T$ is $c_8=0.90$. The other
curves allow one to estimate the sensitivity of the representation of $P_T$
in terms of
a Weibull as a function of the choice of the exponent $c_T$. The curves
have been normalized
by a coefficient $A_T^T$, with
$A_8=40$ for $c_8=0.8$ and $1.0$ and $A_8=360$ for $c_8=0.9$.
}
\end{center}
\end{figure}

\clearpage
\newpage

\begin{figure}
\begin{center}
\epsfig{file=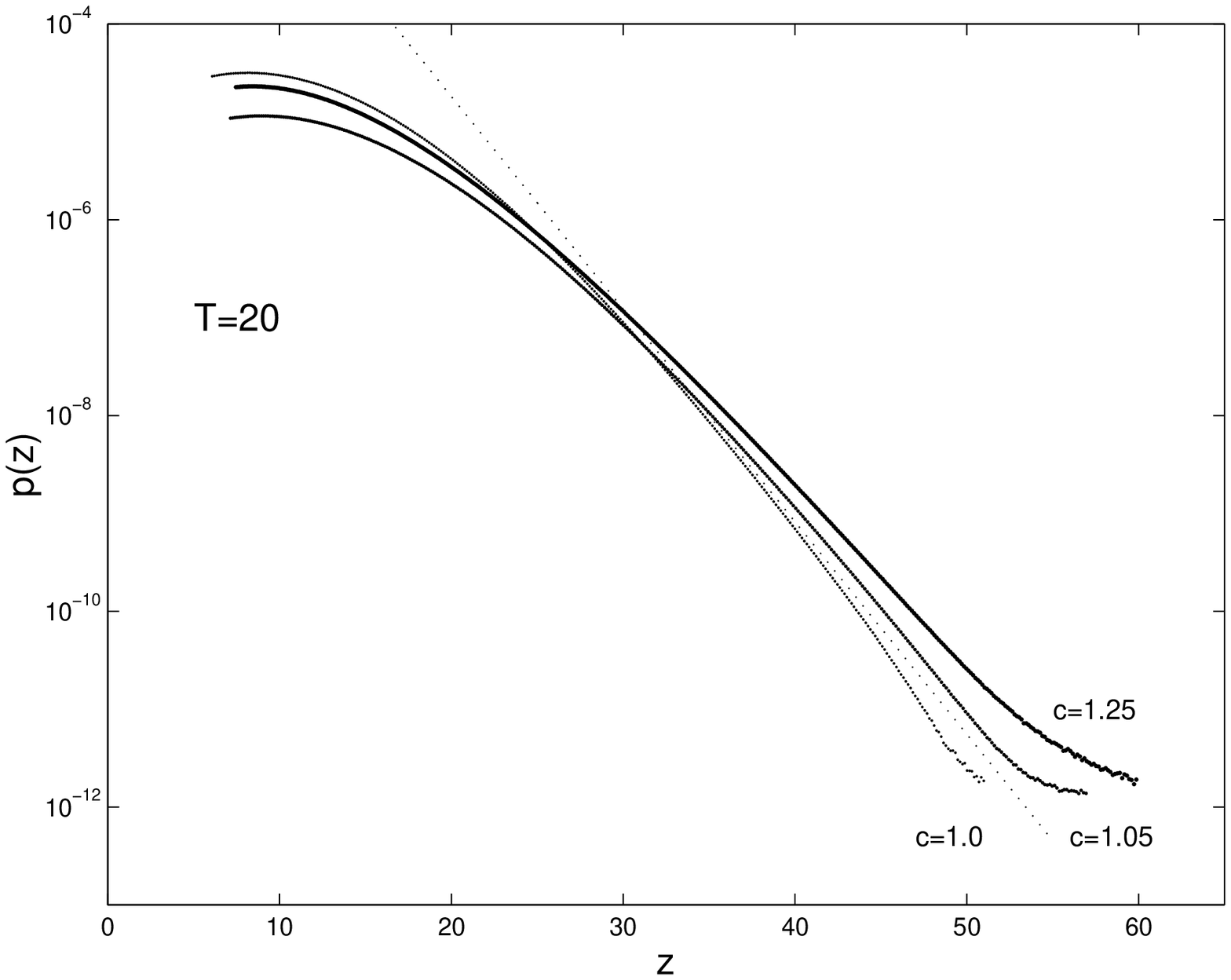}
\caption{\protect\label{figweibullstable3}
Plots of the pdf's $P_T(y)$ as a function of $z \equiv y^2$ so that
a Gaussian (in the $y$ variable) is qualified as a straight line (dashed
line). In turn, by the
construction explained in the text, a straight line
qualifies a Weibull distribution. We show here the case $T=20$ for which
the best
$c_T$ is $c_{20}\approx 1.05$. The other
curves allow one to estimate the sensitivity of the representation of $P_T$
in terms of
a Weibull as a function of the choice of the exponent $c_T$. The curves
have been normalized
by a coefficient $A_T^T$, with
$A_{20}=4.7~10^4$ for $c_{20}=1.0$, $A_{20}=3.5~10^5$ for $c_{20}=1.05$ and
$A_{20}=7~10^7$ for $c_{20}=1.25$.
}
\end{center}
\end{figure}

\clearpage
\newpage

\begin{figure}
\begin{center}
\epsfig{file=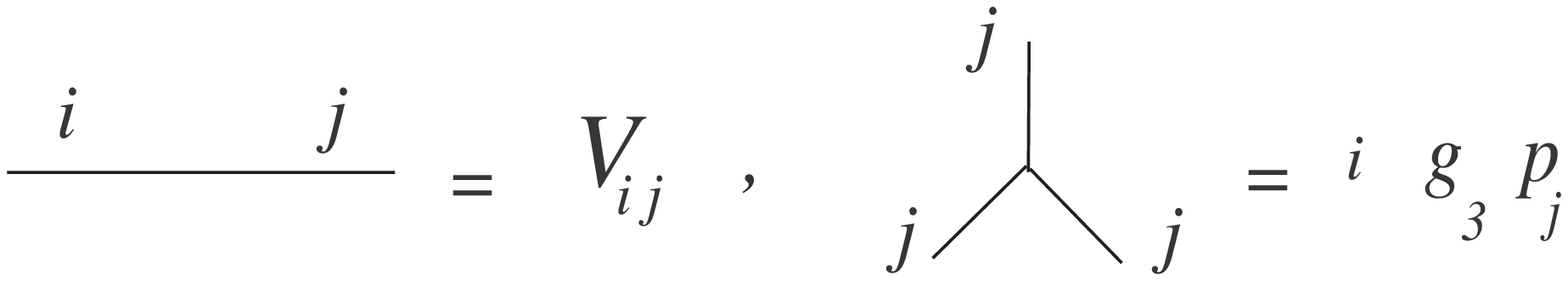,  width=14cm, height=20cm}
\caption{\protect\label{D1} We associate to each factor $V_{ij}$ the
propagator diagram and to each
factor $i g_3 w_j$ the vertex diagram as shown in this figure, where we
have defined
the coupling constant $g_3 = 3! k$.}
\end{center}
\end{figure}

\clearpage
\newpage

\begin{figure}
\begin{center}
\epsfig{file=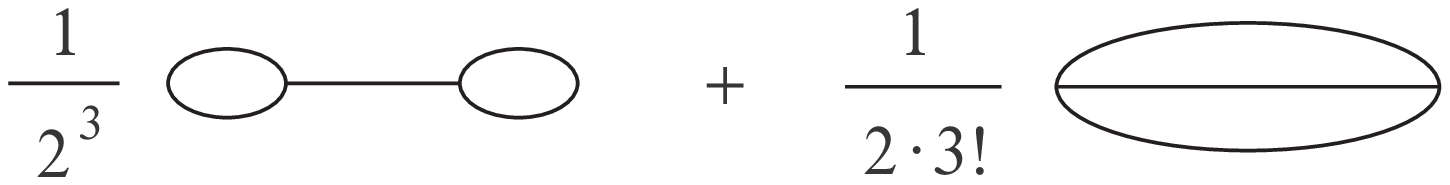,  width=14cm, height=3cm}
\caption{\protect\label{D2} The two contributions in (\ref{qml,klSD}) are
represented
by propagators connecting the vertices as shown in the figure.}
\end{center}
\end{figure}

\clearpage
\newpage

\begin{figure}
\begin{center}
\epsfig{file=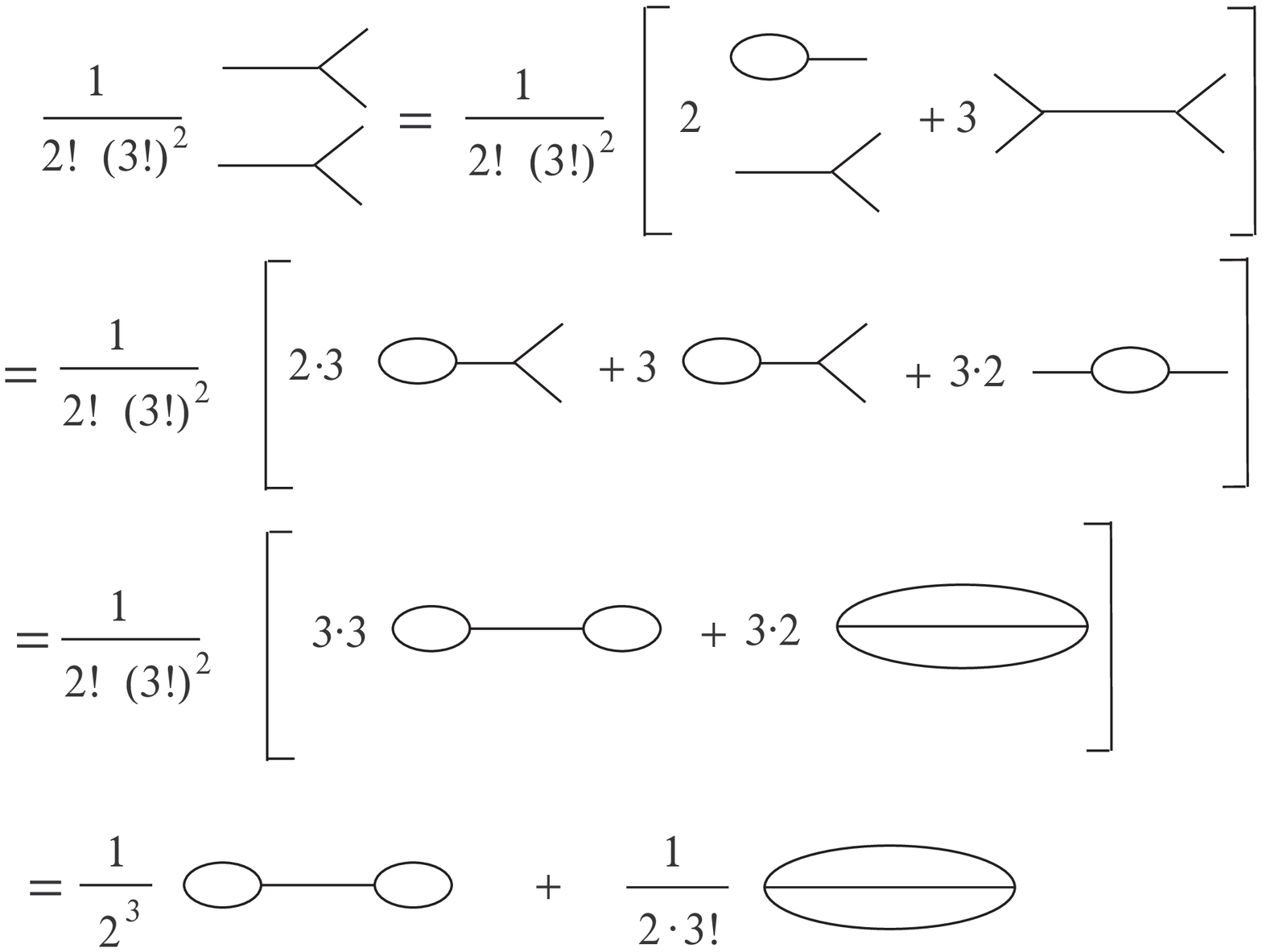,  width=14cm, height=20cm}
\caption{\protect\label{D3} A systematic way to  keep under control the
symmetry factor is to compute
it diagramatically as shown in this figure. The second-order derivative
operator with respect to
$J_i$ and $J_j$ is  represented by the two vertices in the
left hand side of the figure.
The $J$-independent term is given by pairwise combining each leg of each vertex
in all the possible topologically inequivalent way, taking  into account
the multiplicity of each configuration. The figure shows how to proceed.
The coefficient in front of  the vertices of the
left hand side comes from the perturbative expansion.
As a first step, let us consider the first leg of the first  vertex.
We can either contract it with another leg (two possibile
contractions) of the same vertex or with a leg
of the second vertex (three possible contractions). This is summarized in
the first equality of the figure.}
\end{center}
\end{figure}

\clearpage
\newpage

\begin{figure}
\begin{center}
\epsfig{file=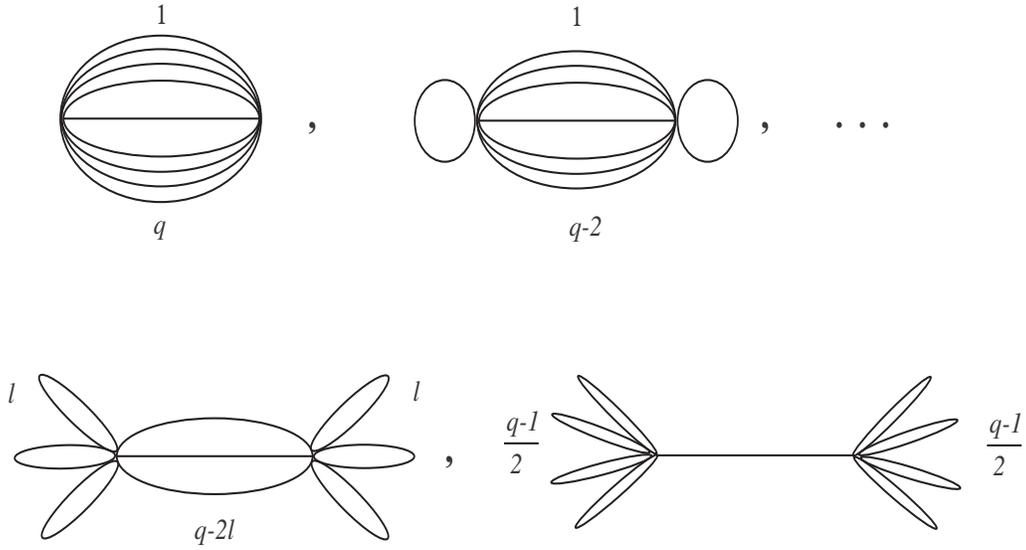,  width=14cm, height=20cm}
\caption{\protect\label{D4} Sequence of all diagrams obtained by combining
pairwise each leg of two vertices  to form all the possible
topologically inequivalent diagrams.
Each diagram is characterized by the number $l$ of loops on each vertex and
the number
$(q-2) l$ of lines connecting the two vertices giving the contribution
(\ref{yhdjqmqmmq})
where each loop around vertex $i$ contributes to a factor $V_{ii}$ and each
propagator
connecting the vertices $i$ and $j$ gives a factor $V_{ij}$.}
\end{center}
\end{figure}

\clearpage
\newpage

\begin{figure}
\begin{center}
\epsfig{file=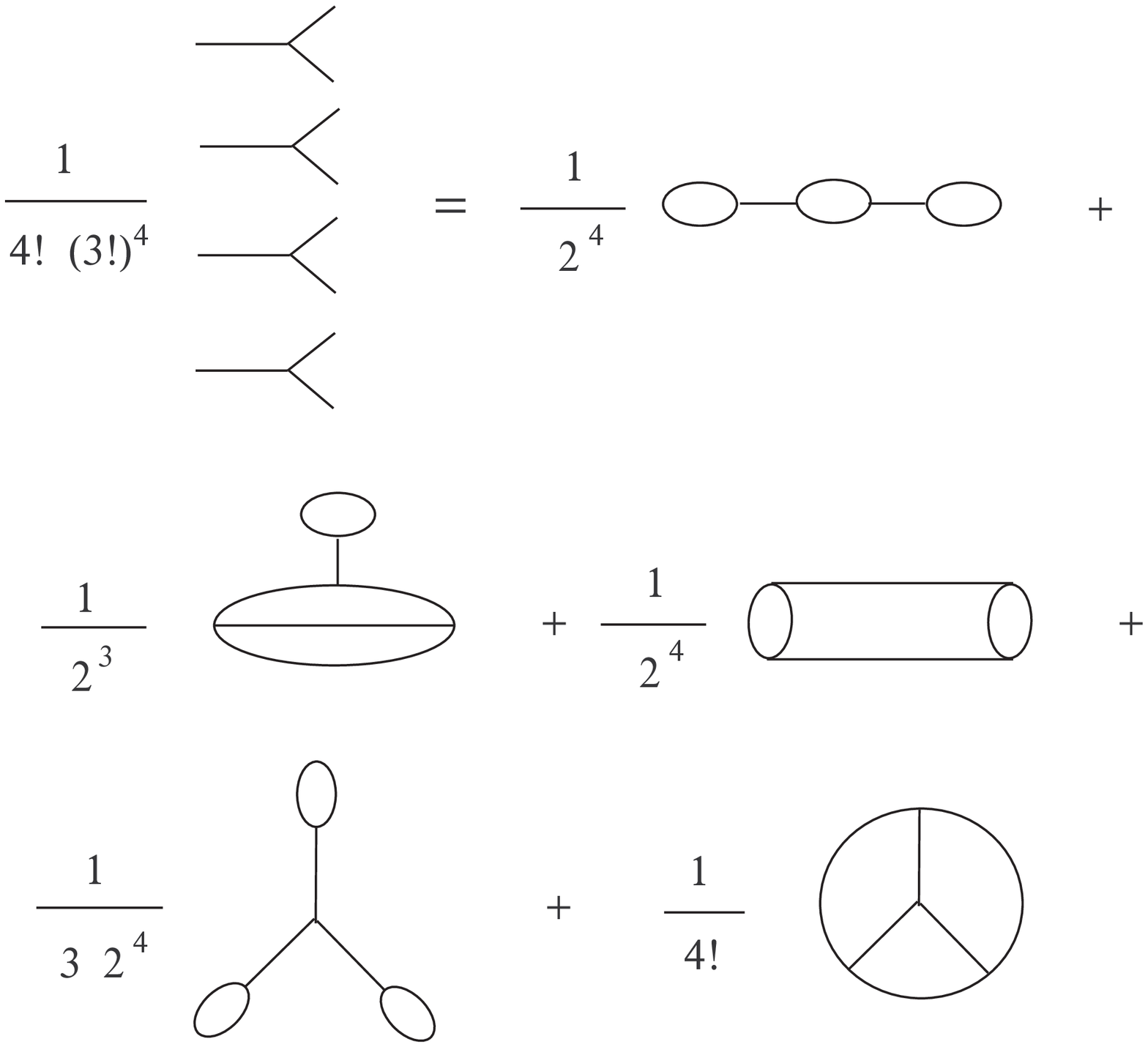,  width=14cm, height=20cm}
\caption{\protect\label{D5} Contraction procedure giving all connected
diagrams of
$4$-th order that contribute directly to the $4$-th cumulant coefficient
$c_4(3)$
(\ref{fjqmmqkqqmm}).}
\end{center}
\end{figure}

\clearpage
\newpage

\begin{figure}
\begin{center}
\epsfig{file=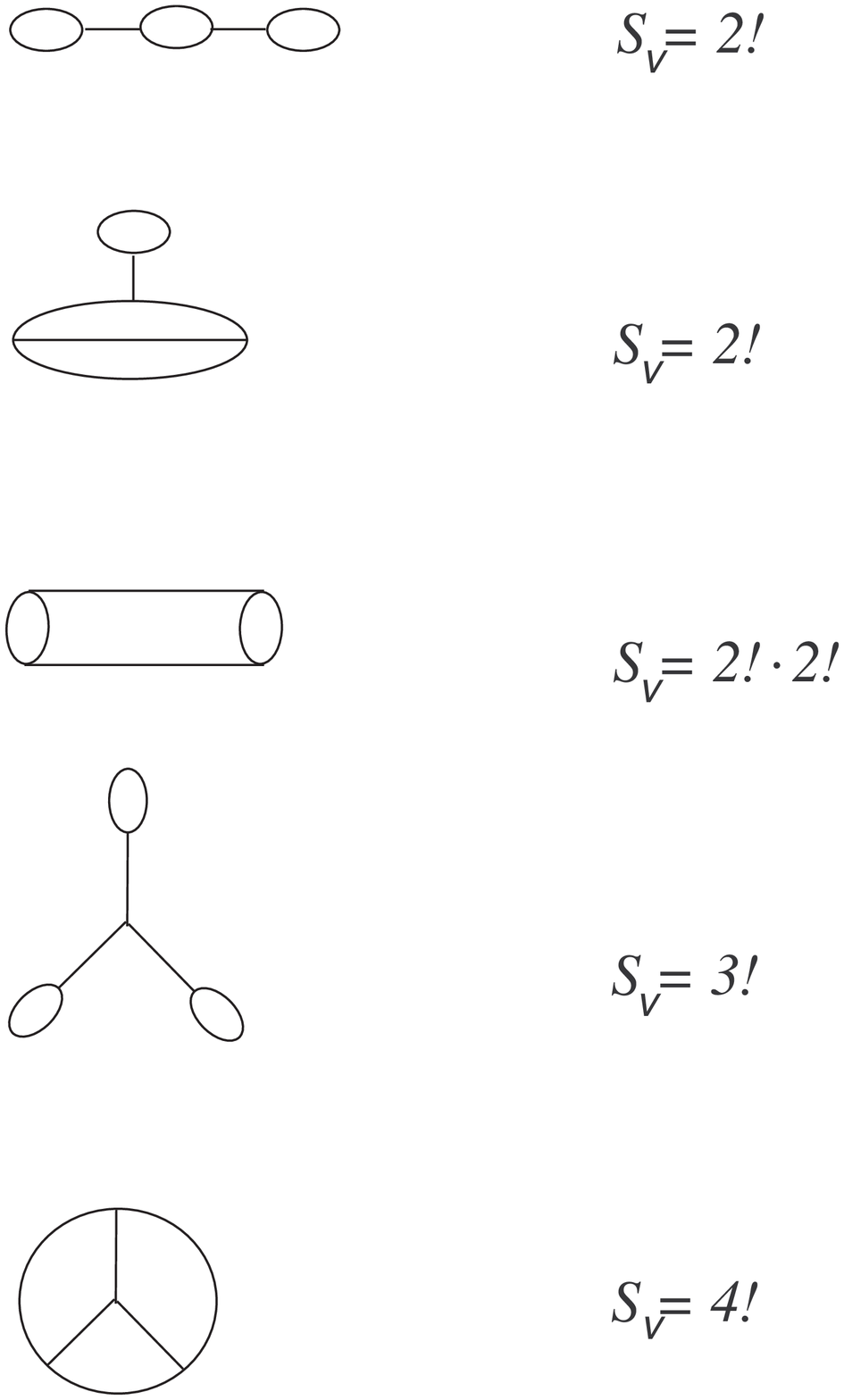,  width=14cm, height=20cm}
\caption{\protect\label{D6} Summary of
 the vertex symmetry factors for the diagrams contributing to $c_4(3)$.}
\end{center}
\end{figure}

\end{document}